%% file: s2let_localisation.tex
\documentclass[3p, sort&compress, a4paper]{elsarticle}

\usepackage{amssymb}
\usepackage{amsmath}
\usepackage{mathtools}
\DeclarePairedDelimiter{\ceil}{\lceil}{\rceil}
\DeclarePairedDelimiter{\floor}{\lfloor}{\rfloor}
\usepackage{graphicx}
\usepackage{subfigure}
\usepackage{color}
\usepackage{booktabs}
\usepackage{grffile}

\usepackage{ifpdf}
\ifpdf
  \usepackage[pdftex,colorlinks=true,breaklinks=true,citecolor=blue]{hyperref}
\else
  \usepackage[ps2pdf,colorlinks=true,citecolor=blue]{hyperref}
\fi

\include{macros}

\renewcommand{\eqn}[1]{Eq.~(#1)}

\renewcommand{\appn}[1]{#1}

\renewcommand{\sectn}[1]{Section~#1}

\renewcommand{\wav}{\ensuremath{\Psi}}
\renewcommand{\swav}{\ensuremath{{}_\spin\Psi}}
\renewcommand{\mmax}{\ensuremath{{N}}}
\renewcommand{\exp}[1]{\ensuremath{{\rm exp}{#1}}}
\newcommand{\expv}{\ensuremath{\mathbb{E}}}
\newcommand{\Toperator}{\ensuremath{\mathcal{T}}}
\renewcommand{\wavsteer}{\ensuremath{\zeta}}
\renewcommand{\wscalemin}{\ensuremath{0}}
\renewcommand{\dilparam}{\ensuremath{{\lambda}}}

\renewcommand{\sshtcode}{{\sc ssht}}
\renewcommand{\sothreecode}{{\sc so3}}
\renewcommand{\stwoletcode}{{\sc s2let}}
\renewcommand{\healpix}{{\sc healpix}}

\begin{document}

%=============================================================================

\begin{frontmatter}

\title{Localisation of directional scale-discretised wavelets on the
  sphere}

%\author[mssl]{Jason~D.~McEwen\corref{corauth}\fnref{fn1}}
\author[mssl]{Jason~D.~McEwen\fnref{fn1}}
\ead{jason.mcewen@ucl.ac.uk}

\author[rome]{Claudio~Durastanti\fnref{fn2}}
\ead{claudio.durastanti@gmail.com}

\author[hwu]{Yves~Wiaux}
\ead{y.wiaux@hw.ac.uk}

\address[mssl]{Mullard Space Science Laboratory, University College
  London, Surrey RH5 6NT, UK}

\address[rome]{University of Tor
  Vergata, Rome, Italy and Ruhr University, Bochum, Germany}

\address[hwu]{Institute of Sensors, Signals, and Systems, Heriot-Watt
  University, Edinburgh EH14 4AS, UK}

%\cortext[corauth]{Corresponding author}

\fntext[fn1]{Supported by the Engineering and Physical Sciences Research Council (grant number EP/M011852/1).}
\fntext[fn2]{Partially supported by E.R.C. Grant n.277742 PASCAL and  DFG-GRK 2131. }

%=============================================================================

\begin{abstract}
  Scale-discretised wavelets yield a directional wavelet framework on
  the sphere where a signal can be probed not only in scale and
  position but also in orientation.  Furthermore, a signal can be
  synthesised from its wavelet coefficients exactly, in theory and
  practice (to machine precision).  Scale-discretised wavelets are
  closely related to spherical needlets (both were developed
  independently at about the same time) but relax the axisymmetric
  property of needlets so that directional signal content can be
  probed.  Needlets have been shown to satisfy important
  quasi-exponential localisation and asymptotic uncorrelation
  properties.  We show that these properties also hold for directional
  scale-discretised wavelets on the sphere and derive similar
  localisation and uncorrelation bounds in both the scalar and spin
  settings.  Scale-discretised wavelets can thus be considered as
  directional needlets. 
\end{abstract}

\begin{keyword}
wavelets; needlets; harmonic analysis on the sphere; cosmic microwave background
\end{keyword}

\end{frontmatter}

%=============================================================================
\section{Introduction}
%=============================================================================

Wavelet methodologies on the sphere are not only of considerable
theoretical interest in their own right but also have important
practical application.  For example, wavelets analyses on the sphere
have led to many insightful scientific studies in the fields of
planetary science \cite[\eg][]{audet:2010, audet:2014}, geophysics
\cite[\eg][]{simons:2011, simons:2011a, charlety:2012, loris:2012}
and cosmology, in particular for the analysis of the cosmic microwave
background (CMB) \cite[\eg][]{barreiro:2000, vielva:2004, mcewen:2005:ng,
  mcewen:2006:ng, mcewen:2008:ng, mcewen:2006:bianchi, vielva:2005,
  mcewen:2006:isw, mcewen:2007:isw2, vielva:2006, wiaux:2008,
  wiaux:2006, lan:2008, pietrobon:2006,
  planck2013-p06, planck2013-p09, planck2013-p09a, planck2013-p20,
  schmitt:2011, bobin:2013, delabrouille:2009} (for a somewhat dated
review see \cite{mcewen:2006:review}), among others.  Of particular
importance in such applications is the scale-space trade-off of the
wavelets adopted, which arises from the extension of the familiar
(Euclidean) Fourier uncertainly principle to the sphere
\cite{narcowich:1996}.  Consequently, characterising the localisation
properties of wavelets on the sphere is of considerable interest.

Many early attempts to extend wavelet transforms to the sphere differ
primarily in the manner in which dilations are defined on the sphere
\cite{narcowich:1996, potts:1995, freeden:1997a,   torresani:1995,
dahlke:1996, holschneider:1996, antoine:1999,   antoine:1998, sanz:2006,
mcewen:2006:cswt2}.   
The construction of Freeden \& Windheuser \cite{freeden:1997a} is
based on singular integrals on the sphere, while Antoine and Vandergheynst
\cite{antoine:1999, antoine:1998} follow a group theoretic approach.  In the
latter construction dilation is defined via the stereographic projection of
the sphere to the plane, leading to a consistent framework that reduces
locally to the usual continuous wavelet transform in the Euclidean limit. An
implementation and technique to approximate functions on the sphere is
developed for this approach in \cite{antoine:2002}.  This construction is
revisited in \cite{wiaux:2005}, independently of the original group theoretic
formalism, and fast algorithms are developmented in
\cite{wiaux:2005b,wiaux:2005c}.

Initial wavelet
constructions were essentially based on continuous methodologies,
which, although insightful, limited practical application to problems
where the exact synthesis of a function from its wavelet coefficients
is not required.  Early discrete constructions \cite{schroder:1995,
  sweldens:1997, barreiro:2000} (and subsequently \cite{mcewen:2008:fsi, mcewen:szip}) that
support exact synthesis were built on particular pixelisations of
the sphere and do not necessarily lead to stable bases
\cite{sweldens:1997}. 
Half-continuous and fully discrete frames based on the continuous framework of \cite{antoine:1999, antoine:1998} were constructed by \cite{bogdanova:2004,bogdanova:2004b} and polynomial frames were constructed by \cite{mhaskar:2000}.
 More recently, a number of a exact discrete wavelet
frameworks on the sphere have been developed, with underlying continuous representations and fast implementations that have been made available publicly, including:
 needlets \cite{narcowich:2006, baldi:2009, marinucci:2008};
directional scale-discretised wavelets \cite{wiaux:2007:sdw,
  leistedt:s2let_axisym, mcewen:2013:waveletsxv}; and the isotropic
undecimated and pyramidal wavelet transforms \cite{starck:2006}.  Each
approach has also been extended to analyse spin functions on the
sphere \cite{geller:2008, geller:2010:sw, geller:2010, geller:2009_bis, mcewen:s2let_spin,
  mcewen:s2let_spin_sccc21_2014, starck:2009} and functions defined on
the three-dimensional ball formed by augmenting the sphere with the
radial line \cite{durastanti:2014, leistedt:flaglets,
  mcewen:flaglets_sampta, lanusse:2012}.

%-----------------------------------------------------------------------------
\subsection{Contribution}

Needlets \cite{narcowich:2006, baldi:2009, marinucci:2008} and
directional scale-discretised wavelets \cite{wiaux:2007:sdw,
  leistedt:s2let_axisym, mcewen:2013:waveletsxv} on the sphere were
developed independently, about the same time, but share many
similarities. Both are essentially constructed by a Meyer-type tiling
of the line defined by spherical harmonic degree $\el$.  Directional
scale-discretised wavelets in addition include a directional component
in the wavelet kernel, yielding a directional wavelet analysis so that
signal content can be probed not only in scale and position but also
in orientation.  Needlets have been shown to satisfy important
quasi-exponential localisation and asymptotic uncorrelation properties
\cite{narcowich:2006, baldi:2009, marinucci:2008, pietrobon:2010,
  geller:2008, geller:2010:sw,  geller:2010}.  
In this article we show that these properties also hold for directional scale-discretised wavelets.  We derive equivalent localisation and uncorrelation
bounds, in both the scalar and spin settings, and show that directional scale-discretised wavelets are characterised by excellent localisation properties in the spatial domain.

More precisely, we prove that for any $\xi \in \realsnz$, there exists strictly positive constants $C_1, C_2\in
\realsnz$, such that the directional scale-discretised wavelet $\wav \in
\ltwo(\sphere)$, defined on the sphere \sphere\ and centred on the North pole, satisfies the localisation bound:
% \begin{equation}
% \bigl\vert \wav ^{(\wscale ) }( \sas ) \bigr\vert \leq 
% \frac{\bigl(L\lambda^{-j}\bigr)^{ 2+\mmax}C_{\xi ,\mmax}}{\bigl( 1+L\lambda^{-j}\theta \bigr) ^{\xi }%
% }\spcend ,
% \end{equation}
\begin{equation}
\bigl\vert \wav ( \sas ) \bigr\vert \leq 
\frac{C_1}
{\bigl( 1+C_2\theta \bigr) ^{\xi }%
}\spcend ,
\end{equation}
where $(\sas)\in\sphere$ denote spherical coordinates, with
colatitude $\saa \in [0,\pi]$ and longitude $\sab \in [0,2\pi)$.
Furthermore, we prove that for Gaussian random fields on the sphere, directional scale-discretised wavelet coefficients are asymptotically uncorrelated.  The correlation of wavelet coefficients corresponding to wavelets at scales $j, j^\prime \in \naturals$ and centred on Euler angles $\eul_1, \eul_2 \in \sothree$, respectively, parameterising the rotation group \sothree, is denoted $\Xi ^{(j j^\prime)}(\eul_1,\eul_2)$.  We show that for any $j, j^\prime \in \naturals$ such that
$\left|j-j^\prime\right|<2$ and for any $\xi \in \realsnz$,
$\xi\geq 2\mmax $ (where $\mmax$ is the azimuthal band-limit of the wavelet), there exists $C_1^{(\wscale)}, C_2^{(\wscale)}\in
\realsnz$ such that the directional wavelet correlation satisfies the bound:
\begin{equation}
\Xi ^{(j j^\prime)}(\eul_1,\eul_2) \leq 
\frac{C_1^{(\wscale)}}
{\bigl( 1+ C_2^{(\wscale)} \eulb \bigr) ^{\xi}}
\spcend,
\end{equation}
where $\eulb\in[0,\pi)$ is an angular separation between $\eul_1$ and $\eul_2$.  For $\left|j-j^\prime\right|\geq 2$, wavelet coefficients are exactly uncorrelated, \ie\ \mbox{$\Xi ^{(j j^\prime)}(\eul_1,\eul_2) =0$}.
We present an overview of these results here only; precise definitions of the quantities involved and more specific bounds, showing the dependence on the parameterisation of the scale-discretised wavelet construction, are presented and derived in the main body of the article.

The characterisation of the localisation properties of directional scale-discretised wavelets presented is of considerable importance for applications, in particular for the analysis of the CMB, which to very good approximation is a realisation of a Gaussian random field on the sphere.   
Directional wavelets are useful for the analysis of the CMB, since, although the CMB is globally isotropic, its peaks are predicted to be elongated \cite{bond:1987}.  Furthermore, weak anisotropic signals with strong directional features are embedded in raw CMB observations (\eg\ due to foreground contamination; \cite{planck2013-p06}).
Numerical experiments using simulated CMB observations are presented to demonstrate both the localisation and uncorrelation properties of scale-discretised wavelets.
All results are also extended to spin scale-discretised wavelets \cite{mcewen:s2let_spin} and so are applicable not only to CMB temperature observations (a scalar signal on the sphere) but also to observations of CMB polarisation (a spin $\pm 2$ signal on the sphere).
In addition to the derivation of the results summarised above, which constitute the main contributions of this article, for the first time, we also explicitly show that scale-discretised wavelets form a tight frame on the sphere and present the detailed derivation of their directional construction.
Since directional scale-discretised wavelets satisfy similar localisation and uncorrelation properties to needlets, and follow a similar construction but extended to a directional analysis, they can thus be considered as directional needlets.

%-----------------------------------------------------------------------------
\subsection{Outline}

The remainder of this article is structured as follows.  In
\sectn{\ref{sec:wavelet_transform}} the directional scale-discretised
wavelet transform on the sphere is reviewed and we show explicitly
that scale-discretised wavelets form a tight frame.  The construction
of scale-discretised wavelets is reviewed in
\sectn{\ref{sec:wavelet_construction}}, where the detailed derivation of their directional construction is elaborated for the first time; their directional
correlation and steerability properties are also reviewed.  The main contributions of this article are presented in \sectn{\ref{sec:localisation_properties}} and
\sectn{\ref{sec:stochastic_properties}}, where the
quasi-exponential localisation and asymptotic uncorrelation properties
of directional scale-discretised wavelets are proved, respectively.  Technical
results and additional
mathematical background are deferred to \appn{\ref{sec:appendix}}.  \appn{\ref{sec:appendix:special_functions}} reviews harmonic analysis on the sphere \sphere\ and rotation group \sothree\ concisely, focusing on definitions and approximations of relevant special functions and their properties, which are used throughout the article. The remaining appendices (\appn{\ref{sec:appendix:general_localisation}}--\appn{\ref{sec:appendix:bound_u}}) present calculations on which the proofs of \sectn{\ref{sec:localisation_properties}} and \sectn{\ref{sec:stochastic_properties}} rely.  Numerical experiments demonstrating the localisation and uncorrelation
properties of directional scale-discretised wavelets are performed in
\sectn{\ref{sec:numerical_experiments}}.  
% Concluding remarks are made in \sectn{\ref{sec:conclusions}}.

%=============================================================================
\section{Scale-discretised wavelet transform}
\label{sec:wavelet_transform}
%=============================================================================

The directional scale-discretised wavelet transform supports the
analysis of oriented spatially localised, scale-dependent features in
signals on the sphere.  In this section we review the
scale-discretised wavelet framework on the sphere
\cite{wiaux:2007:sdw, leistedt:s2let_axisym, mcewen:2013:waveletsxv}, following closely the presentation of \cite{mcewen:2013:waveletsxv}, describing wavelet analysis and synthesis, and admissibility and tight
frame properties.  For clarity we present the scalar setting only,
however the scale discretised wavelet transform on the sphere has been
extended recently to support spin signals \cite{mcewen:s2let_spin}.
The concentration properties of scale-discretised wavelets
derived in subsequent sections of this article hold in both the scalar
and spin settings.

%-----------------------------------------------------------------------------
\subsection{Analysis}

The scale-discretised wavelet transform of a function $\f \in
\ltwo(\sphere)$ on the sphere \sphere\ is defined by the directional
convolution of \f\ with the wavelet $\wav^{(\wscale)} \in
\ltwo(\sphere)$.  The wavelet coefficients $\wcoeff^{\wav^{(\wscale)}} \in
\ltwo(\sothree)$ thus read
\begin{equation}
  \label{eqn:analysis}
  \wcoeff^{\wav^{(\wscale)}}(\eul) \equiv ( \f \circledast \wav^{(\wscale)}) (\eul)
  \equiv \innerp{\f}{\rotarg{\eul}\wav^{(\wscale)}}
  = \int_\sphere \dmu{\sa} \f(\sa) (\rotarg{\eul}\wav^{(\wscale)})^\cconj(\sa)
  \spcend ,
\end{equation}
where $\sa=(\sas) \in \sphere$ denotes spherical coordinates with
colatitude $\saa \in [0,\pi]$ and longitude $\sab \in [0,2\pi)$,
$\dmu{\sa} = \sin\saa \dx\saa \dx\sab$ is the usual rotation invariant
measure on the sphere, and $\cdot^\cconj$ denotes complex conjugation.
The inner product of functions on the sphere is denoted
$\innerp{\cdot}{\cdot}$, while the operator $\circledast$ denotes
directional convolution on the sphere.
The rotation operator is defined by 
\begin{equation}
  (\rotarg{\eul} \wav^{(\wscale)})(\sa) \equiv \wav^{(\wscale)}(\rotmatarg{\eul}^{-1} \vect{\hat{\sa}})
  \spcend ,
\end{equation}
where $\rotmatarg{\eul}$ is the three-dimensional rotation matrix
corresponding to $\rotarg{\eul}$ and $\vect{\hat{\sa}}$ denotes the
Cartesian vector corresponding to $\sa$.  Rotations are specified by elements
of the rotation group $\sothree$, parameterised by the Euler angles
$\eul=(\euls) \in \sothree$, with $\eula \in [0,2\pi)$, $\eulb \in
[0,\pi]$ and $\eulc \in [0,2\pi)$.  
The wavelet transform of \eqn{\ref{eqn:analysis}} thus probes directional
structure in the signal of interest \f, where \eulc\ can be viewed as
the orientation about each point on the sphere $(\sas) = (\eulb,
\eula)$.
The wavelet scale $\wscale \in \naturals$ encodes the angular
localisation of $\wav^{(\wscale)}$, as discussed in more detail
subsequently.  Note that the wavelet scales $\wscale$ are discrete
(hence the name scale-discretised wavelets), which affords the exact
synthesis of a function from its wavelet (and scaling) coefficients.

The wavelet coefficients encode only the detail-information contained
in the signal \f; scaling coefficients must be introduced to represent
the approximation-information of the signal, \ie\ low-frequency signal
content.  The scaling coefficients $\scoeff^\wavs \in \ltwo(\sphere) $
are given by the convolution of \f\ with the axisymmetric scaling
function $\wavs \in \ltwo(\sphere)$ and read
\begin{equation}
  \label{eqn:analysis_scaling}
  \wcoeff^{\wavs}(\sa) \equiv ( \f \odot \wavs) (\sa)
  \equiv \innerp{\f}{\rotarg{\sa}\wavs}
  = \int_\sphere \dmu{\sa\p} \f(\sa\p) (\rotarg{\sa}\wavs)^\cconj(\sa\p)
  \spcend ,
\end{equation}
where $\rotarg{\sa} = \rot_{(\sab,\saa, 0)}$ and the operator $\odot$
denotes axisymmetric convolution on the sphere.  Note that the scaling
coefficients live on the sphere, and not the rotation group \sothree,
since directional structure of the approximation-information of \f\ is
not typically of interest.

%-----------------------------------------------------------------------------
\subsection{Synthesis}

The signal \f\ can be synthesised perfectly from its wavelet and
scaling coefficients by
\begin{equation}
  \label{eqn:synthesis}
  \f(\sa) = \int_\sphere \dmu{\sa\p} 
  \scoeff^\wavs(\sa\p) (\rotarg{\sa\p} \wavs)(\sa)
  +
  \sum_{\wscale=\wscalemin}^\wscalemax \int_\sothree \deul{\eul}
  \wcoeff^{\wav^{(\wscale)}}(\eul) (\rotarg{\eul} \wav^\wscale)(\sa)
  \spcend ,
\end{equation}
where
$\deul{\eul} = \sin\eulb \dx\eula \dx\eulb \dx\eulc$ is the usual
invariant measure on \sothree\ and $\wscalemin$ and $\wscalemax$ are
the minimum and maximum wavelet scales considered, respectively.
We adopt the same convention as \cite{wiaux:2007:sdw} for the wavelet
scales \wscale, with increasing \wscale\ corresponding to larger
angular scales, \ie\ lower frequency content.\footnote{Note that this
  differs to the convention adopted in
  \cite{leistedt:s2let_axisym,mcewen:s2let_spin} where increasing
  \wscale\ corresponds to smaller angular scales but higher frequency
  content.}

Typically, we consider band-limited functions, \ie\ functions such
that their spherical harmonic coefficients $\shc{\f}{\el}{\m} = 0$,
$\forall \el \geq \elmax$, where
$\shc{\f}{\el}{\m} = \innerp{\f}{\shf{\el}{\m}}$ and
$\shf{\el}{\m} \in \ltwo(\sphere)$ are the spherical harmonics
(defined in \appn{\ref{sec:appendix:special_functions}}) with $\el \in
\naturals$ and $\m \in \integers$, such that $\vert \m \vert \leq
\el$.
In practice, for band-limited functions, wavelet analysis and synthesis can be
computed exactly (to machine precision), since one may appeal to
sampling theorems and corresponding exact quadrature rules for the
computation of integrals \cite{mcewen:fssht, mcewen:so3}, and efficiently by
developing fast algorithms \cite{mcewen:2006:fcswt,
  mcewen:2013:waveletsxv, mcewen:s2let_spin, wiaux:2005b, wiaux:2005c}, which scale to very large data-sets
containing tens of millions of samples on the sphere.

%-----------------------------------------------------------------------------
\subsection{Admissibility}

The wavelet admissibility condition under which a function \f\ can be
synthesised perfectly from its wavelet and scaling coefficients
through \eqn{\ref{eqn:synthesis}} is given by the following resolution
of the identity:
\begin{equation}
  \label{eqn:admissibility}
  \frac{4\pi}{2\el+1}   
  \vert \shc{\wavs}{\el}{0} \vert^2 + 
  \frac{8\pi^2}{2\el+1} 
  \sum_{\wscale=0}^\wscalemax
  \summ \vert \shc{\wav}{\el}{\m}^{(\wscale)}\vert^2 = 1 
  \spcend ,  
  \quad \forall\el 
  \spcend ,
\end{equation}
where $\shc{\wavs}{\el}{0} \kron{\m}{0} =
\innerp{\wavs}{\shf{\el}{\m}}$ and $\shc{\wav}{\el}{\m}^{(\wscale)} =
\innerp{\wav^{(\wscale)}}{\shf{\el}{\m}}$ are the spherical harmonic coefficients
of $\wavs$ and $\wav^{(\wscale)}$, respectively, where $\kron{i}{j}$
for $i,j \in \integers$ denotes the Kronecker delta.

%-----------------------------------------------------------------------------
\subsection{Parseval frame}

Scale-discretised wavelets on the sphere satisfy the following Parseval frame property:
\begin{equation}
  \label{eqn:frame_property}
  A \| f \|^2
  \leq 
  \int_\sphere \dmu{\sa} 
  \vert \innerp{\f}{\rotarg{\sa}\wavs} \vert^2  
  +
  \sum_{\wscale=\wscalemin}^\wscalemax \int_\sothree \deul{\eul}
  \vert \innerp{\f}{\rotarg{\eul}\wav^{(\wscale)}} \vert^2  
  \leq 
  B \| f \|^2
  \spcend ,
\end{equation}
with $A=B\in\realsnz$, for any band-limited $f \in \ltwo(\sphere)$,
and where $\| \cdot \|^2 = \innerp{\cdot}{\cdot}$. We adopt a
shorthand integral notation in \eqn{\ref{eqn:frame_property}},
although by appealing to exact quadrature rules \cite{mcewen:fssht,
  mcewen:so3} these integrals may be replaced by finite sums.
We prove the Parseval frame property as follows. 
Firstly, note the harmonic representation of the wavelet and scaling
coefficients given by \cite[see \eg][]{mcewen:2006:fcswt} 
\begin{equation}
  \label{eqn:analysis_harmonic}
  \wcoeff^{\wav^{(\wscale)}}(\eul) 
  = \innerp{\f}{\rotarg{\eul}\wav^{(\wscale)}}
  = \sumlmnbl
  \shc{\f}{\el}{\m}
  \shc{\wav}{\el}{\n}^{(\wscale)\cconj}
  \Dlmnpc  
\end{equation}
and 
\begin{equation}
  \wcoeff^{\wavs}(\sa)
  = \innerp{\f}{\rotarg{\sa}\wavs}
  = \sumlmbl  
  \sqrt{\frac{4\pi}{2\el+1}}
  \shc{\f}{\el}{\m}
  \shcc{\wavs}{\el}{0}
  \shfarg{\el}{\m}{\sa}
  \spcend ,
\end{equation}
respectively, where the Wigner $\dmatbig$-functions \Dlmn\ are the
matrix elements of the irreducible unitary representation of the
rotation group \sothree\ (defined in
\appn{\ref{sec:appendix:special_functions}}).  Substituting these
harmonic expressions and noting the orthogonality of the spherical
harmonics given by \eqn{\ref{eqn:spherical_harmonic_ortho}} and of the
Wigner $\dmatbig$-functions given by \eqn{\ref{eqn:wignerd_ortho}}, it
is straightforward to show that the term of
\eqn{\ref{eqn:frame_property}} bounded between inequalities may be
written
\begin{equation}
  \label{eqn:frame_property_harmonic}
  \sumlmbl
  \frac{4\pi}{2\el+1}   
  \vert \shc{\f}{\el}{\m} \vert^2
  \vert \shc{\wavs}{\el}{0} \vert^2 +   
  \sum_{\wscale=0}^\wscalemax
  \sumlmnbl
  \frac{8\pi^2}{2\el+1} 
  \vert \shc{\f}{\el}{\m} \vert^2 
  \vert \shc{\wav}{\el}{\n}^{(\wscale)}\vert^2
  =
  \| \f \|^2
  \spcend ,
\end{equation}
where the equality of \eqn{\ref{eqn:frame_property_harmonic}} follows
by the admissibility property \eqn{\ref{eqn:admissibility}}.  Thus,
scale-discretised wavelets indeed constitute a Parseval frame with
$A=B=1$, implying the energy of $f$ is conserved in wavelet space.

%=============================================================================
\section{Wavelet construction}
\label{sec:wavelet_construction}
%=============================================================================

Scale-discretised wavelets are constructed to ensure the admissibility
criterion \eqn{\ref{eqn:admissibility}} is satisfied, while also
carefully controlling their angular and directional spatial
localisation, in additional to their harmonic localisation.  Wavelets
are defined in harmonic space in the factorised form:
\begin{equation}
  \label{eqn:wav_factorized}
  \shc{\wav}{\el}{\m}^{(\wscale)} \equiv 
  \sqrt{\frac{2\el+1}{8\pi^2}} \:
  \wavker^{(\wscale)}(\el) \: 
  \shc{\wavsteer}{\el}{\m}
  \spcend,
\end{equation}
in order to control their angular and directional localisation
separately, respectively through the kernel
$\wavker^{(\wscale)} \in \ltwo(\reals^{+})$ and directionality
component $\wavsteer \in \ltwo(\sphere)$, with harmonic coefficients
$\shc{\wavsteer}{\el}{\m} = \innerp{\wavsteer}{\shf{\el}{\m}}$
($\wavker^{(\wscale)}$ and $\wavsteer$ are defined 
explicitly in \sectn{\ref{sec:construction:kernel}} and \sectn{\ref{sec:construction:directional}}, respectively).
Without loss of generality, the directionality component is
normalised to impose
\begin{equation}
  \label{eqn:directionality_normalisation}
  \summ \vert \shc{\wavsteer}{\el}{\m} \vert^2 = 1
  \spcend, 
\end{equation}
for all values of \el\ for which $\shc{\wavsteer}{\el}{\m}$ are
non-zero for at least one value of \m.  The angular localisation
properties of the wavelet $\wav^{(\wscale)}$ are then controlled by
the kernel $\wavker^{(\wscale)}$, while the directionality component
$\wavsteer$ controls the directional properties of the wavelet (\ie\
the behaviour of the wavelet with respect to the azimuthal variable
\sab, when centered on the North pole).  In the remainder of this
section we describe the construction of the wavelet kernel, the
wavelet steerability property, and, for the first time, the explicit construction of the
directionality component.

%-----------------------------------------------------------------------------
\subsection{Kernel construction}
\label{sec:construction:kernel}

\begin{figure}
\centering
\includegraphics[width=0.75\textwidth]{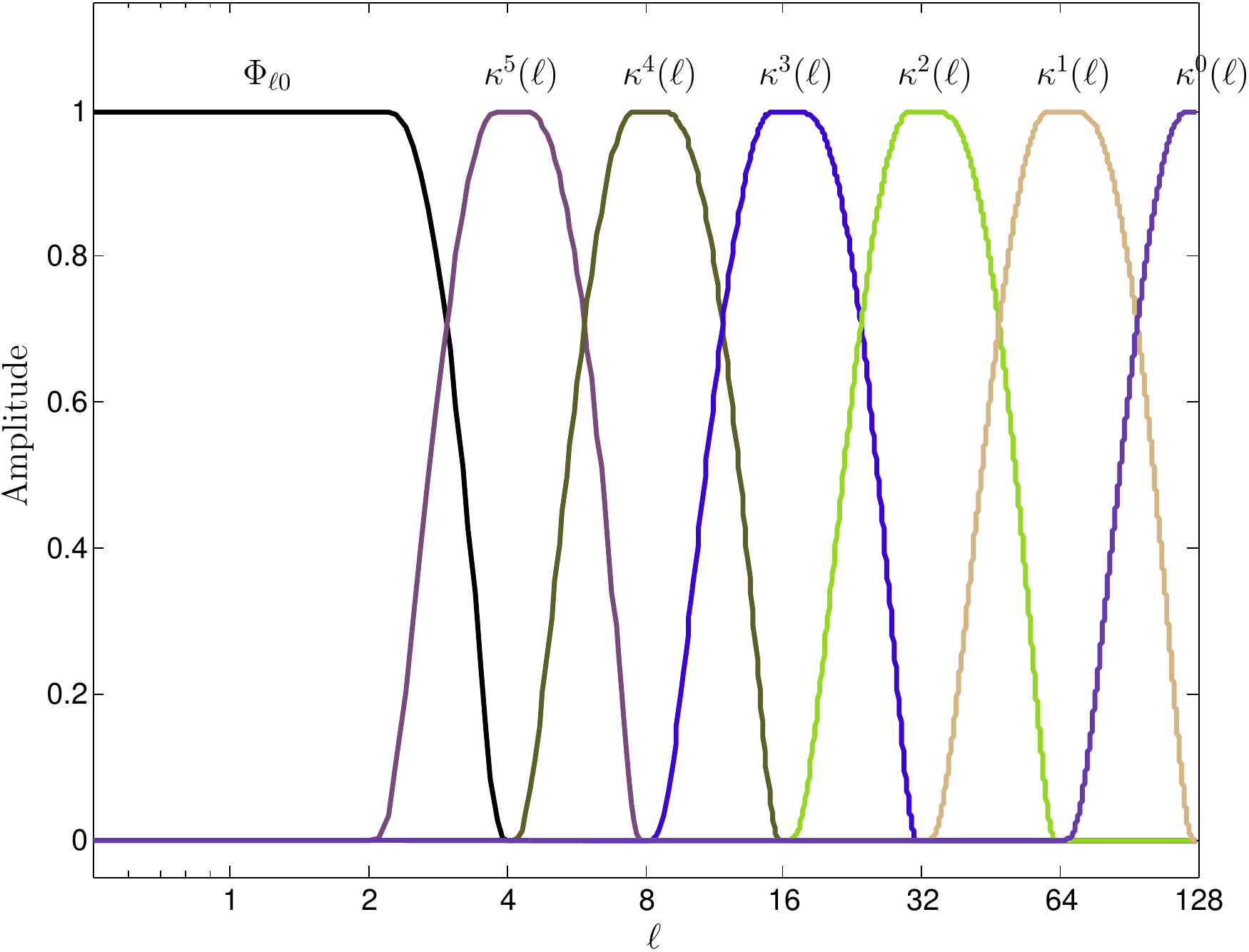}
\caption{Scale-discretised wavelet tiling in harmonic space
  ($\elmax=128$, $\dilparam=2$, $\nmax=3$,
  $\wscalemax=5$).} 
\label{fig:tiling}
\end{figure}

The kernel $\wavker^{(\wscale)}(t)$ is a positive real function, with
argument $t \in \reals^{+}$, although $\wavker^{(\wscale)}(t)$ is
evaluated only for natural arguments $t=\el$ in
\eqn{\ref{eqn:wav_factorized}}.  The kernel controls the angular
localisation of the wavelet and is constructed to be a smooth function
with compact support, as follows.  Consider the smooth, infinitely
differentiable (Schwartz) function with compact support
$t \in [\dilparam^{-1}, 1]$, for dilation parameter
$\dilparam \in \realsnz$, $\dilparam>1$:
\begin{equation}
  \label{eqn:schwartz_function}
  s_\dilparam(t) 
  \equiv s\biggl( \frac{2\dilparam}{\dilparam-1} (t-\dilparam^{-1})-1\biggr)
  \spcend,
  \textrm{\quad with \quad } s(t) \equiv \Biggl\{ \begin{array}{ll} 
  \ 
  {\rm exp}\bigl(-(1-t^2)^{-1}\bigr), & t\in[-1,1] \\ \  0, & t \notin [-1,1]\end{array} \spcend.
\end{equation}
Define the smoothly decreasing function $k_\dilparam$ by
\begin{equation}
  k_\dilparam(t) \equiv \frac{\int_{t}^1\frac{{\rm d}t^\prime}{t^\prime}s_\dilparam^2(t^\prime)}{\int_{\dilparam^{-1}}^1\frac{{\rm d}t^\prime}{t^\prime}s_\dilparam^2(t^\prime)}, 
\end{equation}
which is unity for $t<\dilparam^{-1}$, zero for $t>1$, and is smoothly
decreasing from unity to zero for $t \in [\dilparam^{-1},1]$.  Define
the wavelet kernel generating function by
\begin{equation}
  \wavker_\dilparam(t) \equiv \sqrt{ k_\dilparam(\dilparam^{-1} t) - k_\dilparam(t) }
  \spcend,
\end{equation}
which has compact support $t \in [\dilparam^{-1}, \dilparam]$ and
reaches a peak of unity at $t=1$.  The scale-discretised wavelet
kernel for scale \wscale\ is then defined by
\begin{equation}
  \wavker^{(\wscale)}(\el) \equiv  \wavker_\dilparam(\dilparam^{\wscale} \elmax^{-1} \el)
  \spcend,
\end{equation}
which has compact support on
$\el \in \bigl[\floor{\dilparam^{-(1+\wscale)} \elmax},
\ceil{\dilparam^{1-\wscale} \elmax} \bigr]$,
where $\floor{\cdot}$ and $\ceil{\cdot}$ are the floor and ceiling
functions respectively, and reaches a peak of unity at
$\dilparam^{-\wscale}\elmax$.
With this construction the kernel functions tile the harmonic line, as
illustrated in \fig{\ref{fig:tiling}}. 
Note that needlets are constructed by a similar Meyer-like tiling of
the line defined by spherical harmonic degree $\el$
\cite{narcowich:2006, baldi:2009, marinucci:2008}, where the 
function $s(t)$ of \eqn{\ref{eqn:schwartz_function}} is also used but
minor differences in the kernel construction mean that the needlet and
scale-discretised wavelet kernels differ slightly
\cite[see][\fig{1}]{leistedt:s2let_axisym}.

The maximum possible wavelet scale $\wscalemax_\elmax(\dilparam)$ is
given by the lowest integer $\wscale$ for which the kernel peak occurs
at or below $\el=1$, \ie\ by the lowest integer value such that
$\dilparam^{-\wscalemax_\elmax(\dilparam)} \elmax \leq 1$, yielding
$\wscalemax_\elmax(\dilparam) = \ceil{\log_\dilparam(\elmax)}$.  All
wavelets for $\wscale > \wscalemax_\elmax(\dilparam)$ would be
identically null as their kernel would have compact support in
$\el \in (0, 1)$.  The maximum scale to be probed by the wavelets
$\wscalemax$ can be chosen within the range
$0\leq\wscalemax\leq\wscalemax_\elmax(\dilparam)$.  For
$\wscalemax=\wscalemax_\elmax(\dilparam)$ the wavelets probe the
entire frequency content of the signal of interest \f\ except its
mean, encoded in $\shc{\f}{0}{0}$ and incorporated in the scaling
coefficients.

To represent the signal content not probed by the wavelets the scaling
function \wavs\ is required, as discussed previously.  Recall that the
scaling function $\wavs$ is chosen to be axisymmetric; hence, we
define the harmonic coefficients of the scaling function by
\begin{equation}
  \shc{\wavs}{\el}{\m} 
  \equiv 
  \sqrt{\frac{2\el+1}{4\pi}} \:
  \sqrt{ k_\dilparam(\dilparam^{\wscalemax} \elmax^{-1} \el)} \:
  \kron{\m}{0}
  \spcend,
\end{equation}
in order to ensure the scaling function probes the signal content not
probed by the wavelets.

For the wavelets, scaling function and wavelet scale parameter ranges
outlined above, the admissibility criterion
\eqn{\ref{eqn:admissibility}} is satisfied.  Although the precise
construction of the directionality component has not yet been defined,
provided it is normalised according to
\eqn{\ref{eqn:directionality_normalisation}}, admissibility holds.

%-----------------------------------------------------------------------------
\subsection{Steerability}

A function on the sphere is steerable if an azimuthal rotation of the
function can be written as a linear combination of weighted basis
functions.  By imposing an azimuthal band-limit \nmax\ on the
directionality component such that $\shc{\wavsteer}{\el}{\m}=0$,
$\forall \el,\m$ with $\vert \m \vert \geq \nmax$, we recover wavelets
that are steerable \cite{wiaux:2007:sdw, freeman:1991}.  Moreover, if
$T \in \naturals$ of the harmonic coefficients
$\shc{\wavsteer}{\el}{\m}$ are non-zero for a given $\m$ for at least
one $\el$, then the number of basis functions $M \in \naturals$
required to steer the wavelet directionality component satisfies
$M \geq T$ and the optimal number $M=T$ can be chosen.  Furthermore,
if $\wavsteer$ exhibits an azimuthal band-limit, then it can be
steered using basis functions that are in fact rotations of itself:
\begin{equation}
  \label{eqn:steerability_wav}
  \wavsteer_\eulc(\sa) = 
  \sum_{\eulci=0}^{M-1} \steerinterp(\eulc - \eulciang) \: 
  \wavsteer_{\eulciang}(\sa)
  \spcend,
\end{equation}
which extends to the wavelets $\wav^{(\wscale)}$ also, where
$\wavsteer_\eulc \equiv \rot_{(0,0,\eulc)} \wavsteer$ and $g \in
\naturals$.
The rotation angles $\eulciang \in [0, 2\pi)$ and interpolating
function $\steerinterp \in \ltwo(\reals)$ are defined subsequently.
Note that the interpolating function is independent of the
directionality component $\wavsteer$ of the wavelet.  Due to the
linearity of the wavelet transform, the steerability property is
transferred to the wavelet coefficients themselves, yielding
\begin{equation}
  \label{eqn:steerability_wcoeff}
  \wcoeff^{\wav^{(\wscale)}}(\euls) = 
  \sum_{\eulci=0}^{M-1} \steerinterp(\eulc - \eulciang) \:
  \wcoeff^{\wav^{(\wscale)}}(\eula,\eulb,\eulciang)  
  \spcend.
\end{equation}

Before proving the steerability property of
\eqn{\ref{eqn:steerability_wav}}, we consider additional azimuthal
symmetries that the directionality component, and thus wavelets, are
designed to satisfy.
For $\nmax-1$ odd (even), the wavelets are constructed to
exhibit odd (even) symmetry under a reflection of \sab: 
\begin{equation}
  \label{eqn:symmetry_ref}
  \wavsteer(\saa, -\sab) = (-1)^{N-1} \: \wavsteer(\saa, \sab)
  \spcend,
\end{equation}
which for real functions on the sphere implies the harmonic
coefficients $\shc{\wavsteer}{\el}{\m}$ are purely real for $N-1$ even
and purely imaginary for $N-1$ odd.  In addition, for $\nmax-1$ odd
(even), the wavelets are constructed to exhibit odd (even) symmetry
under an azimuthal rotation by $\pi$:
\begin{equation}
  \label{eqn:symmetry_rot}
  \wavsteer(\saa, \sab+\pi) = (-1)^{N-1} \: \wavsteer(\saa, \sab)
  \spcend,
\end{equation}
which implies the harmonic coefficients $\shc{\wavsteer}{\el}{\m}$ are
zero for $\m$ odd when $\nmax-1$ is even and zero for $\m$ even when
$\nmax-1$ is odd. This symmetry is exploited to optimise the
number of basis functions required to steer the wavelet to $M=N$.

Returning to the steerability relation of
\eqn{\ref{eqn:steerability_wav}}, we prove this expression by proving the
equivalent harmonic space representation:
\begin{equation}
  \label{eqn:steerability_proof_1}
  \shc{(\wavsteer_{\eulc})}{\el}{\m} = 
  \sum_{\m\p=-K}^K
  \sum_{\n=-\el}^\el 
  \steerinterp_{\m\p} \:
  \exp{(\img \m\p \eulc)} \:
  \dlmn(0) \:
  \shc{\wavsteer}{\el}{\n}
  \sum_{\eulci=0}^{M-1} \exp{(-i (\m\p + \n) \eulciang)}
  \spcend,
\end{equation}
where $\steerinterp_\m$ are the Fourier coefficients of the
interpolating function, $K \in \naturals$ is the (as yet
unconstrained) band-limit of the interpolating function, and $\dlmn$
are the Wigner $d$-functions (defined in
\appn{\ref{sec:appendix:special_functions}}).  Performing a rotation
in harmonic space of the left-hand-side of
\eqn{\ref{eqn:steerability_proof_1}}, and noting the orthogonality of
the final summation of \eqn{\ref{eqn:steerability_proof_1}} for the
equiangular sampling $\eulciang = \eulci \pi / M$, with
$\m\p,\n < \floor{M}$, and thus $K=\floor{M}-1$, one finds that the
steerability relation is satisfied provided
$\steerinterp_{\m} = 1/M=1/N$ over the domain where
$\shc{\wavsteer}{\el}{\m}$ is non-zero and zero elsewhere, where we
have exploited the symmetry relation of \eqn{\ref{eqn:symmetry_rot}}.

\begin{figure}
\centering
\subfigure[$\eulc_0 = 0^\circ$]{\includegraphics[viewport = 130 80 415 365, clip=true, width=.23\textwidth]{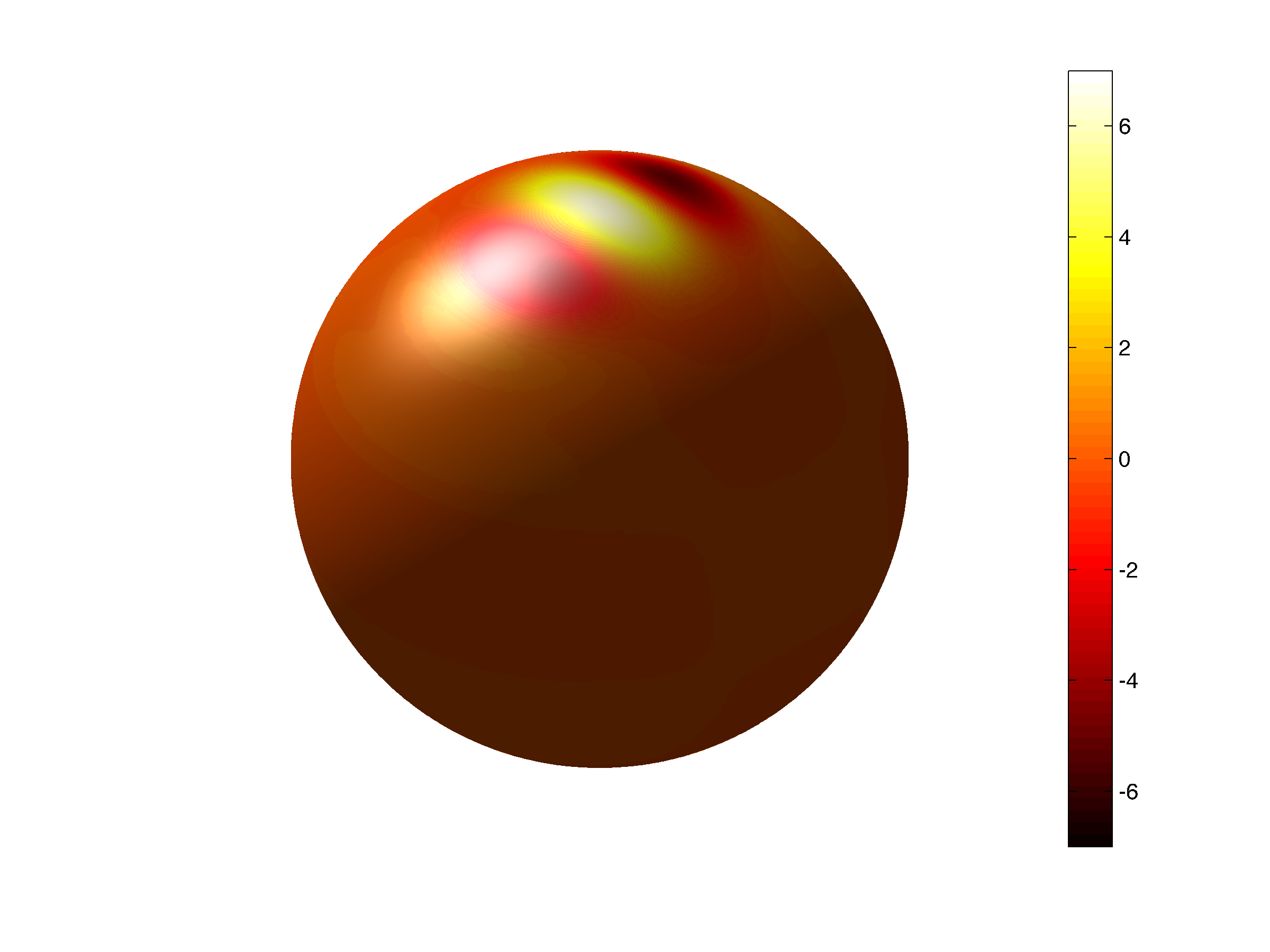}}
\subfigure[$\eulc_1 = 60^\circ$]{\includegraphics[viewport = 130 80 415 365, clip=true, width=.23\textwidth]{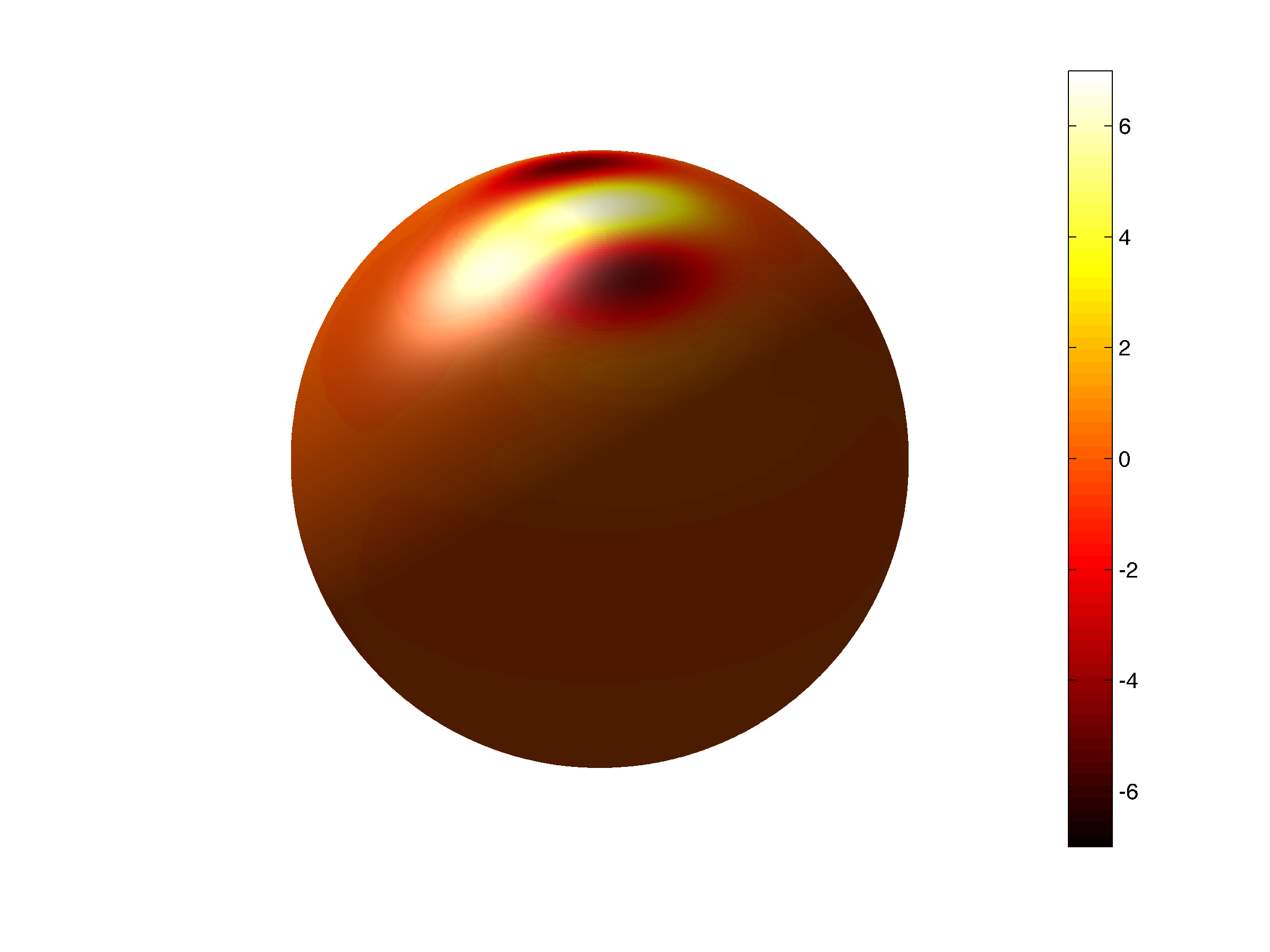}}
\subfigure[$\eulc_2 = 120^\circ$]{\includegraphics[viewport = 130 80 415 365, clip=true, width=.23\textwidth]{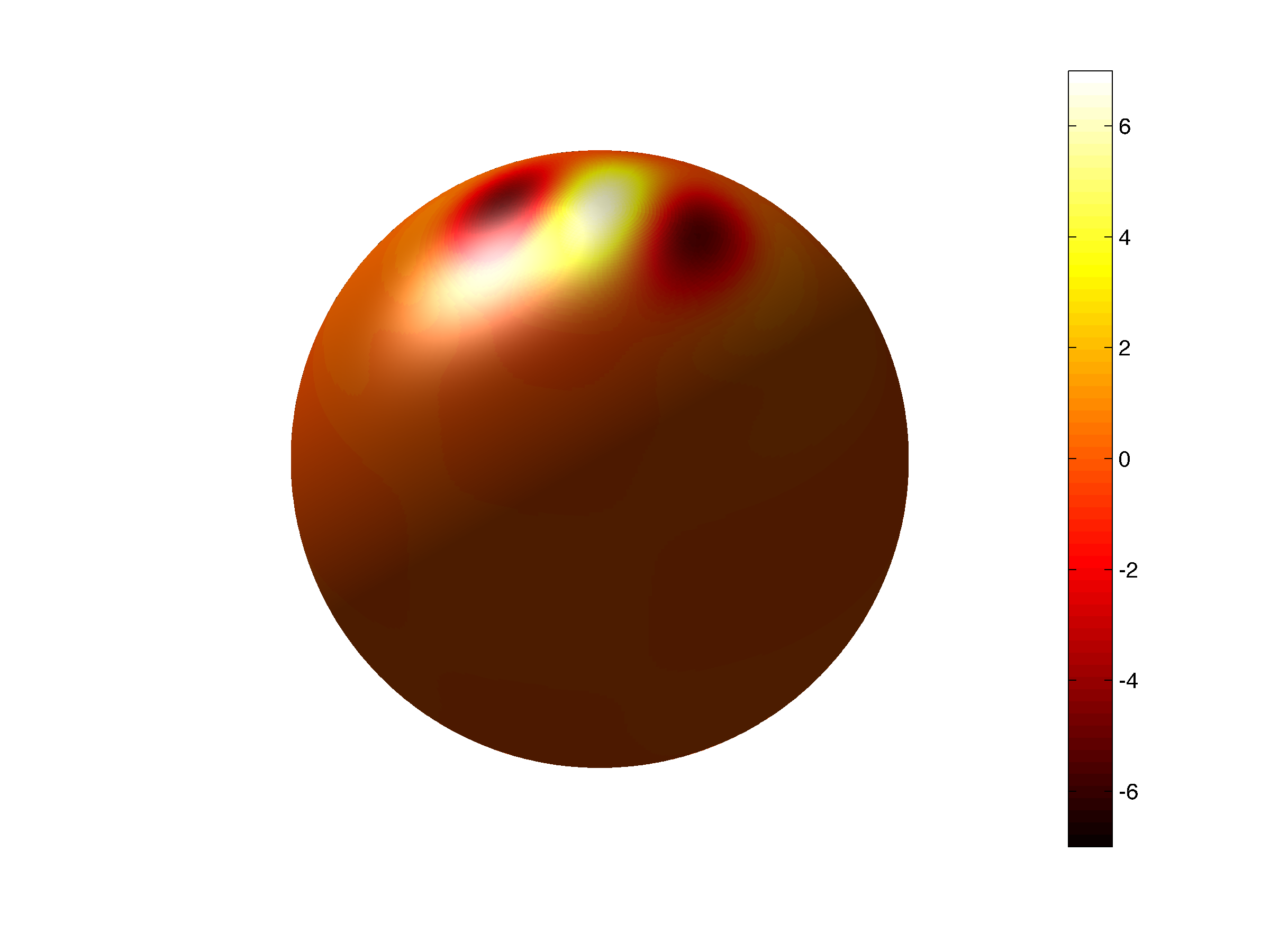}}
\subfigure[Steered to $\eulc=30^\circ$]{\includegraphics[viewport = 130 80 415 365, clip=true, width=.23\textwidth]{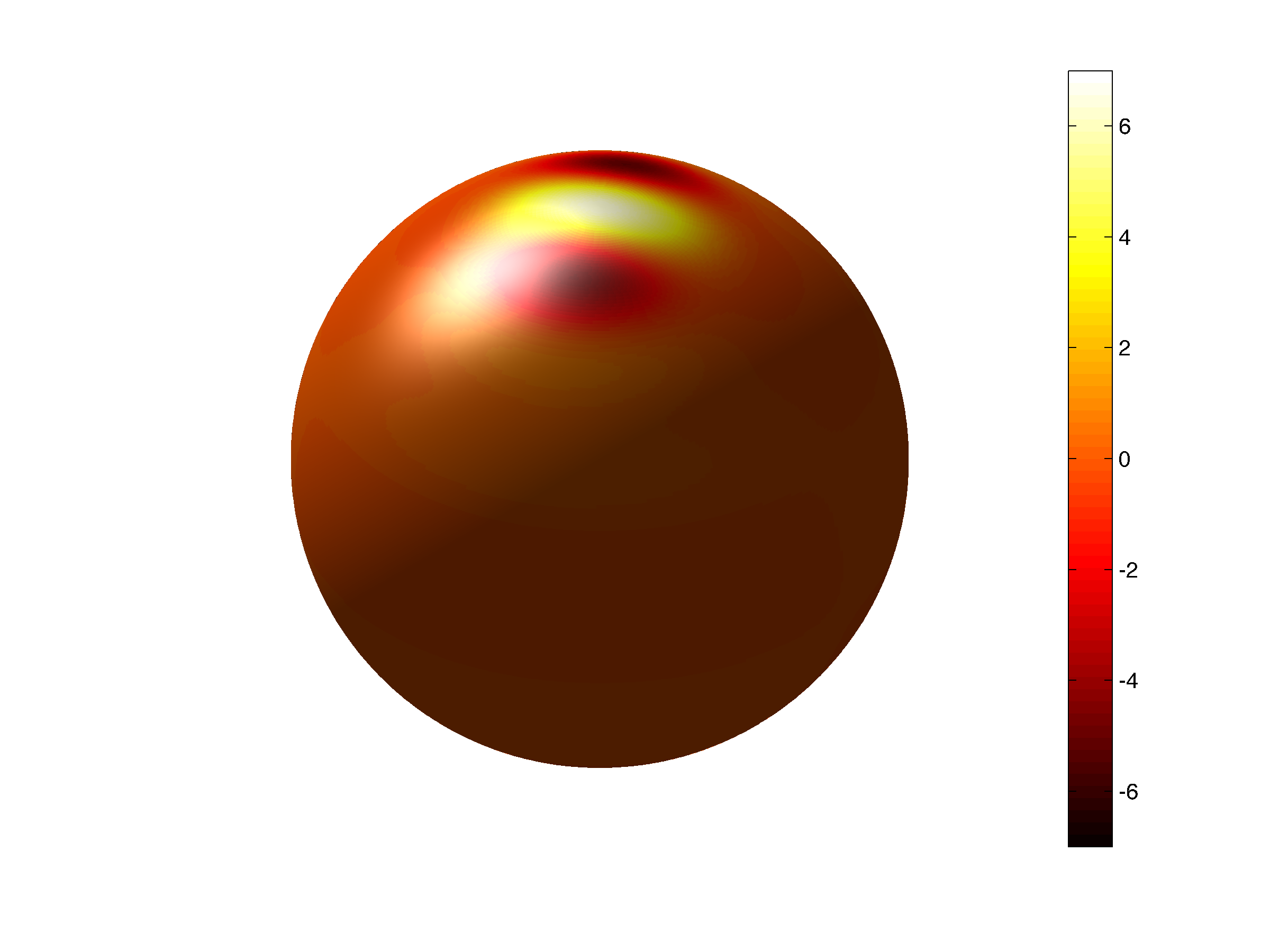}}\\
\includegraphics[viewport = 75 25 525 60, clip=true, width=.6\textwidth]{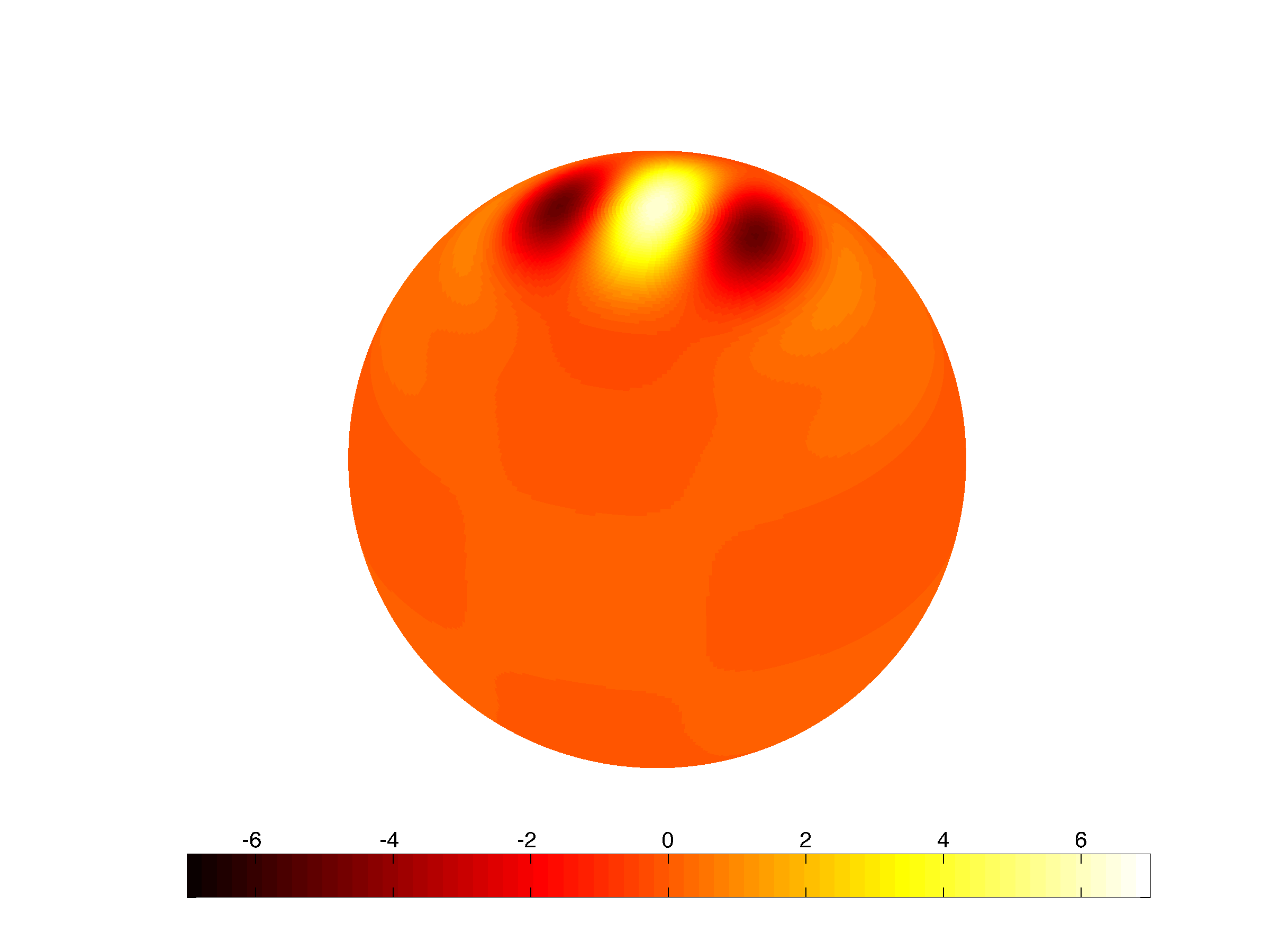}
\caption{Demonstration of wavelet steerability property
  ($\elmax = 256,\ \dilparam=2,\ \mmax=3,\ \wscale=5$).
%  ($\elmax = 256,\ \dilparam=2,\ \mmax=3,\ \wscalemax=8,\ \wscale=3$).
  The wavelet in panel (d) is steered by constructed from a weighted
  sum of the wavelets shown in panels (a)--(c) through
  \eqn{\ref{eqn:steerability_wav}}.}
\label{fig:steerability}
\end{figure}

A steered wavelet for $\mmax=3$ and its basis functions, given by
rotated versions of the wavelet, are plotted in
\fig{\ref{fig:steerability}}.  The wavelet can be steered to any
continuous orientation $\eulc$ by taking weighted sums of its three
basis functions.

%-----------------------------------------------------------------------------
\subsection{Directional construction}
\label{sec:construction:directional}

\begin{figure}
\centering
\subfigure[$\mmax=1,\ \wscale=6$]{\includegraphics[width=.23\textwidth]{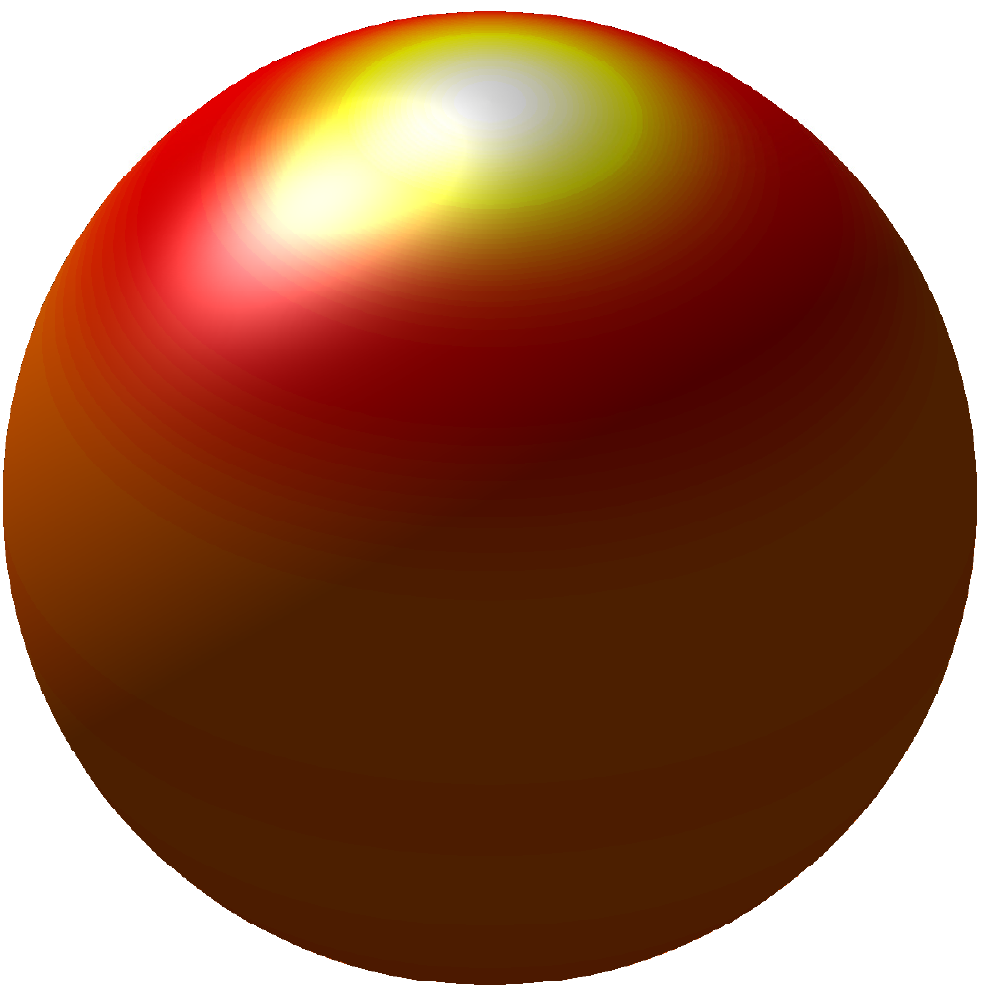}}
%{Figures/wav/s2dw_wavelet_L0256_B2.0_N03_j02}}
\subfigure[$\mmax=1,\ \wscale=5$]{\includegraphics[width=.23\textwidth]{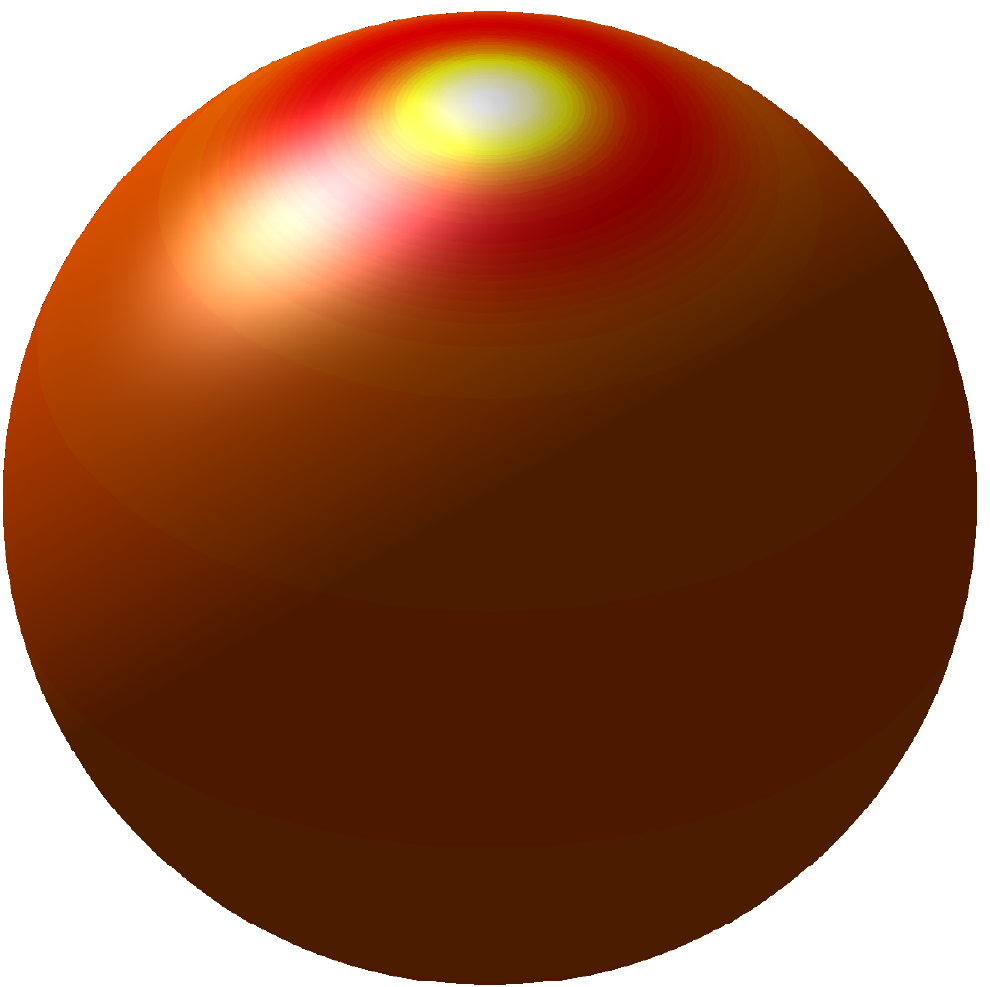}}
\subfigure[$\mmax=1,\ \wscale=4$]{\includegraphics[width=.23\textwidth]{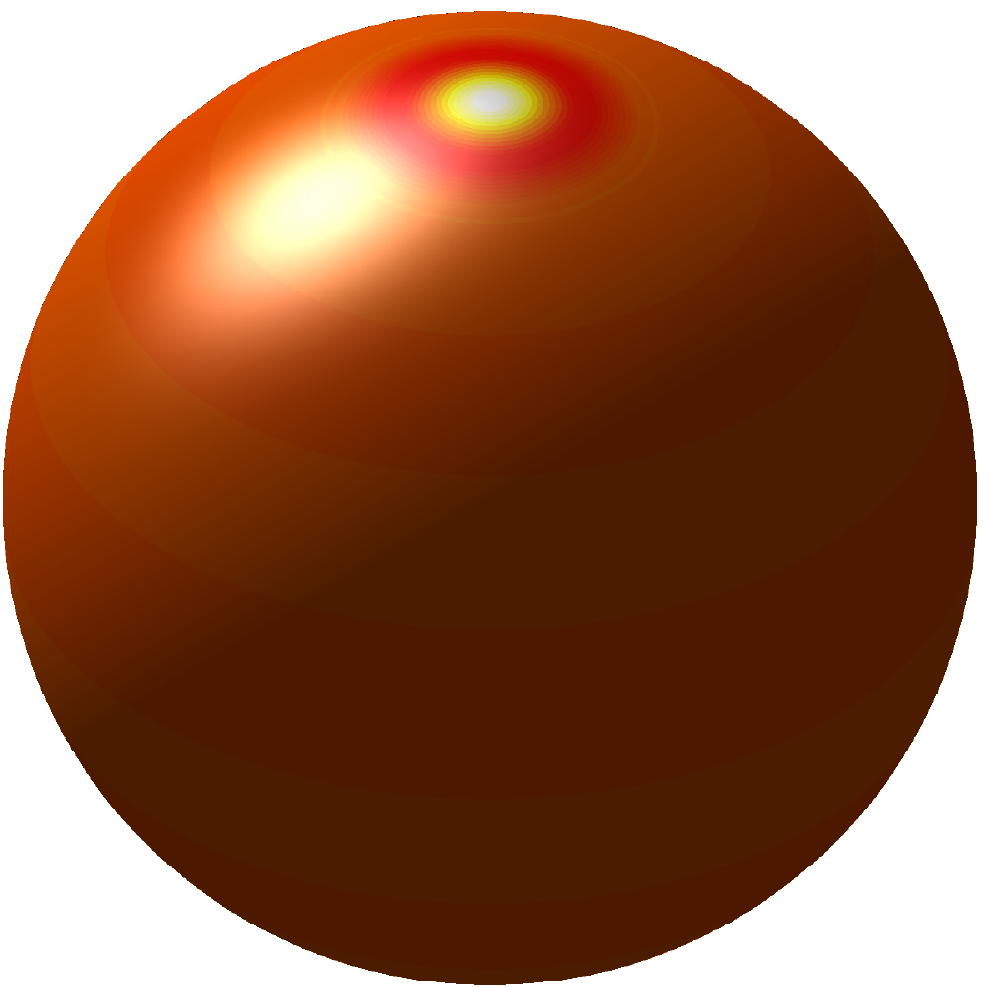}}
\subfigure[$\mmax=1,\ \wscale=3$]{\includegraphics[width=.23\textwidth]{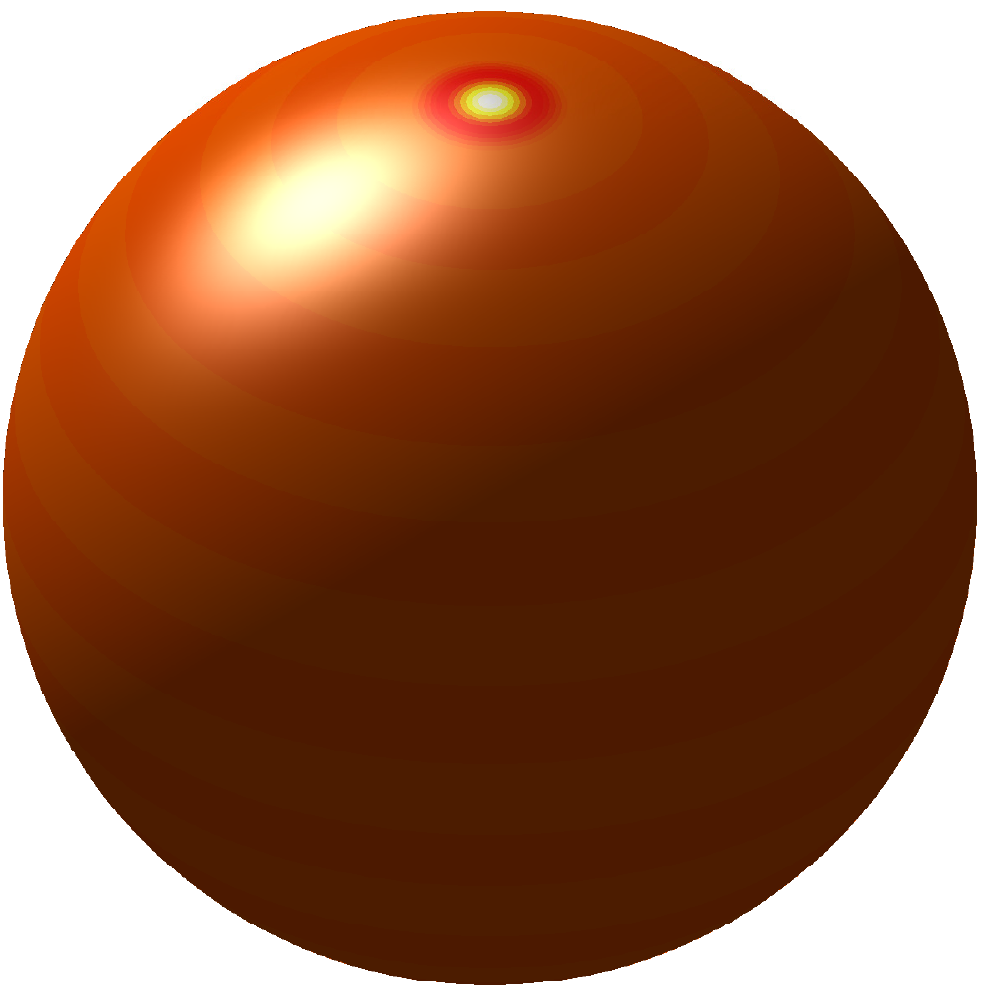}}
\subfigure[$\mmax=2,\ \wscale=6$]{\includegraphics[width=.23\textwidth]{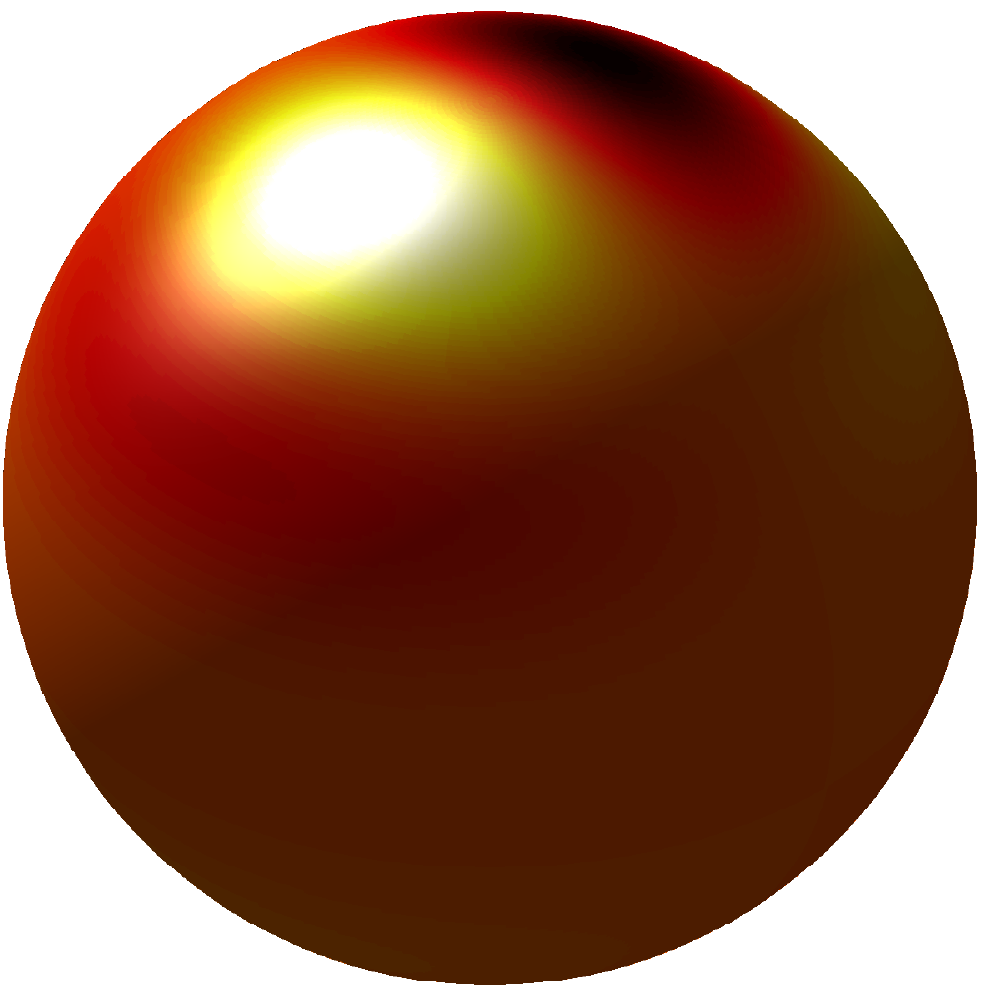}}
\subfigure[$\mmax=2,\ \wscale=5$]{\includegraphics[width=.23\textwidth]{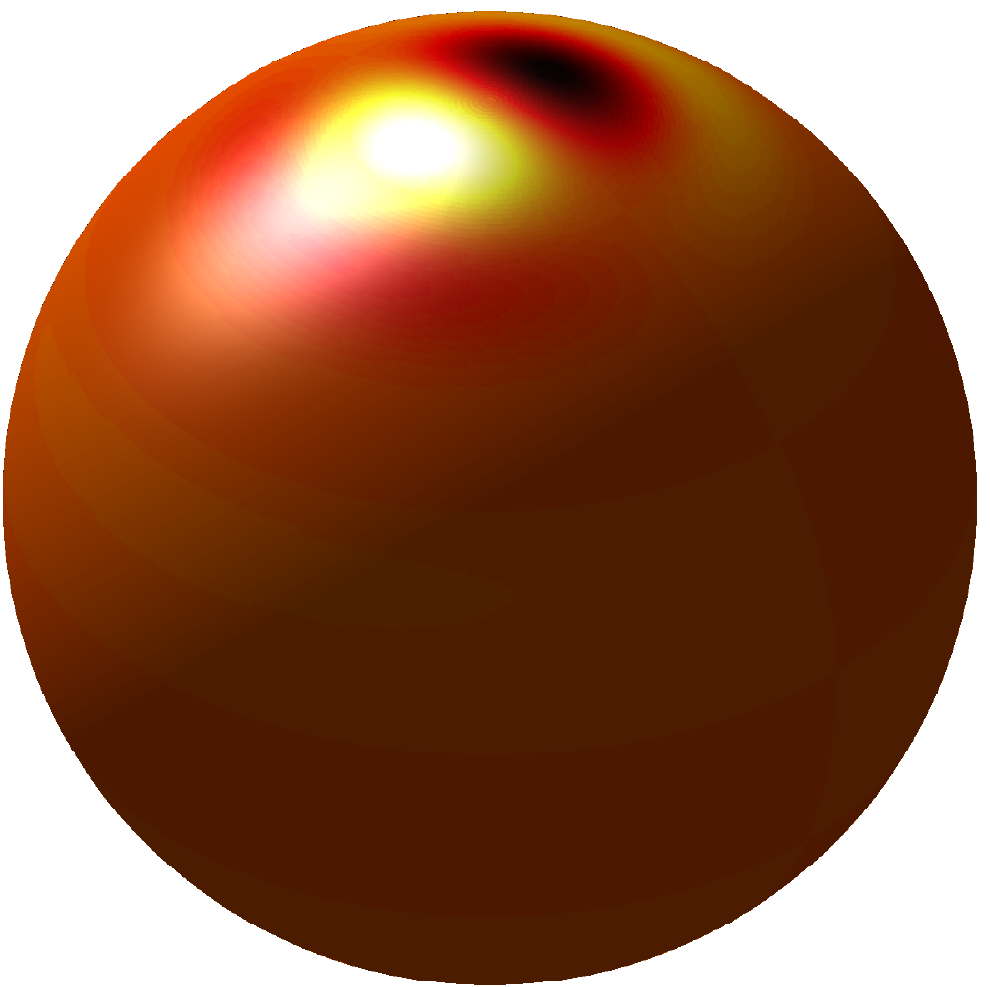}}
\subfigure[$\mmax=2,\ \wscale=4$]{\includegraphics[width=.23\textwidth]{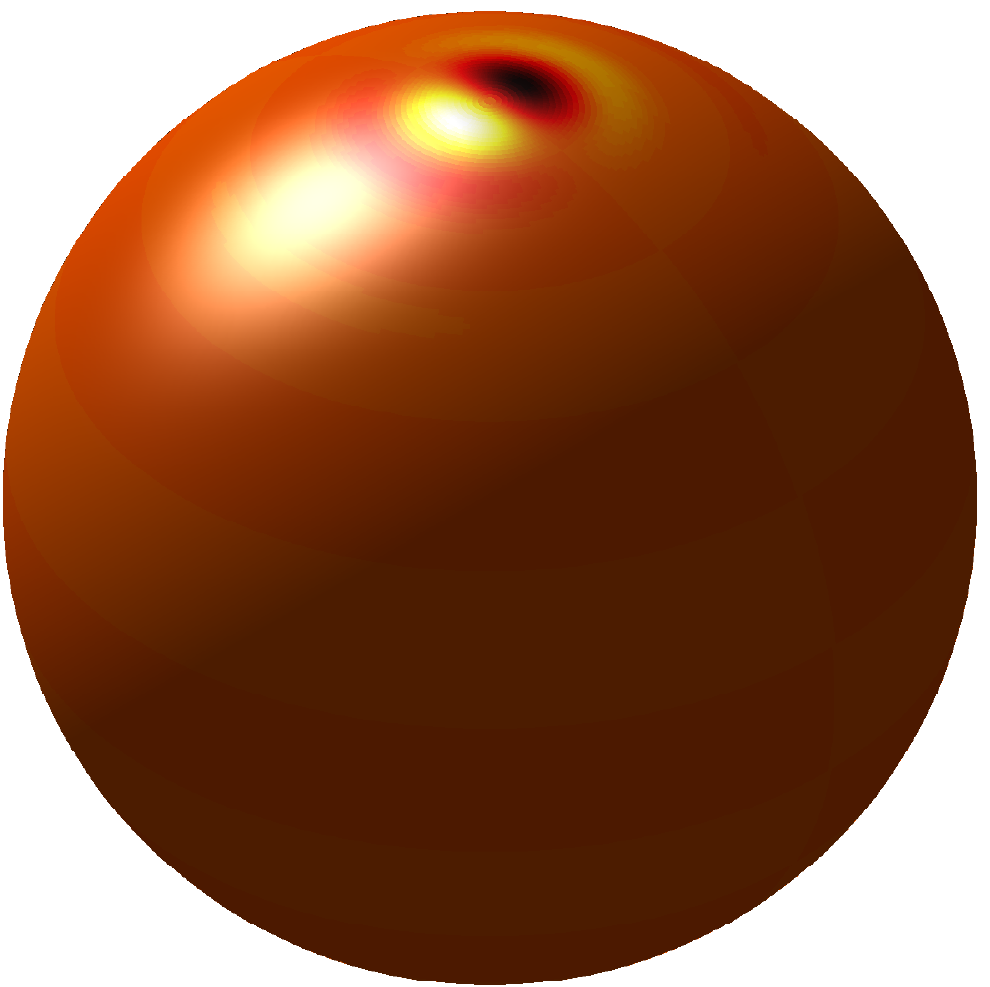}}
\subfigure[$\mmax=2,\ \wscale=3$]{\includegraphics[width=.23\textwidth]{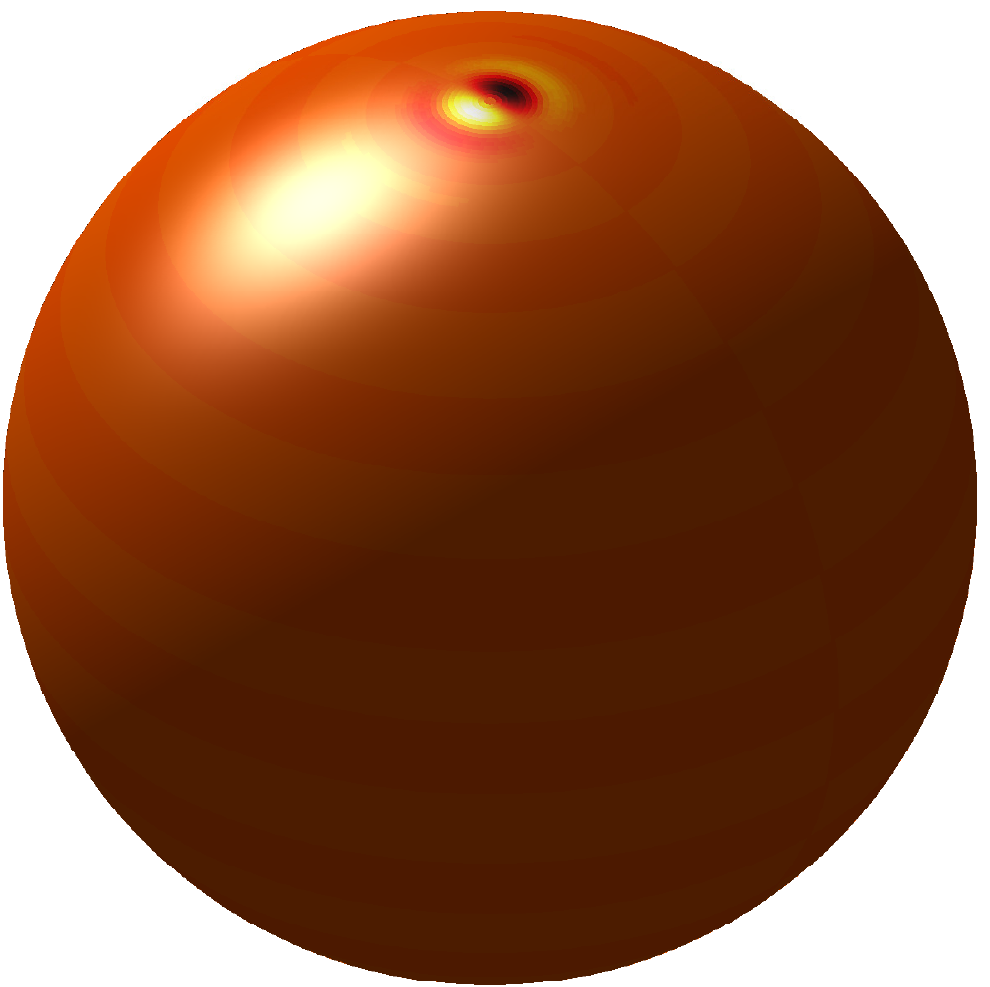}}
\subfigure[$\mmax=3,\ \wscale=6$]{\includegraphics[width=.23\textwidth]{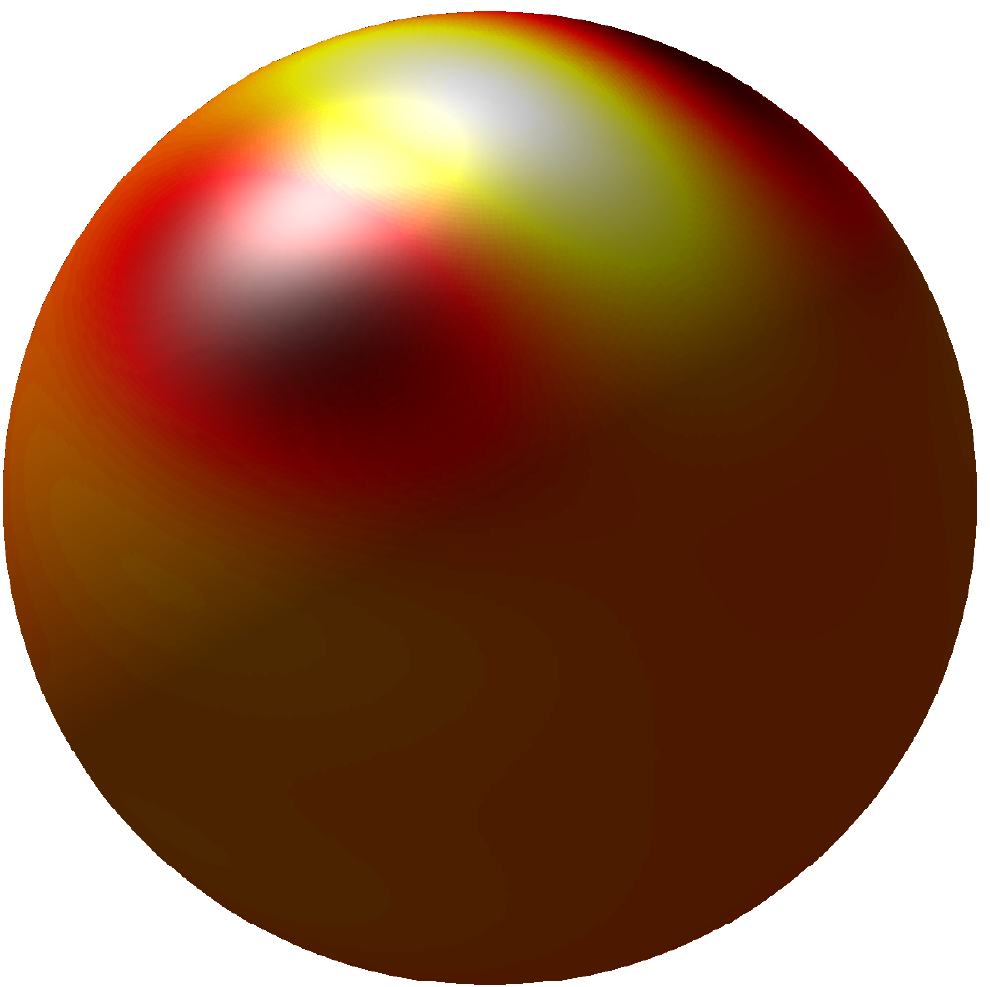}}
\subfigure[$\mmax=3,\ \wscale=5$]{\includegraphics[width=.23\textwidth]{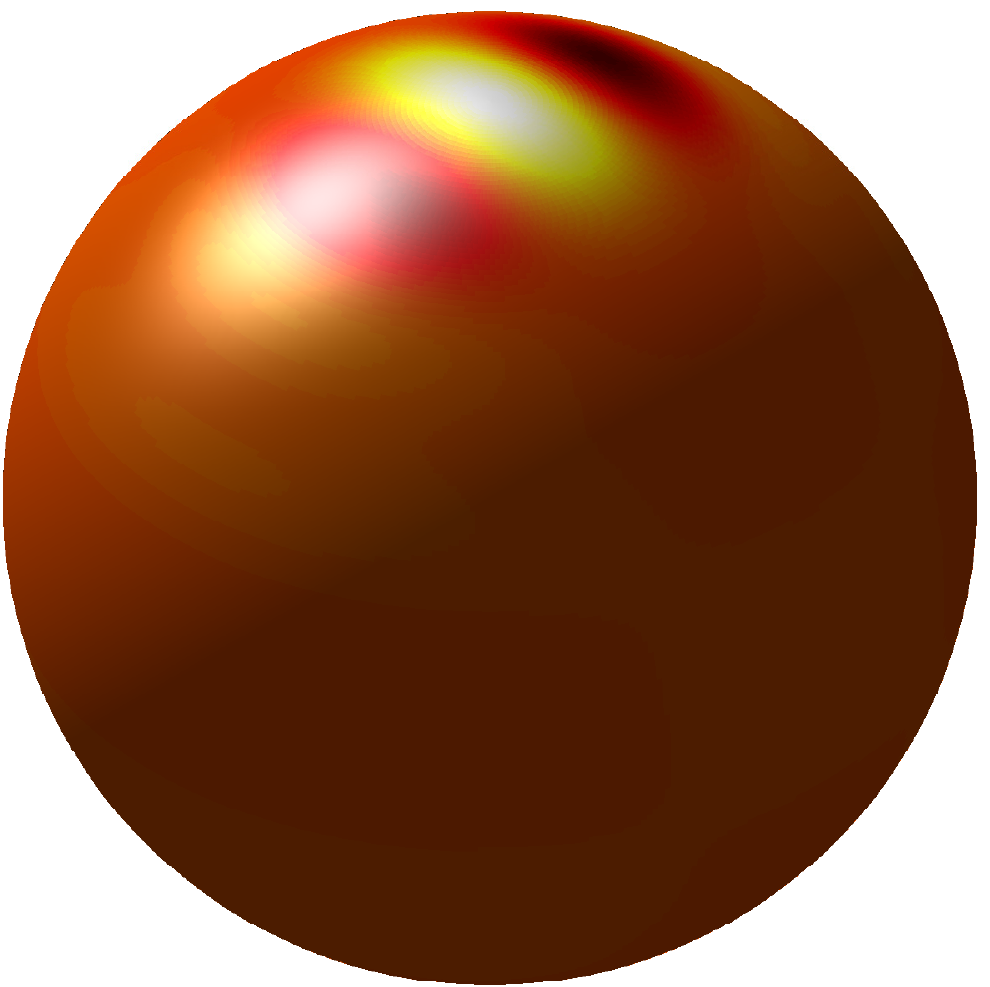}}
\subfigure[$\mmax=3,\ \wscale=4$]{\includegraphics[width=.23\textwidth]{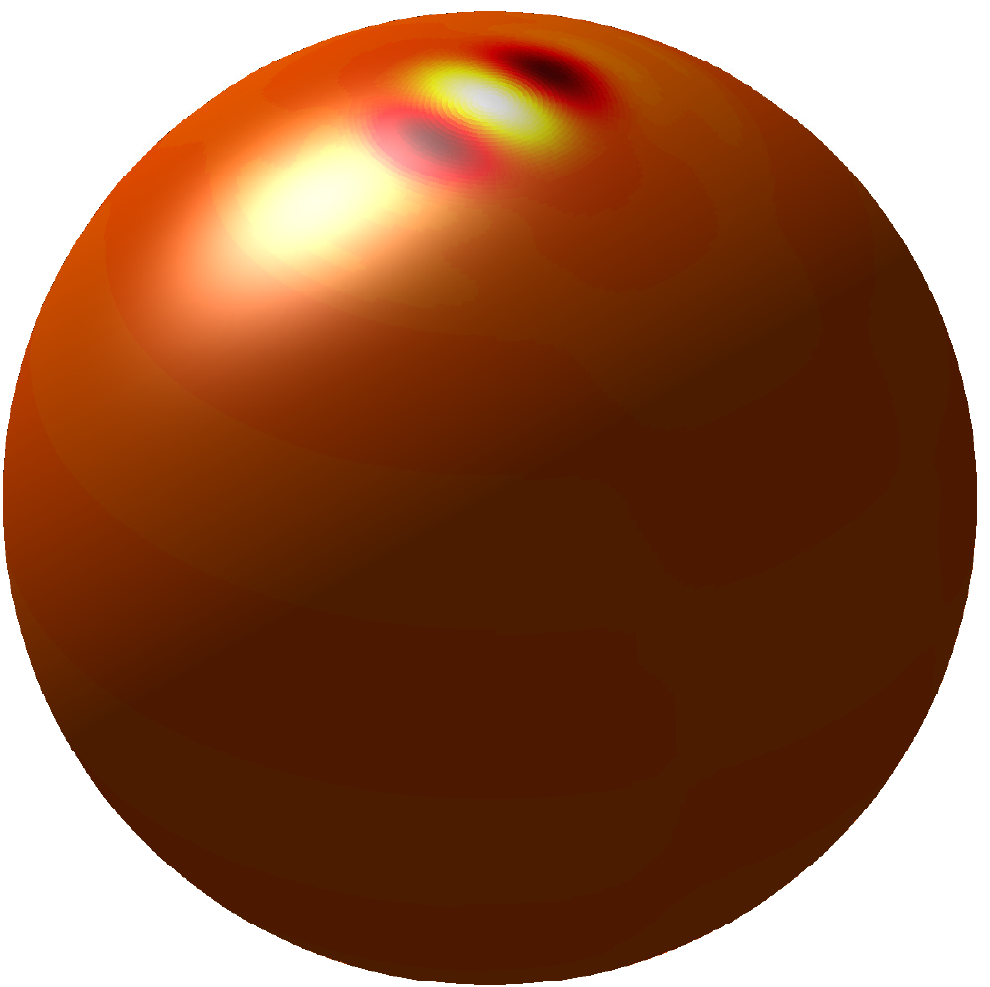}}
\subfigure[$\mmax=3,\ \wscale=3$]{\includegraphics[width=.23\textwidth]{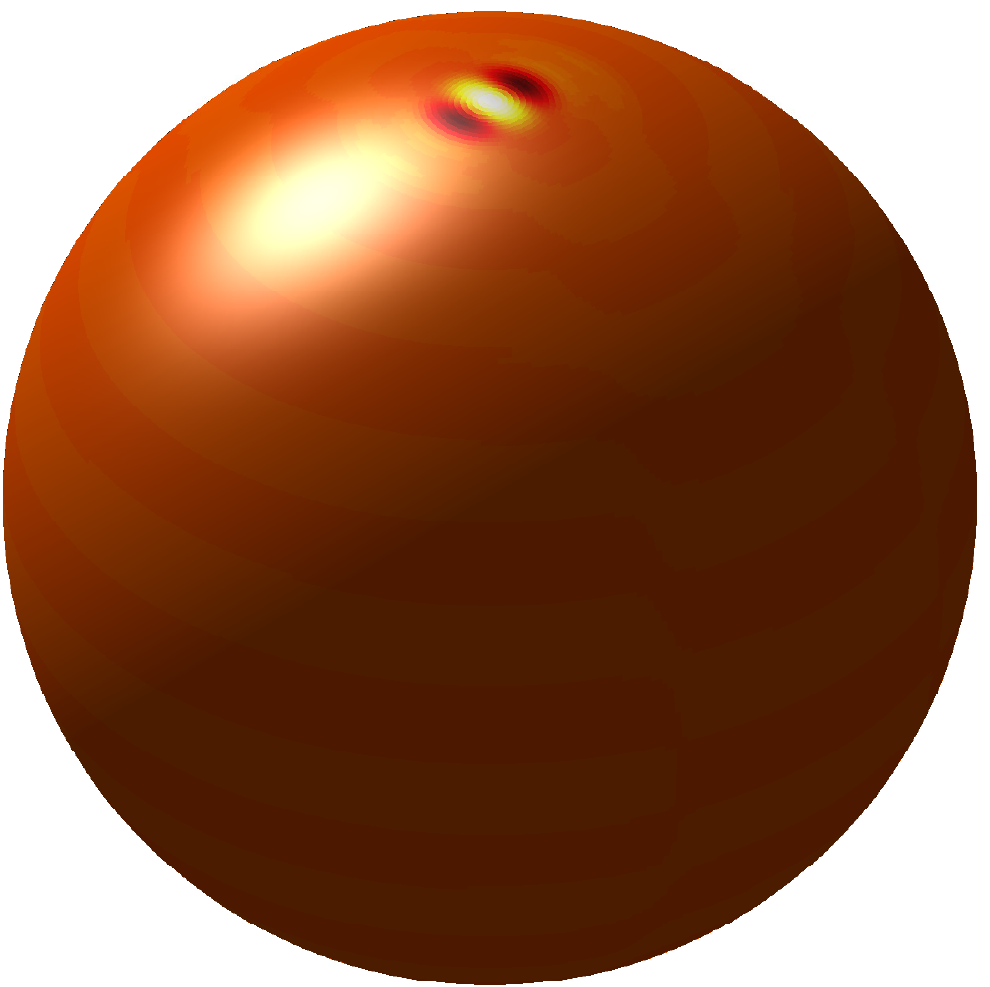}}
\subfigure[$\mmax=4,\ \wscale=6$]{\includegraphics[width=.23\textwidth]{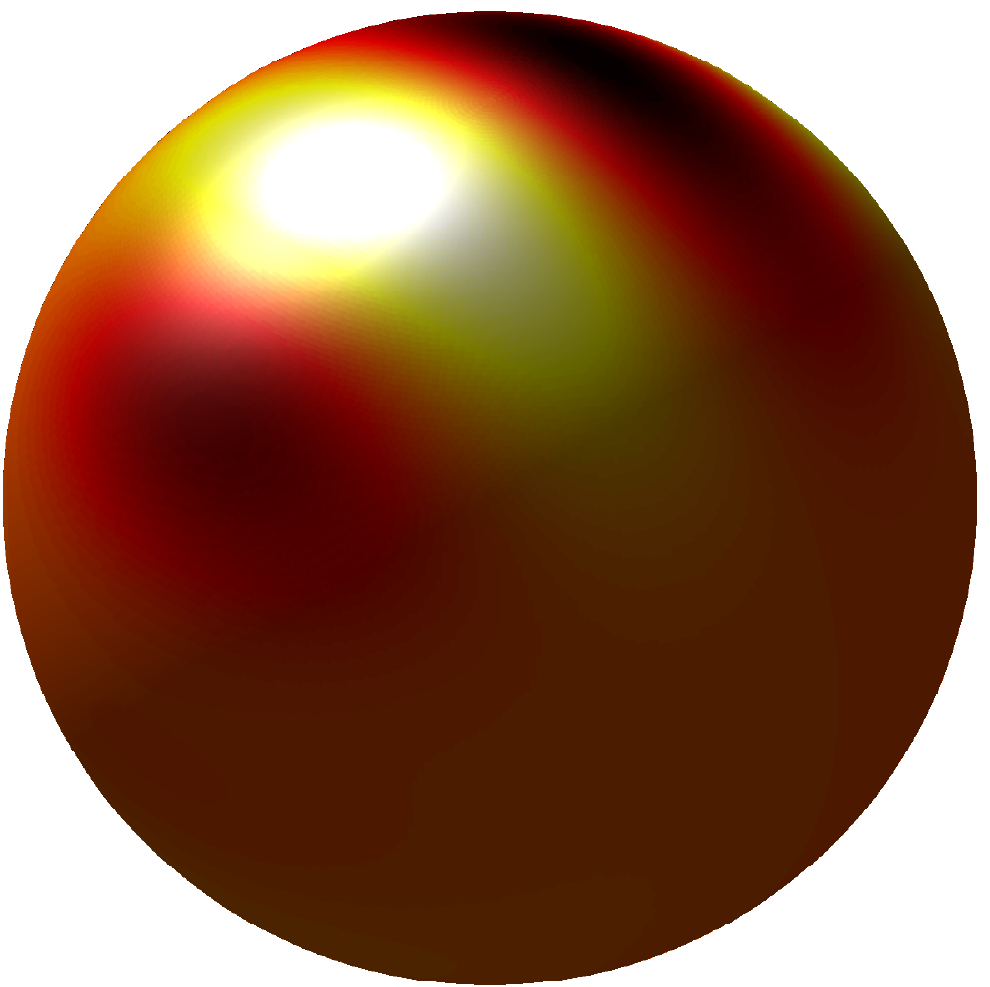}}
\subfigure[$\mmax=4,\ \wscale=5$]{\includegraphics[width=.23\textwidth]{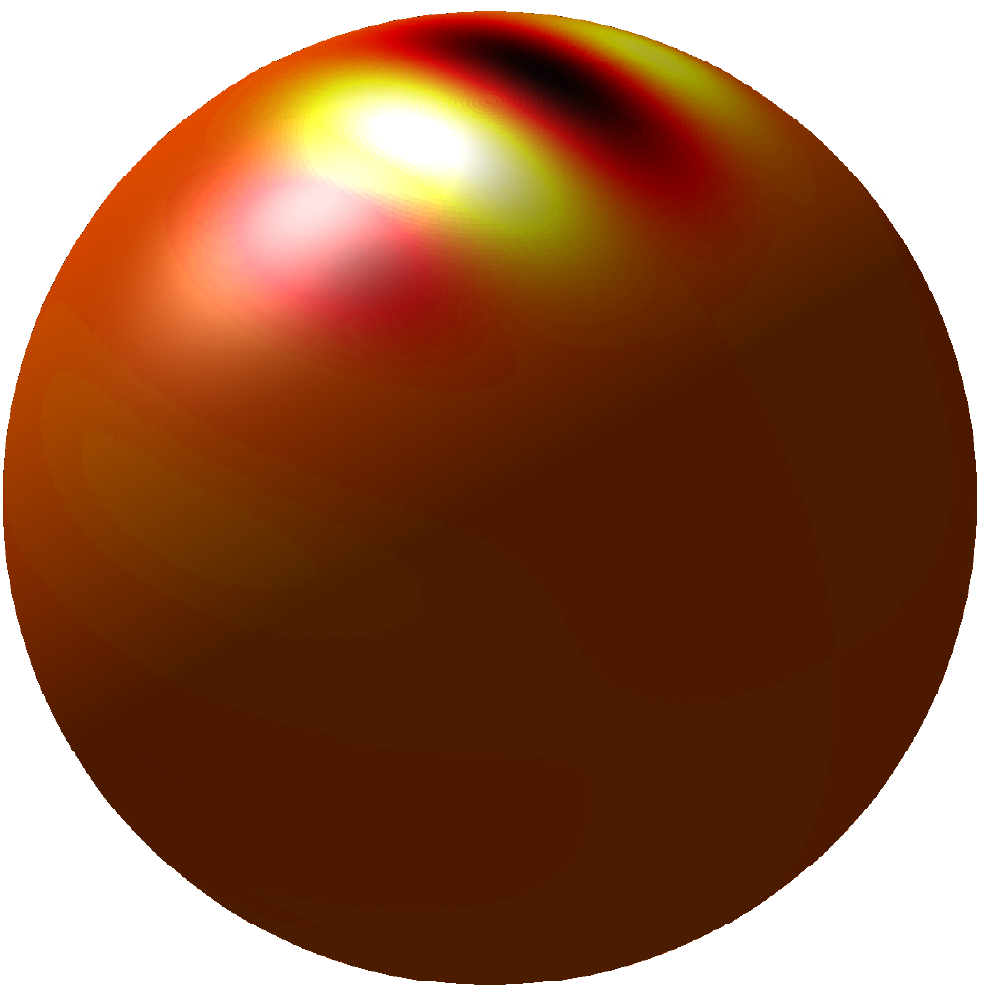}}
\subfigure[$\mmax=4,\ \wscale=4$]{\includegraphics[width=.23\textwidth]{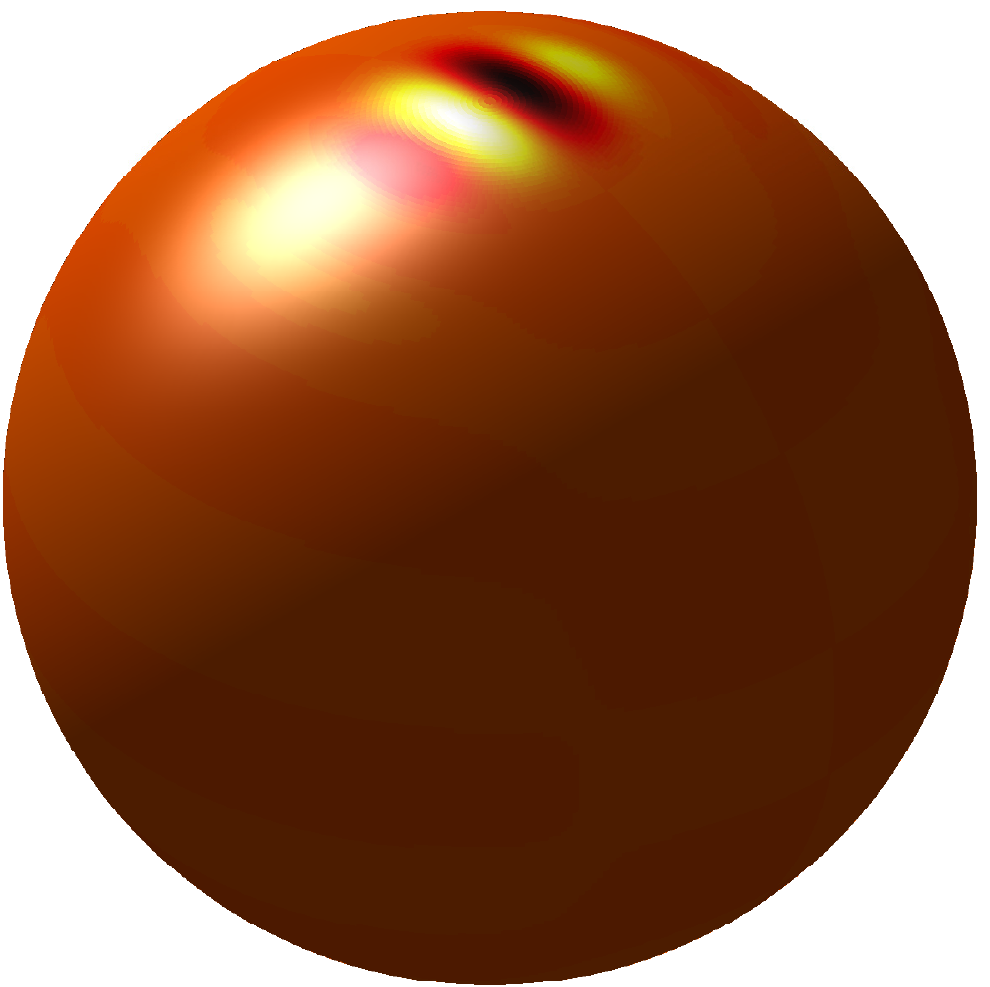}}
\subfigure[$\mmax=4,\ \wscale=3$]{\includegraphics[width=.23\textwidth]{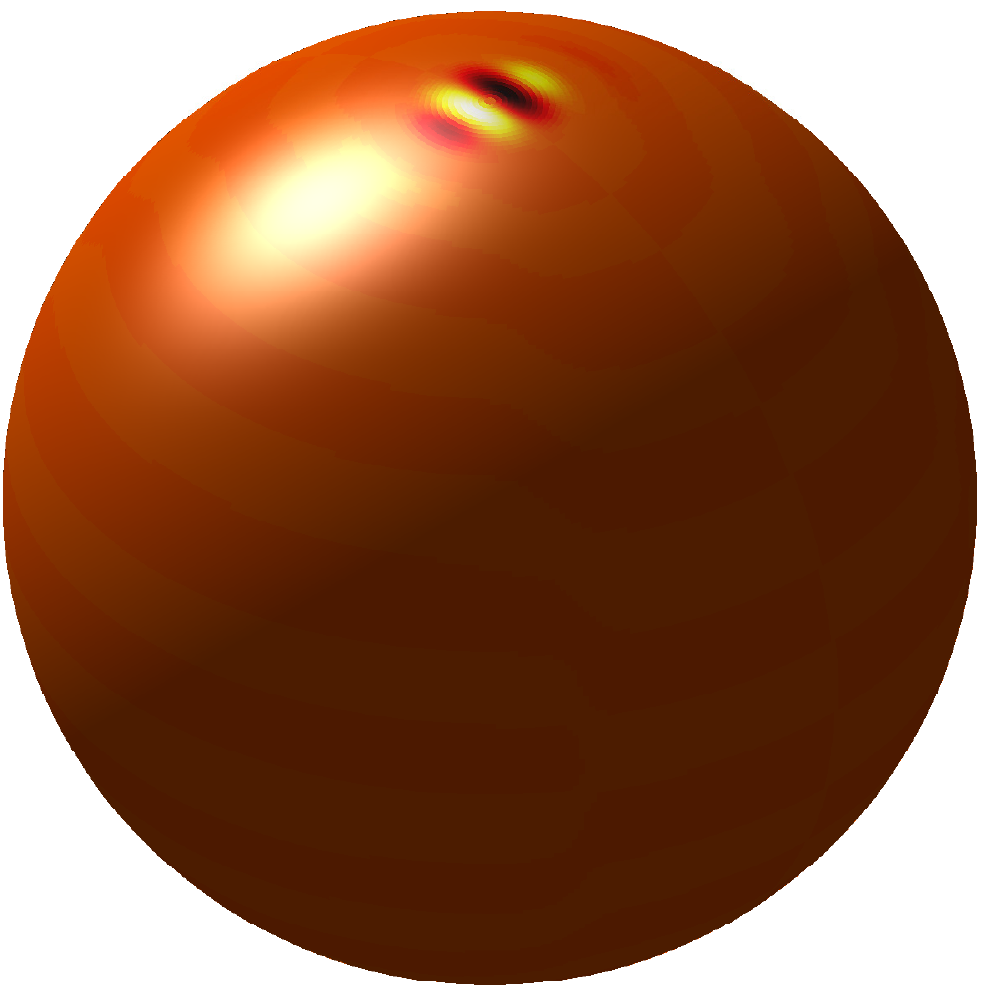}}
\includegraphics[viewport = 75 25 525 60, clip=true, width=.6\textwidth]{Figures/wavg_colorbar}
\caption{Directional scale-discretised wavelets
  ($\elmax = 256,\ \dilparam=2,\ \wscalemax=8$).  Wavelet scale
  $\wscale$ varies across columns, while azimuthal band-limit $\mmax$
  values across rows.}
\label{fig:wavelets}
\end{figure}

\begin{figure}
\subfigure[Odd $N-1$]{\includegraphics[width=.45\textwidth]{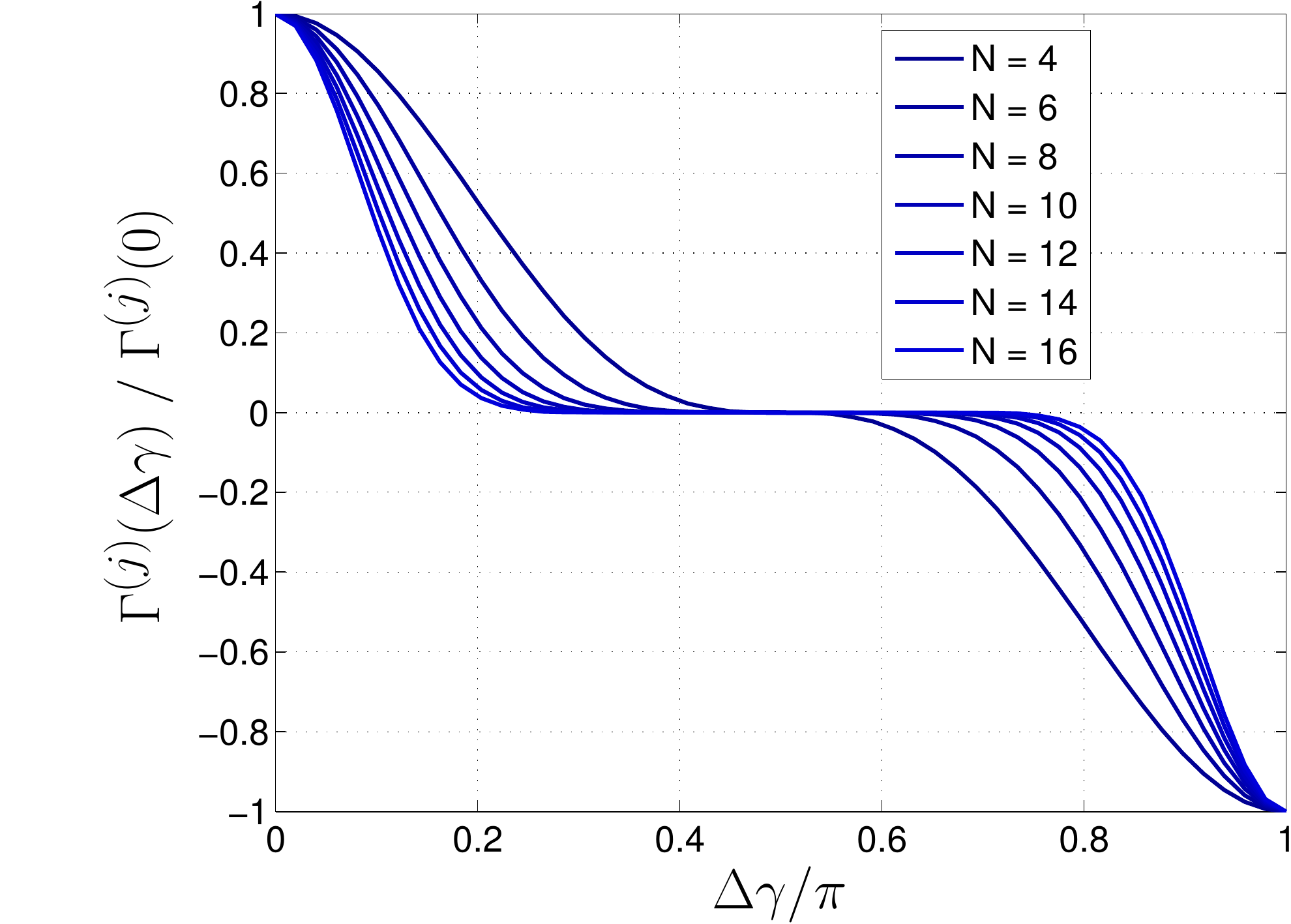}}
\subfigure[Even $N-1$]{\includegraphics[width=.45\textwidth]{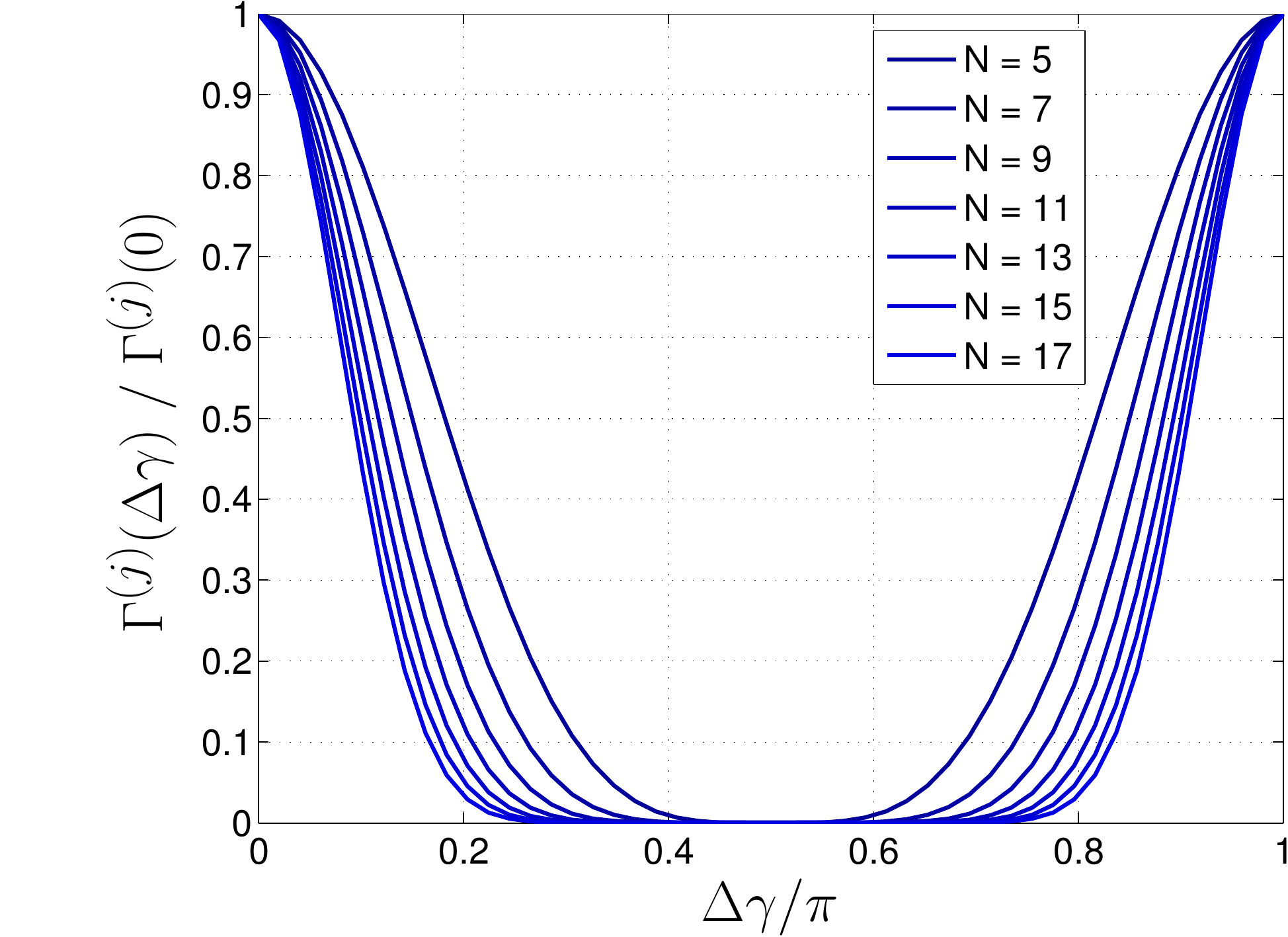}}
\caption{Directional auto-correlation for even and odd $\mmax-1$.  As
  \nmax\ increases the directional auto-correlation function becomes more
  peaked and the associated wavelet more directional.}
\label{fig:directional_correlation}
\end{figure}

The directionality component is constructed to carefully control the
directional localisation of the wavelet, while steerability is
achieved by imposing an azimuthal band-limit $\mmax$ (as described
above).  Directional localisation is controlled by imposing a specific form
for the directional auto-correlation of the wavelet, where the directional
auto-correlation is defined by 
\begin{equation}
  \Gamma^{{(\wscale)}}(\Delta \eulc=\eulc\p - \eulc) 
  \equiv \innerp{\wav^{(\wscale)}_\eulc}{\wav^{(\wscale)}_{\eulc\p}}
  % = \suml
  % \sum_{\m=-\min(N-1,\el)}^{\min(N-1,\el)}
  % \vert \shc{\wav}{\el}{\m}^{(\wscale)} \vert^2 \exp{(\img
  % \m \Delta \eulc)}
  = \suml \: \bigl\vert \wavker^{(\wscale)}(\el) \bigr\vert^2
  \sum_{\m=-\min(N-1,\el)}^{\min(N-1,\el)} 
  \bigl\vert \shc{\wavsteer}{\el}{\m} \bigr \vert^2 \:
  \exp{(\img \m \Delta \eulc)}
  \spcend ,
\end{equation}
where $\wav^{(\wscale)}_\eulc \equiv \rot_{(0,0,\eulc)} \wav^{(\wscale)}$.
The peakedness of the directional auto-correlation function can be
considered as a measure of the directionality of the wavelet: the more
peaked the directional auto-correlation, the more directional the wavelet
\cite{wiaux:2005, wiaux:2007:sdw}.  To control the directional
localisation of the wavelet precisely, we seek a directional
auto-correlation function of the form: 
\begin{equation}
  \Gamma^{{(\wscale)}}(\Delta \eulc)   
  = \suml \:
  \bigl\vert \wavker^{(\wscale)}(\el) \bigr\vert^2 \:
  \cos^{p}(\Delta\eulc)
  \spcend.
\end{equation}
The directionality component of the wavelet is then defined to satisfy
\begin{equation}
  \cos^{p}(\Delta\eulc)
  =
  \sum_{\m=-\min(N-1,\el)}^{\min(N-1,\el)}
  \: \bigl\vert \shc{\wavsteer}{\el}{\m} \bigr\vert^2 \:
  \exp{(\img \m \Delta \eulc)}
  \spcend,
\end{equation}
for $p\in\naturals$, where it is apparent that the modulus squared of
the spherical harmonic coefficients of the directionality component
identifies with the Fourier coefficients of $\cos^{p}(\Delta\eulc)$.
By noting Euler's formula and performing a binomial expansion,
$\cos^n(\phi)$ can be written
\begin{equation}
  \cos^{n}(\phi) 
  = \frac{1}{2^n}
  \sum_{m=0}^n \:
  \biggl( {{n}\atop{m}} \biggr) \:
  \exp{\bigl(\img (n-2m) \phi\bigr)}
  \spcend ,
\end{equation}
for $\phi \in [0, 2\pi)$, from which we recover the following Fourier
series expansions for even and odd exponents, respectively:
\begin{equation}
  \cos^{2n}(\phi) 
  = \frac{1}{2^{2n}}
  \sum_{\substack{m=-2n,\\ 
          m\:\text{even}}}^{2n} 
  \biggl( {{2n}\atop{n-m/2}} \biggr) \:
  \exp{(\img m \phi)}
\end{equation}
and
\begin{equation}
  \cos^{2n+1}(\phi) 
  = \frac{1}{2^{2n+1}}
  \sum_{\substack{m=-(2n+1),\\ 
          m\:\text{odd}}}^{2n+1} 
  \biggl( {{2n+1}\atop{n-(m-1)/2}} \biggr) \:
  \exp{(\img m \phi)}
  \spcend.
\end{equation}
Associating the Fourier coefficients of cosine raised to a power with
the harmonic coefficients of the directionality component of the
wavelet, the following coherent expression is recovered for both
even and odd exponents:
\begin{equation} 
  \shc{\wavsteer}{\el}{\m}
  = \eta \: \upsilon \:
  \sqrt{\frac{1}{2^{p}} \biggl( {{p}\atop{(p-m)/2}} \biggr)}
  \spcend,
\end{equation}
where 
\begin{equation}
  \eta =
    \begin{cases} 
      1, & \text{if $\mmax - 1$ even} \\ 
      \img, & \text{if $\mmax - 1$ odd}
    \end{cases}
    \spcend,
\end{equation}
\begin{equation}
  \upsilon = [1 - (-1)^{\mmax+\m}]/2=
    \begin{cases} 
      0, & \text{if $\mmax + \m$ even} \\ 
      1, & \text{if $\mmax + \m$ odd}
    \end{cases}
\end{equation}
and
\begin{equation}
  p = \min \{ \mmax-1, \el - [1 + (-1)^{\mmax+\el}]/2 \}=
    \begin{cases} 
      \min (\mmax-1, \el - 1 ), & \text{if $\mmax + \el$ even} \\ 
      \min (\mmax-1, \el ), & \text{if $\mmax + \el$ odd}
    \end{cases}
    \spcend.
\end{equation}
The factors $\eta$ and $\upsilon$ are introduced to ensure the
symmetries given by \eqn{\ref{eqn:symmetry_ref}} and
\eqn{\ref{eqn:symmetry_rot}}, respectively, are satisfied.  The
exponent $p$ is defined to achieve the greatest directionality
supported by the azimuthal band-limit available at a given $\el$.  For
wavelets with support within $\el \geq \mmax$, \ie\
$\wscale \leq \wscalemax_\mmax = \floor{
  \log_\dilparam(\elmax/\mmax)-1 }$,
(which is usually the case since $\mmax$ is typically chosen to be
relatively small and the scaling function is used to represent the
approximation-information of the signal), the parameter $p$ is given
by $\mmax-1$ and becomes independent of $\el$.
Note that the directional component normalisation of
\eqn{\ref{eqn:directionality_normalisation}} is satisfied since
$\sum_k \bigl( {{n}\atop{k}} \bigr) = 2^n$.
Example wavelets are plotted in \fig{\ref{fig:wavelets}}, while
directional auto-correlation functions are plotted in
\fig{\ref{fig:directional_correlation}}.

%=============================================================================
\section{Localisation properties}
\label{sec:localisation_properties}
%=============================================================================

We show in this section that directional scale-discretised wavelets
$\wav^{(\wscale) }$ are characterised by an excellent localisation
property in the spatial domain.  More precisely, for any
$\xi \in \realsnz$, there exists strictly positive
$C_{\xi ,\mmax}\in \realsnz$, such that the local concentration of a
directional wavelet centred on the North pole reads:
\begin{equation}
\bigl\vert \wav ^{(\wscale ) }( \sas ) \bigr\vert \leq 
\frac{\bigl(L\lambda^{-j}\bigr)^{ 2+\mmax}C_{\xi ,\mmax}}{\bigl( 1+L\lambda^{-j}\theta \bigr) ^{\xi }%
}\spcend .  \label{localisation}
\end{equation}
For the sake of simplicity, we present here the outline of the
proof of this property; all the mathematical technicalities are
extensively described in \appn{\ref{sec:appendix}}.

Firstly, observe the following decomposition of directional
scale-discretised wavelets in terms of spherical harmonics:
\begin{equation}
\wav ^{(\wscale) }( \omega ) = \suml \summ \kappa
^{(\wscale) }( \ell ) \: \wavsteer_{\ell m}\:\sqrt{\frac{2\ell +1}{8 \pi^2 }%
}\: 
  \shfarg{\el}{\m}{\omega} \label{dirwave}
  \spcend .
\end{equation}
For our purposes, we recall a general result in mathematical analysis,
namely Theorem 2.2 of \cite{geller:2009}, which states that the
spatial concentration properties of such wavelet constructions are
conserved under the action of $C^{\infty}$-differential operators, up
to a polynomial term depending on the degree of the operator. A full
statement of the theorem can be found in
\appn{\ref{sec:appendix:general_localisation}}.  In order to apply 
this result, we rewrite directional wavelets in terms of some proper
differential operator.

Let us start by assuming, for the sake of simplicity, that
$\wavsteer_{\ell m}$ does not depend on the multipole $\el$ but just
on the azimuthal angle.  For wavelet scales
$\wscale \leq \wscalemax_\mmax = \floor{
  \log_\dilparam(\elmax/\mmax)-1 }$
such that the harmonic support of the wavelet lies within
$\ell \geq N$ (the standard setting), this assumption is satisfied
directly (as discussed in \sectn{\ref{sec:construction:directional}}).
For $\ell <N$, straightforward calculations lead to:
\begin{equation}
\wavsteer_{\ell m}\leq \left( \frac{\ell }{\sqrt{2}}\right)
^{-\frac{1}{4}}\leq 1, 
\text{ for }\ell >1
\spcend. 
\end{equation}
Consequently, from now on we will write
$\wavsteer_{\m}\equiv \wavsteer_{\el \m}$.  

Let us define the following $C^{\infty }$-differential operator on
$\theta $ of order $m$ on the sphere:
\begin{equation}\label{TT}
\Toperator_{m}\equiv (\sin \theta )^{m}\frac{\partial ^{m}}{\partial (\cos \theta )^{m}}%
=\sum_{k=0}^{m}a_{k}\sin ^{k}\theta \frac{\partial ^{k}}{\partial \theta ^{k}%
}
\spcend .
\end{equation}
Following \eqn{\ref{Toper}} in
\appn{\ref{sec:appendix:special_functions}}, we can rewrite the
wavelet as
\begin{equation}
\wav ^{(\wscale) }( \sas) 
=\sum_{m=-\mmax}^{\mmax}\wavsteer_{m} \: \exp{(\img m\sab)}\left(-1\right)^m\sum_{\ell =(L\lambda ^{-( 1+j)
}\vee m)}^{L\lambda ^{1-j}}\kappa ^{( j) }( \ell ) \:
\frac{2\ell +1}{2^{5/2}\pi^{3/2} }\sqrt{\frac{( \ell -m) !}{( \ell
+m) !}} \:
\Toperator_{m} P_{\ell}(\cos \theta )
\spcend ,
\end{equation}
where $P_{\ell}(\cdot)$ are the Legendre polynomials (defined in
\appn{\ref{sec:appendix:special_functions}}) and we have adopted the shorthand
notation $a \vee b = \max(a,b)$.
Straightforward manipulations lead to the following bound,
for any $\sab$:
\begin{equation}
\bigl\vert \wav ^{(\wscale) }( \sas ) \bigr\vert \leq
( 2\mmax+1) \max_{m=-\mmax,...,\mmax}
\:
\bigl \vert \wavsteer_{m} \bigr\vert
\:
\bigl\vert Q_{m}^{( j) }( \cos \theta ) \bigr\vert 
\spcend,
\end{equation}
where 
\begin{equation}
Q_{m}^{( j) }( \cos \theta ) \equiv \sum_{\ell =(L\lambda
^{-( 1+j) }\vee m)}^{L\lambda ^{1-j}}\kappa ^{( j)
}( \ell ) \: \frac{2\ell +1}{2^{7/2}\pi^{5/2} } \: \ell ^{-m} \: \Toperator_\m P_{\ell }(\cos \theta
)
\spcend.
\end{equation}
We have employed the bound of \eqn{\ref{glm}} above, which is derived in
\appn{\ref{sec:appendix:bound_el_m}}.

For the sake of simplicity, let $\varepsilon _{j}=\lambda ^{j}L^{-1}$
and define the function $b_{m,\varepsilon_\wscale }$ mapping $\mathbb{R}$ to
$\realsnn$ as
\begin{equation}
b_{m,\varepsilon_\wscale }(x)\equiv \kappa ^{( j) }( x) \: x^{-m}, \text{ for }
x>0 \spcend .
\end{equation}
As in \cite{narcowich:2006}, we extend this function to the negative
axis simply by taking
$b_{m,\varepsilon_\wscale }(-\vert x \vert )=b_{m,\varepsilon_\wscale
}(\vert x\vert )$. Furthermore, because the function
$\kappa ^{( j) }$ assumes the value $0$ in the interval
$( -\dilparam^{-(1+\wscale)}\elmax,\dilparam^{-(1+\wscale)}\elmax) $,
it also holds that 
$b_{m,\varepsilon_\wscale }(x)=0$ for $x\in ( -\dilparam^{-(1+\wscale)}\elmax,0) \cup (
0,\dilparam^{-(1+\wscale)}\elmax) $. Moreover, we extend for
continuity $b_{m,\varepsilon_\wscale }( 0) =0$. 

We then obtain:
\begin{equation}\label{ququ}
Q_{m}^{( j) }(\cos \theta ) =\frac{1}{\left(2\pi\right)^{5/2}}\Toperator_{m} U^{\left(\wscale\right)}_{m}( \cos \theta ) 
\spcend ,
\end{equation}%
where  
\begin{equation} \label{Kkernel}
U^{\left(\wscale\right)}_{m}( \cos \theta ) \equiv \sum_{\ell =(L\lambda ^{-( 1+j)
}\vee m)}^{L\lambda ^{1-j}}b_{m,\varepsilon_\wscale }( \ell ) \Bigl( \ell
+\frac{1}{2}\Bigr) P_{\ell }( \cos \theta ) 
\spcend .
\end{equation}
As proved in \appn{\ref{sec:appendix:bound_u}}, for
any $\xi \in \realsnz$, there exists
$C_{\xi} \in \realsnz$ such that $U^{\left(\wscale\right)}_{m}( \cos \theta )$ is bounded above:
\begin{equation}
\bigl\vert U^{\left(\wscale\right)}_{m}( \cos \theta ) \bigr\vert
\leq \frac{C_{\xi}/\varepsilon_\wscale ^{2}}{\Bigl(
1+\bigl\vert \frac{\theta }{\varepsilon_\wscale }\bigr\vert \Bigr)
^{\xi}}
\spcend .
\end{equation}
According to \eqn{\ref{gellermayeli1}} of
\appn{\ref{sec:appendix:general_localisation}} it follows that 
\begin{equation}
\bigl\vert Q^{(\wscale)}_{m}(\cos \theta )\bigr\vert \leq \frac{(L\lambda^{-j} )^{
2+m}C_{\xi }}{\bigl( 1+L\lambda ^{-j}\theta \bigr) ^{\xi }}
\spcend ,
\end{equation}
and, therefore, we obtain 
\begin{align}
\bigl\vert \wav ^{( j) }( \sas ) \bigr\vert  
% &\leq
% \sum_{m=-M}^{M}\bigl\vert \wavsteer_{m}\bigr\vert \: \bigl\vert \exp{(\img m\sab)
% }\bigr\vert \:
% \bigl\vert Q^{(\wscale)}_{m}(\cos \theta )\bigr\vert  \\
&\leq ( 2\mmax+1) \max_{m=-\mmax,...,\mmax} \: \bigl\vert \wavsteer_{m}
\bigr\vert \: \frac{(L\lambda^{-j}) ^{
2+\mmax}C_{\xi }}{\bigl( 1+L\lambda ^{-j}\theta \bigr) ^{\xi }} \\
&\leq \frac{(L\lambda^{-j}) ^{ 2+\mmax }C_{\xi ,\mmax}}{\bigl( 1+L\lambda
^{-j}\theta \bigr) ^{\xi }}
\spcend ,
\end{align}%
as claimed. 

If we consider spin scale-discretised wavelets
\cite{mcewen:s2let_spin}, denoted by $\swav^{(\wscale)}(\omega)$,
$\omega \in \sphere$, the following decomposition in terms of spin
spherical harmonics $ \sshfarg{\el}{\m}{\omega}{s}$ holds:
\begin{equation}
\swav ^{(\wscale) }( \omega ) = \suml \summ \kappa
^{(\wscale) }( \ell ) \: \wavsteer_{\ell m}\:\sqrt{\frac{2\ell +1}{8 \pi^2 }%
}\:\sshfarg{\el}{\m}{\omega}{s} 
    \label{dirwavespin} 
\spcend.
\end{equation}
Details and properties concerning spin spherical harmonics are
extensively discussed in the
\appn{\ref{sec:appendix:special_functions}}. Considering $s>0$, and using
\eqn{\ref{spintoscale1}}, we obtain
\begin{equation}
\swav ^{(\wscale) }( \omega ) = \suml \summ \kappa
^{(\wscale) }( \ell ) \: \wavsteer_{\ell m}\:\sqrt{\frac{2\ell +1}{8 \pi^2 }%
}\ \biggl[ \frac{(\el-s)!}{(\el+s)!} \biggr]^{1/2} 
  \spinup^s
  \shfarg{\el}{\m}{\omega}
  \spcend
%    \label{dirwavespin} 
\spcend.
\end{equation}
Observe that $ \bigl[{(\el-s)!}/{(\el+s)!}\bigr]^{1/2}$ is bounded by $\ell^{-s}$,
as given by \eqn{\ref{glm}} and shown in
\appn{\ref{sec:appendix:bound_el_m}}. Noting
\eqn{\ref{eqn:spherical_harmonic_function}}, \eqn{\ref{eqn:associated_legendre_function}} and \eqn{\ref{Toper}},
straightforward calculations, entirely analogous to the scalar case
and therefore omitted for the sake of brevity, lead to the
following inequality:
\begin{equation}
\bigl\vert \swav ^{(\wscale) }( \sas ) \bigr\vert \leq
( 2\mmax+1) \max_{m=-\mmax,...,\mmax}
\:
\bigl \vert \wavsteer_{m} \bigr\vert
\:
\bigl\vert {}_{s}Q_{m}^{( j) }( \cos \theta ) \bigr\vert 
\spcend,
\end{equation}
where 
\begin{equation}
{}_{s}Q_{m}^{( j) }( \cos \theta ) \equiv \sum_{\ell =(L\lambda
^{-( 1+j) }\vee m)}^{L\lambda ^{1-j}}\kappa ^{( j)
}( \ell ) \: \frac{2\ell +1}{2^{7/2}\pi^{5/2} } \: \ell ^{-(m+s)} \: \spinup^s \Toperator_\m P_{\ell }(\cos \theta
)
\spcend.
\end{equation}
Similar manipulations of ${}_{s}Q_{m}^{( j) }( \cos \theta )$ to those
presented above (see again \ref{sec:appendix:general_localisation}),
lead to the following result.  For any $\xi \in \realsnz$, there
exists a strictly positive $C_{\xi ,\mmax,s}\in \realsnz$ such that
\begin{align}
\bigl\vert {}_{s}\wav ^{( j) }( \sas ) \bigr\vert  
&\leq \frac{(L\lambda^{-j}) ^{2+\mmax+s}C_{\xi ,\mmax,s}}{\bigl( 1+L\lambda
^{-j}\theta \bigr) ^{\xi }}
\spcend .
\end{align}

%=============================================================================
\section{Stochastic properties}
\label{sec:stochastic_properties}
%=============================================================================

We study in this section the stochastic properties of zero-mean
homogeneous and isotropic Gaussian random fields on the sphere $\f$
when decomposed by directional scale-discretised wavelets.  The stochastic field
\f\ is characterised by its power spectrum $\noisecl_\el$, with
\begin{equation}
  \expv(\shc{\f}{\el}{\m} \shcc{\f}{\el\p}{\m\p} ) 
  = \noisecl_\el \: 
  \kron{\el}{\el\p}
  \kron{\m}{\m\p}
\end{equation}
and
\begin{equation}
  \expv(\shc{\f}{\el}{\m} ) = 0 
  \spcend .
\end{equation}
Specifically, we study the correlation of directional scale-discretised
wavelet coefficients given by
\begin{equation}
  \label{eqn:correlation}
  \Xi^{(\wscale\wscale\p)}(\eul_1,\eul_2) \equiv
  \frac{ \expv \Bigl[ \wcoeff^{\wav^{(\wscale)}}(\eul_1) \:
    \wcoeff^{\wav^{(\wscale\p)}}{}^{\cconj}(\eul_2) \Bigr]}
  {\sqrt{\expv \biggl[ \Bigl\vert\wcoeff^{\wav^{(\wscale)}}(\eul_1)\Bigr\vert^2
      \biggr]}
    \sqrt{\expv \biggl[
      \Bigl\vert\wcoeff^{\wav^{(\wscale\p)}}(\eul_2)\Bigr\vert^2 \biggr]}
  }
  \spcend .
\end{equation}
For notational convenience, we also introduce the covariance
\begin{equation}
  \xi^{(\wscale\wscale\p)}(\eul_1,\eul_2)
  \equiv
  \expv \Bigl[ \wcoeff^{\wav^{(\wscale)}}(\eul_1) \:
  \wcoeff^{\wav^{(\wscale\p)}}{}^{\cconj}(\eul_2) \Bigr]  
  \spcend ,
\end{equation}
such that 
\begin{equation}
  \Xi^{(\wscale\wscale\p)}(\eul_1,\eul_2) 
  = 
  \frac{\xi^{(\wscale\wscale\p)}(\eul_1,\eul_2)}
  {\sqrt{\xi^{(\wscale\wscale)}(\eul_1,\eul_1) \: \xi^{(\wscale\p\wscale\p)}(\eul_2,\eul_2)}}
  \spcend .
\end{equation}

Recall the harmonic representation of the scale-discretised wavelet
transform of \eqn{\ref{eqn:analysis_harmonic}}.
%, repeated here for convenience:
% \begin{equation}
%   \wcoeff^{\wav^{(\wscale)}}(\eul) 
%   = \innerp{\f}{\rotarg{\eul}\wav^{(\wscale)}}
%   = \sumlmn
%   \shc{\f}{\el}{\m}
%   \shc{\wav}{\el}{\n}^{(\wscale)\cconj}
%   \Dlmnpc  
%   \spcend .
% \end{equation}
Noting this expansion, the wavelet covariance may be written
\begin{align}
  \xi^{(\wscale\wscale\p)}(\eul_1,\eul_2)
  & =
  \suml \sumn \sum_{\n\p=-\el}^\el
  \noisecl_\el \:
  \shc{\wav}{\el}{\n}^{(\wscale) \cconj} \:
  \shc{\wav}{\el}{\n\p}^{(\wscale\p)}
  \summ 
  \dmatbig_{\m \n}^{\el \cconj}(\eul_1) \:
  \dmatbig_{\m \n\p}^{\el}(\eul_2) \\
  & =
  \suml \sumn \sum_{\n\p=-\el}^\el
  \noisecl_\el \:
  \shc{\wav}{\el}{\n}^{(\wscale) \cconj} \:
  \shc{\wav}{\el}{\n\p}^{(\wscale\p)} \:
  \dmatbig_{\n\n\p}^{\el}(\eul) 
  \spcend ,
\end{align}
where the second line follows from \eqn{\ref{eqn:wignerd_add_conj}}
and $\rotmatarg{\eul} = \rotmatarg{\eul_1}^{-1} \rotmatarg{\eul_2}$
(see \appn{\ref{sec:appendix:special_functions}} for further
details).  For the case where $\eul_1 = \eul_2$ and $\wscale\p = \wscale$, the covariance
reduces to 
\begin{equation}
  \label{eqn:covariance_analytic}
  \xi^{(\wscale\wscale)}(\eul,\eul) 
  = 
  \suml \sumn
  \noisecl_\el \:  
  \bigl \vert \shc{\wav}{\el}{\n}^{(\wscale)} \bigr \vert^2
  \spcend ,
\end{equation}
by \eqn{\ref{eqn:wignerd_add_conj_same_ang}} of
\appn{\ref{sec:appendix:special_functions}} (as also shown in
\cite{mcewen:2006:isw}).

Expressing the wavelet by its kernel and directionality component, the
wavelet covariance reads:
\begin{equation}
  \xi^{(\wscale\wscale\p)}(\eul_1,\eul_2)
  =
  \suml \sumn \sum_{\n\p=-\el}^\el
  \noisecl_\el \:
  \frac{2\el+1}{8\pi^2} \:
  \wavker^{(\wscale)}(\el) \:
  \wavker^{(\wscale\p)}(\el) \:
  \shc{\wavsteer}{\el}{\n} \:
  \shc{\wavsteer}{\el}{\n\p} \:
  \dmatbig_{\n\n\p}^{\el}(\eul) 
\end{equation}
Furthermore, using \eqn{\ref{Toper}}, we obtain
\begin{equation}
  \xi^{(\wscale\wscale\p)}(\eul_1,\eul_2)
  =  \xi_{(+)}^{(\wscale\wscale\p)}(\eul_1,\eul_2) +  \xi_{(-)}^{(\wscale\wscale\p)}(\eul_1,\eul_2) ,
\end{equation}
where 
\begin{align}
  \xi_{(+)}^{(\wscale\wscale\p)}(\eul_1,\eul_2)
=&
  \suml \sumn \sum_{\n\p=0}^\el
  \noisecl_\el \:
  \frac{2\el+1}{2^{7/2}\pi^{5/2} }
  \wavker^{(\wscale)}(\el) 
  \wavker^{(\wscale\p)}(\el) 
  \shc{\wavsteer}{\el}{\n}
  \shc{\wavsteer}{\el}{\n\p} \nonumber \\
  &\times (-1)^\n
  \sqrt{\frac{(\el-\n\p)! (\el-\n)!}{(\el+\n\p)! (\el+\n\p)!}} \:
  \exp{\bigl(\img (-\n\eula + \n\p \eulc)\bigr)} \:
  \spindown^{\n\p} \:
  \Toperator_{n} \:
  \leg{\el}{\cos\eulb} \label{xi1}   
\end{align}
and
\begin{align}
  \xi_{(-)}^{(\wscale\wscale\p)}(\eul_1,\eul_2)
  =&
  \suml \sumn \sum_{\n\p=-\el}^{-1}
  \noisecl_\el \:
  \frac{2\el+1}{2^{7/2}\pi^{5/2} }
  \wavker^{(\wscale)}(\el) 
  \wavker^{(\wscale\p)}(\el) 
  \shc{\wavsteer}{\el}{\n}
  \shc{\wavsteer}{\el}{\n\p} \nonumber \\
  &\times 
  \sqrt{\frac{(\el-|\n\p|)! (\el-\n)!}{(\el+|\n\p|)! (\el+\n\p)!}} \:
  \exp{\bigl(\img (-\n\eula + \n\p \eulc)\bigr)} \:
  \spindown^{|\n\p|} \:
  \Toperator_{n} \:
  \leg{\el}{\cos\eulb} , \label{xi2}
\end{align}
where in \eqn{\ref{xi1}} we applied \eqn{\ref{wigner2}}, while in
\eqn{\ref{xi2}} we applied \eqn{\ref{wigner3}}. In both those formulas
and henceforth, $\eulb$ is the Euler angle associated to the
resultant rotation $\eul=(\euls)$, as stated above. These
expression show that, up to a complex exponential factor, both
$\xi_{(+)}^{(\wscale\wscale\p)}(\eul_1,\eul_2)$ and
$\xi_{(-)}^{(\wscale\wscale\p)}(\eul_1,\eul_2)$ depend on the absolute
value of $\n \p$.

Following, for instance, \cite{baldi:2009}, we introduce some mild
regularity conditions on $C_{\ell }$ (see also
\cite{marinucci:2011:book}). Assume there exists $R \in\naturals$,
$\alpha \in \reals$, $\alpha\geq 2$ and a sequence of functions
$\bigl\{g^{(\lambda)}_j(\cdot)\bigr\}$ such that we can rewrite
\begin{equation}
C_\ell =\ell^{-\alpha} g_j^{(\lambda)}(\lambda^j L^{-1} \ell  )>0
\spcend ,
\end{equation}
for $\el \in \bigl[\floor{\dilparam^{-(1+\wscale)} \elmax},
\ceil{\dilparam^{1-\wscale} \elmax} \bigr]$ and for $j \in \naturals$,
$0<g_j^{(\lambda)}<\infty$, while for $r=1,\ldots,R$, there exists
$c_r \in \reals^+$ such that
\begin{equation}
\sup_{j \in \naturals} \: \sup_{u\in [L \lambda^{-1}, L \lambda]} \:
\Bigl\vert\frac{d^r}{du^r}g_j^{(\lambda)}(u)\Bigr\vert \leq c_r
\spcend .
\end{equation}
These conditions guarantee the boundedness and smoothness of $C_\ell$
and are useful in the context of practical applications. For instance,
these conditions encompass standard cosmological models describing the
CMB, where the CMB is modelled as a realisation of a Gaussian random
field on the sphere and where $C_\ell$ can be modelled approximately by
inverse polynomials (\cf\ \cite{dodelson:2003}).  Note also that, as stated in
\cite{baldi:2009}, the sequence $\bigl\{g^{(\lambda)}_j(\cdot)\bigr\}$
belongs to the Sobolev space $W^{R,\infty}$. Because it follows
immediately that there should exist $g_1, g_2, \alpha \in \realsnn$,
$\alpha \geq 2$, such that
$g_1 \ell^{-\alpha} \leq C_\ell \leq g_1 \ell^{-\alpha}$, for the sake
of the simplicity and without losing any generality, we assume
henceforth $C_\ell = g_1 \ell^{-\alpha}$ (see, again, \cite{baldi:2009}).

Consider now the variance: in order to compute its lower bound, observe that, for $\ell$
sufficiently large, the following integral approximation holds
\begin{align}
  \suml 
 (\lambda^\wscale L^{-1} \ell)^{-\alpha} \:
  (\lambda^\wscale L^{-1})^2 \left(2\el+1\right) \:
  \bigl(\wavker^{(\wscale)}(\el)\bigr)^2  \:
 &  =  \sum_{\ell = \floor{\dilparam^{-(1+\wscale)} \elmax}}^{\ceil{\dilparam^{1-\wscale} \elmax} } \:
 (\lambda^\wscale L^{-1} \ell)^{-\alpha} \:
 (\lambda^\wscale L^{-1})^2 \left(2\el+1\right) \:
  \bigl(\kappa_\lambda(\lambda^j L^{-1} \el)\bigr)^2  \: \nonumber\\
  & \simeq 2\int_{\lambda^{-1} }^{\lambda}x^{1-\alpha} \:
    \bigl(\wavker_{\lambda}(x)\bigr)^2 \dx x 
  \spcend , 
\end{align}
where
$0< C_1 \leq 2\int_{\lambda^{-1} }^{\lambda
}x^{1-\alpha} \: \bigl(\wavker_{\lambda}(x)\bigr)^2 \dx x \leq C_2
<\infty$,
and $C_1, C_2 \in \realsnn$ (\cf\ Lemma 3 in \cite{baldi:2009}). Recalling
$\sumn \bigl \vert \zeta_{\n}\bigr \vert^2=1$, straightforward manipulations
lead us to the following inequality:
 \begin{equation}%\label{den}
 \xi^{(\wscale\wscale)}(\eul,\eul)
% &  = 
%   \suml \sumn 
%   \noisecl_\el \:
%   \frac{2\el+1}{8\pi^2} \:
%   (\wavker^{(\wscale)}(\el))^2 \:
%   \bigl\vert  \shc{\wavsteer}{\el}{\n} \bigr\vert^2 \nonumber \\
   = 
  \suml 
  \noisecl_\el \:
  \frac{2\el+1}{8\pi^2} \:
  \bigl(\wavker^{(\wscale)}(\el)\bigr)^2 
  \geq  C_{1} ( L \lambda^{-\wscale})^{(2-\alpha)} .
\end{equation}
As far as the correlation $\xi^{(\wscale\wscale\p)}(\eul_1,\eul_2)$ is
concerned, observe that
$\wavker^{(\wscale)}(\el) \wavker^{(\wscale\p)}(\el)=0$ if
$|\wscale-\wscale\p|>1$, so that
$\xi^{(\wscale\wscale\p)}(\eul_1,\eul_2)$ is different from $0$ only
if $\wscale \p =\wscale$ or if $\wscale \p = \wscale \pm1$.

Let us consider $\wscale = \wscale \p$ and define a sequence of
functions $\phi_{\lambda, 1}^{(\wscale)}$ given by
$\phi_{\lambda, 1}^{(\wscale)}(u)=
\bigl(\wavker^{(\wscale)}(u)\bigr)^2 g_j^{(\lambda)}(u)
u^{-\alpha-n-n^{\prime}}$.
Assuming again that $\shc{\wavsteer}{\el}{\n} = \zeta_{n}$, we have
\begin{equation}
  \xi_{(+)}^{(\wscale \wscale)}(\eul_1,\eul_2)
 \leq 
 \sum_n \sum_{\n\p}  \zeta_{n}
  \zeta_{n^\prime}  \sum_l
  \frac{2\el+1}{2^{7/2}\pi^{5/2} } \:
\phi_{\lambda, 1}^{(\wscale)}(\el) \:
  \spindown^{\n\p} 
  \Toperator_{n} 
  \leg{\el}{\cos\eulb} \spcend. 
\end{equation}
A similar result is attained as far as
$\xi_{(-)}^{(\wscale\wscale)}(\eul_1,\eul_2)$ is concerned.  Recalling
that both $n$ and $n^\prime$ are bounded by $\mmax$, we apply again
the same techniques used to achieve localisation in Section
\ref{sec:localisation_properties}, where
$\phi_{\lambda, 1}^{(\wscale)}(\el)$ plays the same role as
$b_{m,\epsilon_j}\left(\ell\right)$ in \eqn{\ref{ququ}}, here
omitted for the sake of brevity. These considerations lead to the
following result.  For any $\xi ^{\prime } \in \reals^+$,
$\xi ^{\prime }\geq 2( 1+\mmax) $, there exists $C_{\xi ^{\prime },M}$
such that 
\begin{equation}\label{num}
\xi ^{(j j)}(\eul_1,\eul_2) \leq \frac{%
(L\lambda^{-j}) ^{2( 1+\mmax)-\alpha}C_{\xi ^{\prime },\mmax}}{\bigl( 1+L\lambda
^{-j}\eulb \bigr) ^{\xi ^{\prime }}}
\spcend.
\end{equation}
Combining \eqn{\ref{den}} and \eqn{\ref{num}}, we attain the following bound. For any
$\xi ^{\prime \prime } \in \reals^+$, $\xi ^{\prime \prime}\geq 2\mmax $, there
exists $C_{\xi ^{\prime \prime },M}$ such that
\begin{equation}\label{corr1}
\Xi ^{(j j)}(\eul_1,\eul_2) \leq \frac{%
(L\lambda^{-j}) ^{2\mmax}C_{\xi ^{\prime \prime },\mmax}}{\bigl( 1+L\lambda
^{-j}\eulb \bigr) ^{\xi ^{\prime \prime }}}
\spcend.
\end{equation}

Likewise, let us suppose $j^{\prime}= j-1$, remarking that the case
$j^{\prime}= j+1$ is entirely analogous. Let us define a sequence of
functions $\phi_{\lambda, 2}^{(\wscale)}$ given by
$\phi_{\lambda, 1}^{(\wscale)}(u)=
\wavker^{(\wscale)}(u)\wavker^{(\wscale-1)}(u) g_j^{(\lambda)}(u)
u^{-\alpha-n-n^{\prime}}$.
Again recalling that $\shc{\wavsteer}{\el}{\n} = \zeta_{n}$, we have
\begin{equation}
  \xi_{(+)}^{(\wscale,\wscale-1)}(\eul_1,\eul_2)
 \leq 
 \sum_n \sum_{\n\p}  \zeta_{n}
  \zeta_{n^\prime}  \sum_l
  \frac{2\el+1}{8\pi^2} \:
\phi_{\lambda, 2}^{(\wscale)}(\el)\:
  \spindown^{\n\p} 
  \Toperator_{n} 
  \leg{\el}{\cos\eulb}.
\end{equation}
Straightforward calculations, similar to the case $j=j^\prime$, provide
an identical bound for $\Xi ^{(j,j^\prime)}(\eul_1,\eul_2)$ . 

Therefore, for the sake of the clarity, we state directly the final
result: for any $j, j^\prime \in \naturals$ such that
$\left|j-j^\prime\right|<2$ and for any $\xi_{0} \in \reals^+$,
$\xi _{0}\geq 2\mmax $, there exists $C_{\xi _{0},M}$ such that
\begin{equation}%\label{corr2}
\Xi ^{(j j^\prime)}(\eul_1,\eul_2) \leq \frac{%
(L\lambda^{-j}) ^{2\mmax}C_{\xi_{0},\mmax}}{\bigl( 1+L\lambda
^{-j}\eulb \bigr) ^{\xi_{0}}}
\spcend.
\end{equation} 

As far as spin scale-discretised wavelets are concerned
\cite{mcewen:s2let_spin}, consider an isotropic spin $s$ random field
\mbox{${}_s f \in \sphere$} and its corresponding spherical harmonic
coefficients ${}_s f_{\ell m}=\langle f_s , {}_s Y_{\el m}\rangle$. As
proved in Theorem 7.2 of \cite{geller:2010:sw}, it holds that
$\mathbb{E}({}_s f_{\el m} {}_sf_{\el^\prime m^\prime}^*)=C_\ell
\kron{\el}{\el\p} \kron{\m}{\m\p}$,
where $C_\el$ is the spin power spectrum, which is invariant with
respect to the choice of the system of coordinates over $\sphere$ (see
also \cite{geller:2010}). Therefore, the upper bound established for
the correlation between spin directional scale-discretised wavelet
coefficients is entirely analogous to the one developed for the scalar
case. Indeed, the spin wavelet correlation becomes
\begin{align}
{}_s  \xi^{(\wscale\wscale\p)}(\eul_1,\eul_2)
  & =
  \suml \sumn \sum_{\n\p=-\el}^\el
  \noisecl_\el \:
  {}_s \shc{\wav}{\el}{\n}^{(\wscale) \cconj} \:
  {}_s \shc{\wav}{\el}{\n\p}^{(\wscale\p)} \:
  \dmatbig_{\n\n\p}^{\el}(\eul) 
  \spcend ,
\end{align}
and the wavelet variance is bounded as 
\begin{equation}\label{den}
{}_s \xi^{(\wscale\wscale)}(\eul,\eul)
   = 
  \suml 
  \noisecl_\el \:
  \frac{2\el+1}{8\pi^2} \:
  \bigl(\wavker^{(\wscale)}(\el)\bigr)^2 
  \geq  C_{1} ( L \lambda^{-\wscale})^{(2-\alpha)} ,
\end{equation}
because
$\sum_{m=-\el}^{\el} {}_s Y_{\el m}\left(\sa\right) {}_s Y_{\el
  m}^*\left(\sa\right)=\frac{2 \el +1 }{4 \pi}$,
for any $\sa \in \sphere$. Straightforward calculations lead to the
following result.  For any $j, j^\prime \in \naturals$ such that
$\left|j-j^\prime\right|<2$ and for any $\xi_{0}^\prime \in \reals^+$,
$\xi _{0}^{\prime}\geq 2\mmax $, there exists $C_{\xi _{0}^\prime,M}$ such that
\begin{equation}%\label{corr2}
{}_s \Xi ^{(jj^\prime)}(\eul_1,\eul_2) \leq \frac{%
(L\lambda^{-j}) ^{2\mmax}C_{\xi_{0}^\prime,\mmax}}{\bigl( 1+L\lambda
^{-j}\eulb \bigr) ^{\xi_{0}^\prime}}
\spcend.
\end{equation} 

%=============================================================================
\section{Numerical experiments}
\label{sec:numerical_experiments}
%=============================================================================

\begin{figure}
\centering
\includegraphics[viewport= 75 110 520 335,clip = true, width=.6\textwidth]{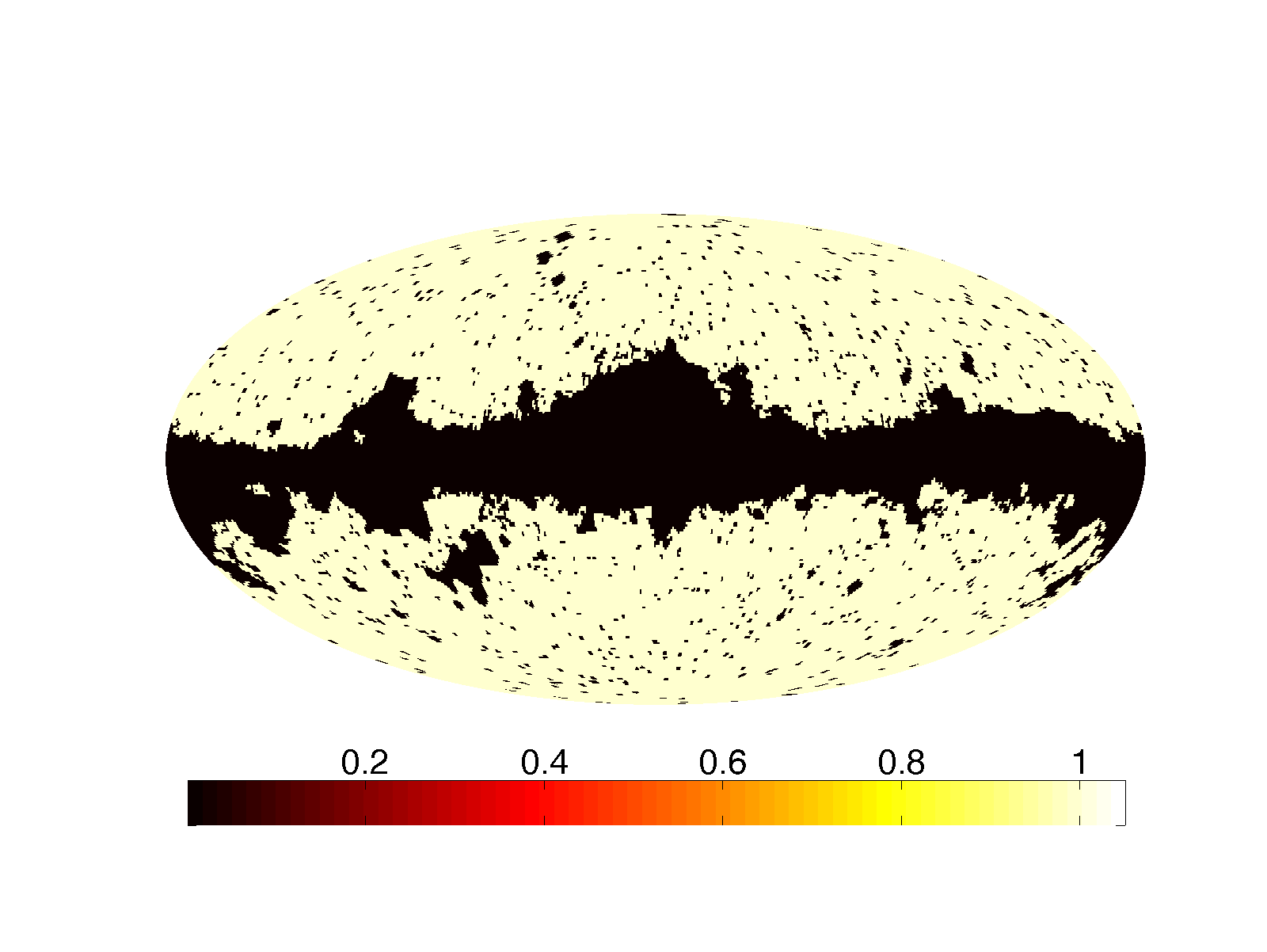}\\
\includegraphics[viewport= 75 55 520 110,clip = true, width=.6\textwidth]{Figures/mask}
\caption{Binary (WMAP9 KQ75) mask plotted using the Mollweide projection showing regions of the sky where the
  CMB is accurately observable (ones of mask) and unobservable (zeros
  of mask).}
\label{fig:mask}
\end{figure}

We perform numerical experiments to study the localisation and
correlation properties of directional scale-discretised wavelet
coefficients of simulations of homogenous and isotropic Gaussian
random fields on the sphere.  Specifically, we simulate realisations
of the the CMB, which, in the standard Lambda Cold Dark Matter
($\Lambda$CDM) cosmological model, is assumed to be a realisation of a
Gaussian random field on the sphere.  We assume a power spectrum
$C_\el$ specified by the $\Lambda$CDM cosmological model that best
fits observations of the CMB made by NASA's \wmaptext\ (\wmap)
\cite{hinshaw:2013} (combined with other cosmological data: we adopt
the full 9-year WMAP$+$BAO$+$H0 best-fit 6 parameter $\lambda$CDM
model).\footnote{Available at: \url{http://lambda.gsfc.nasa.gov}}
Our Milky Way galaxy obscures our view of the CMB, hence real
observations are made over incomplete sky coverage.  We study
statistical properties of wavelet coefficients in the presence of
incomplete coverage on the sphere, adopting the \wmap\ KQ75 mask
\cite{bennett:2013} (see \fig{\ref{fig:mask}}).
To compute directional scale-discretised wavelet transforms on the
sphere we use the \stwoletcode\footnote{\url{http://www.s2let.org}}
code \cite{leistedt:s2let_axisym, mcewen:s2let_spin}, which in turn
relies on the \sothreecode\footnote{\url{http://www.sothree.org}}
\cite{mcewen:so3} and \sshtcode\footnote{\url{http://www.spinsht.org}}
\cite{mcewen:fssht} codes, all of which are open-source and publicly
available.

% With bar
%viewport= 75 55 520 335

% Without bar
% viewport= 75 110 520 335

%=============================================================================
\subsection{Localisation}
%=============================================================================

\begin{figure}
\centering
\subfigure[$\eulc=0^\circ$]{\includegraphics[viewport= 75 110 520 335,clip = true, width=.6\textwidth]{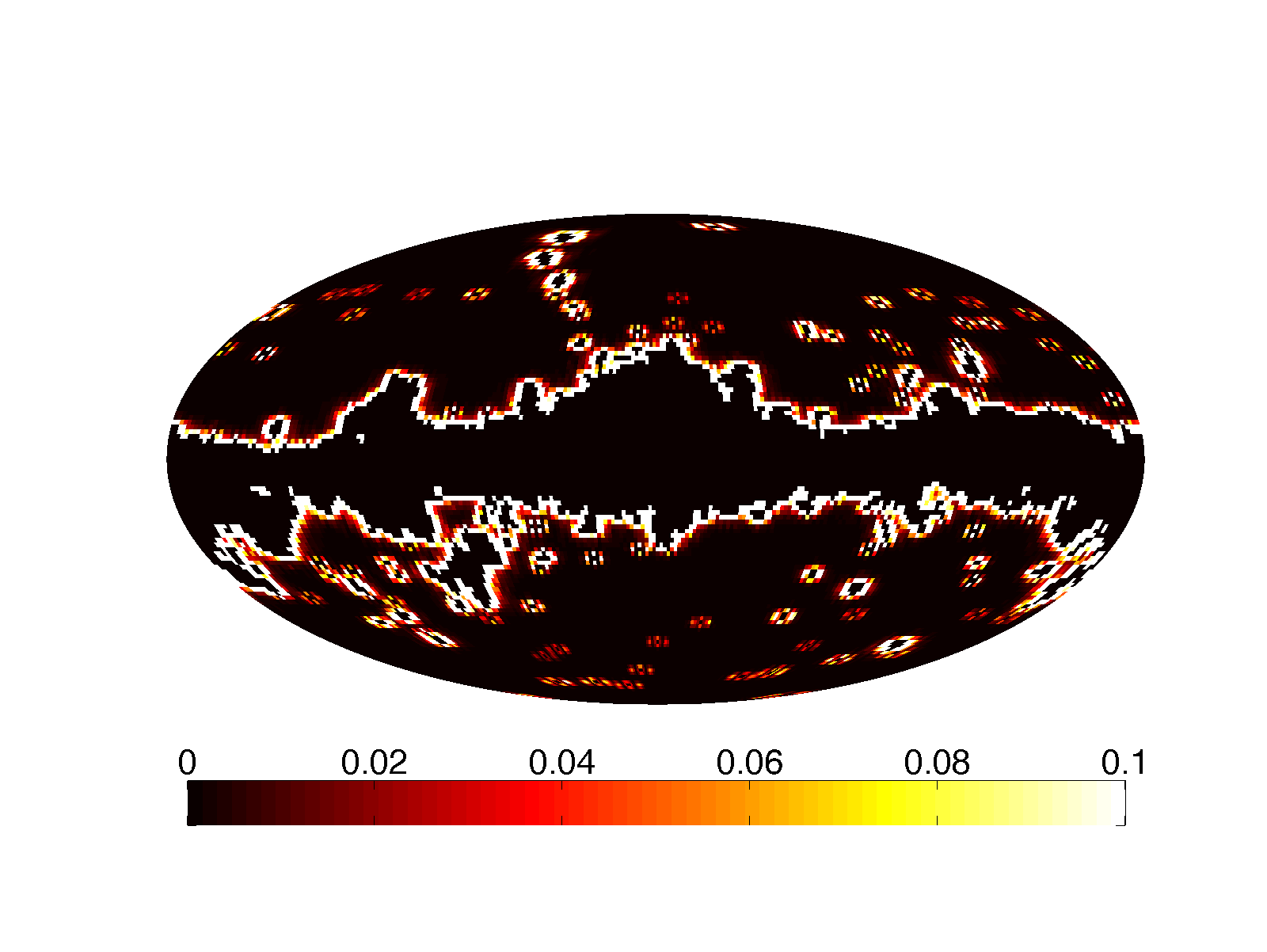}}\\
\subfigure[$\eulc=60^\circ$]{\includegraphics[viewport= 75 110 520 335,clip = true, width=.6\textwidth]{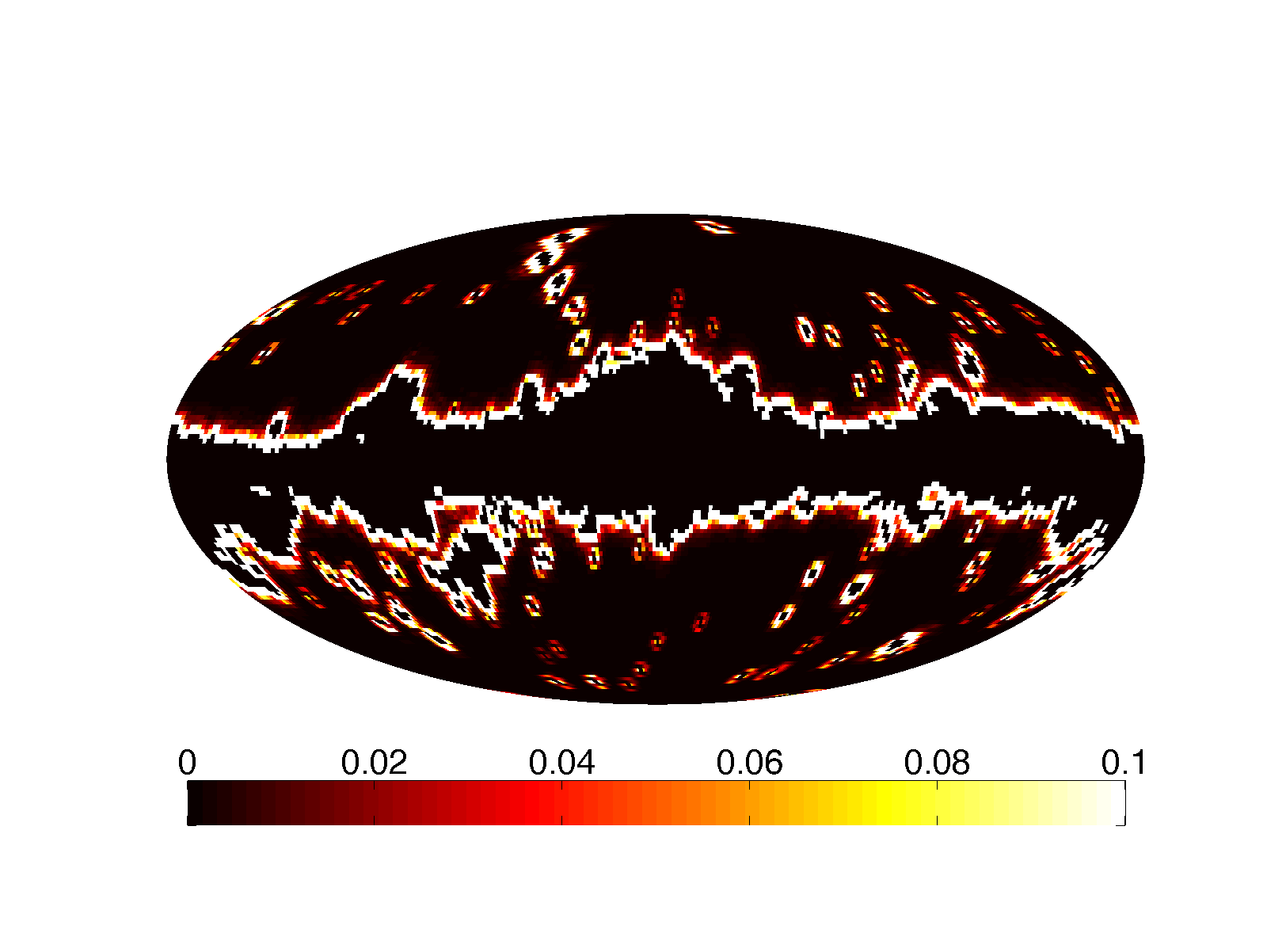}}\\
\subfigure[$\eulc=120^\circ$]{\includegraphics[viewport= 75 110 520 335,clip = true, width=.6\textwidth]{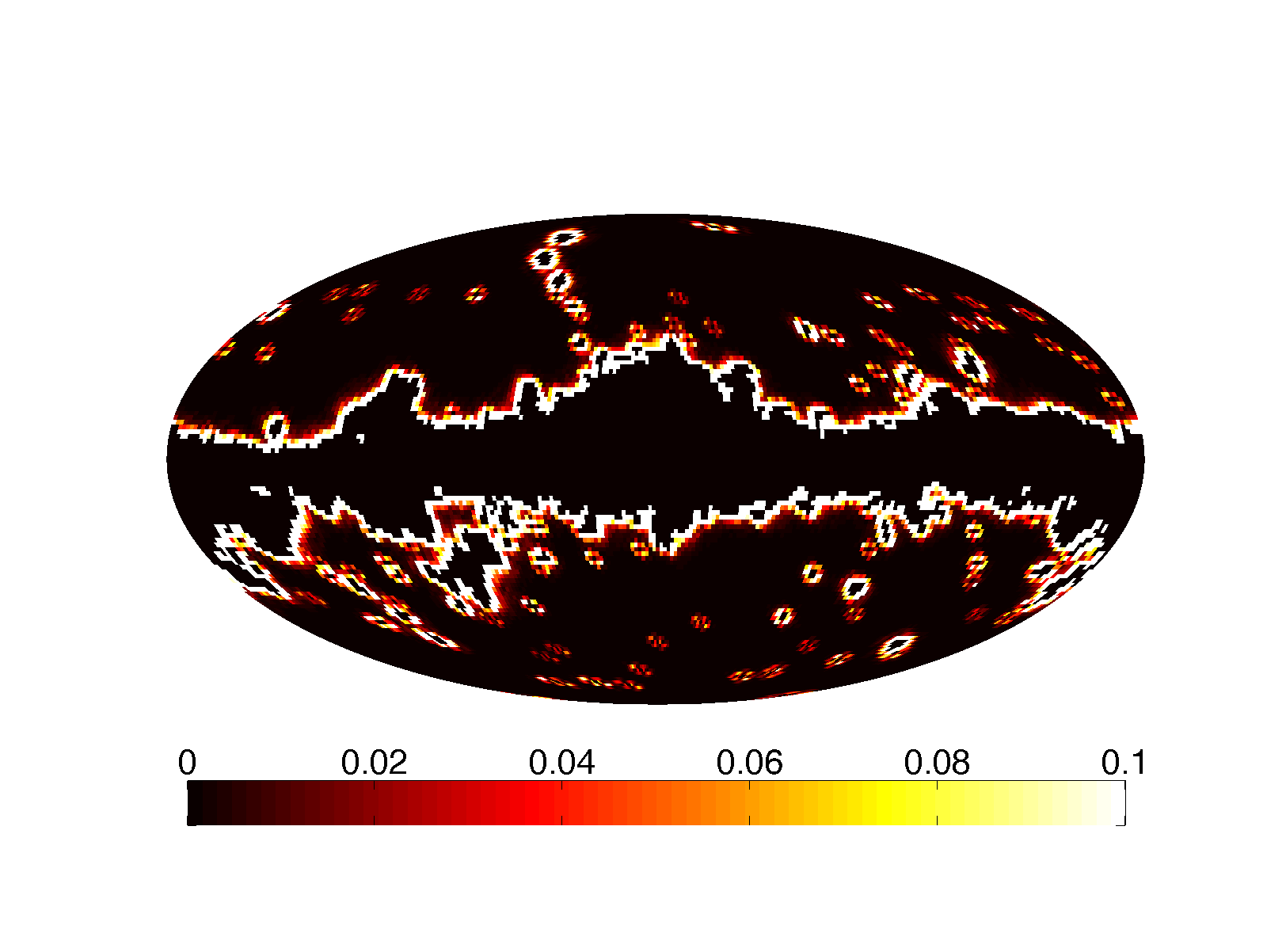}}\\
% \subfigure[a]{\includegraphics[viewport= 75 110 520 335,clip = true, width=.4\textwidth]{Figures/loc_L0128_J007_Jmin002_B2.00e+00_N003_nsim030_j006_n004}}
% \subfigure[a]{\includegraphics[viewport= 75 110 520 335,clip = true, width=.4\textwidth]{Figures/loc_L0128_J007_Jmin002_B2.00e+00_N003_nsim030_j006_n005}}
\includegraphics[viewport= 75 55 520 110,clip = true, width=.6\textwidth]{Figures/loc_L0128_J007_Jmin002_B2.00e+00_N003_nsim030_j006_n001}
\caption{Localisation statistic $\Delta^{{(\wscale)}}(\eul)$, plotted
  using a Mollweide projection for each orientation \eulc, computed from %30
 Monte Carlo simulations 
($\elmax=128,\ \dilparam=2,\ \mmax=3,\ \wscale=2$).}
%($\elmax=128,\ \dilparam=2,\ \mmax=3,\ \wscalemax=7,\ \wscalemin=2,\ \wscale=5$).}
\label{fig:localisation_1}
\end{figure}

\begin{figure}
\centering
\subfigure[$\eulc=0^\circ$]{\includegraphics[viewport= 75 110 520 335,clip = true, width=.6\textwidth]{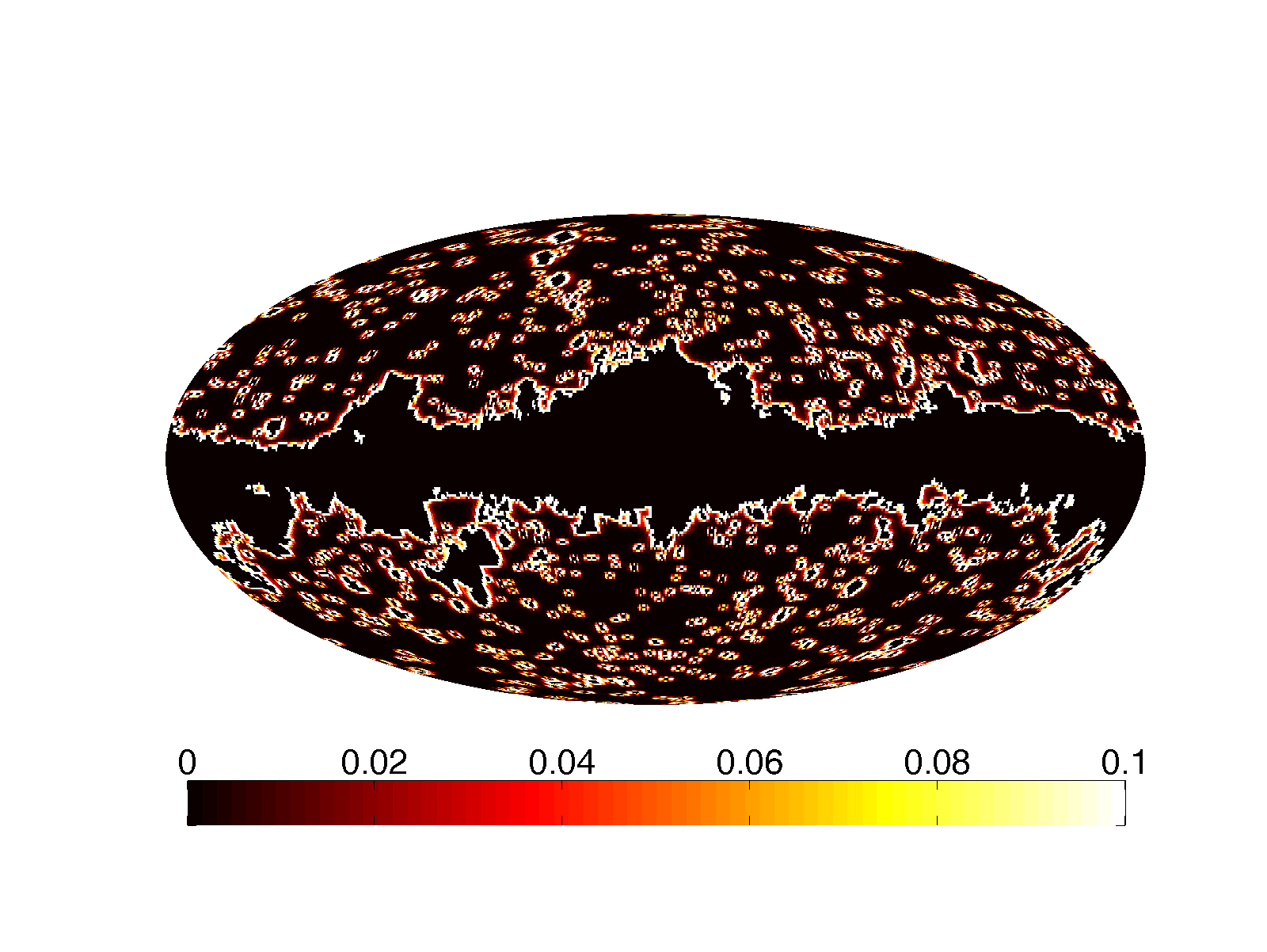}}\\
\subfigure[$\eulc=60^\circ$]{\includegraphics[viewport= 75 110 520 335,clip = true, width=.6\textwidth]{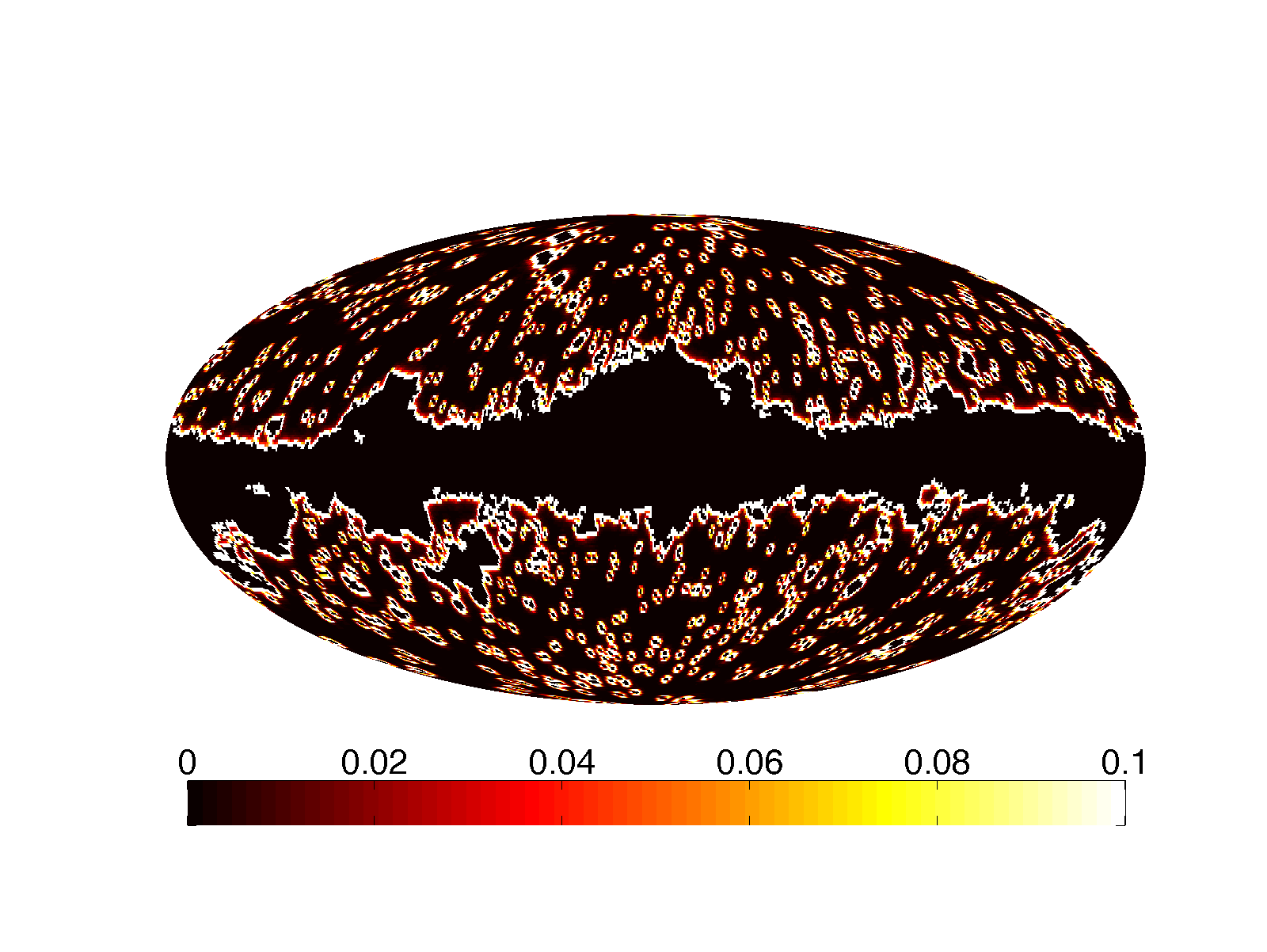}}\\
\subfigure[$\eulc=120^\circ$]{\includegraphics[viewport= 75 110 520 335,clip = true, width=.6\textwidth]{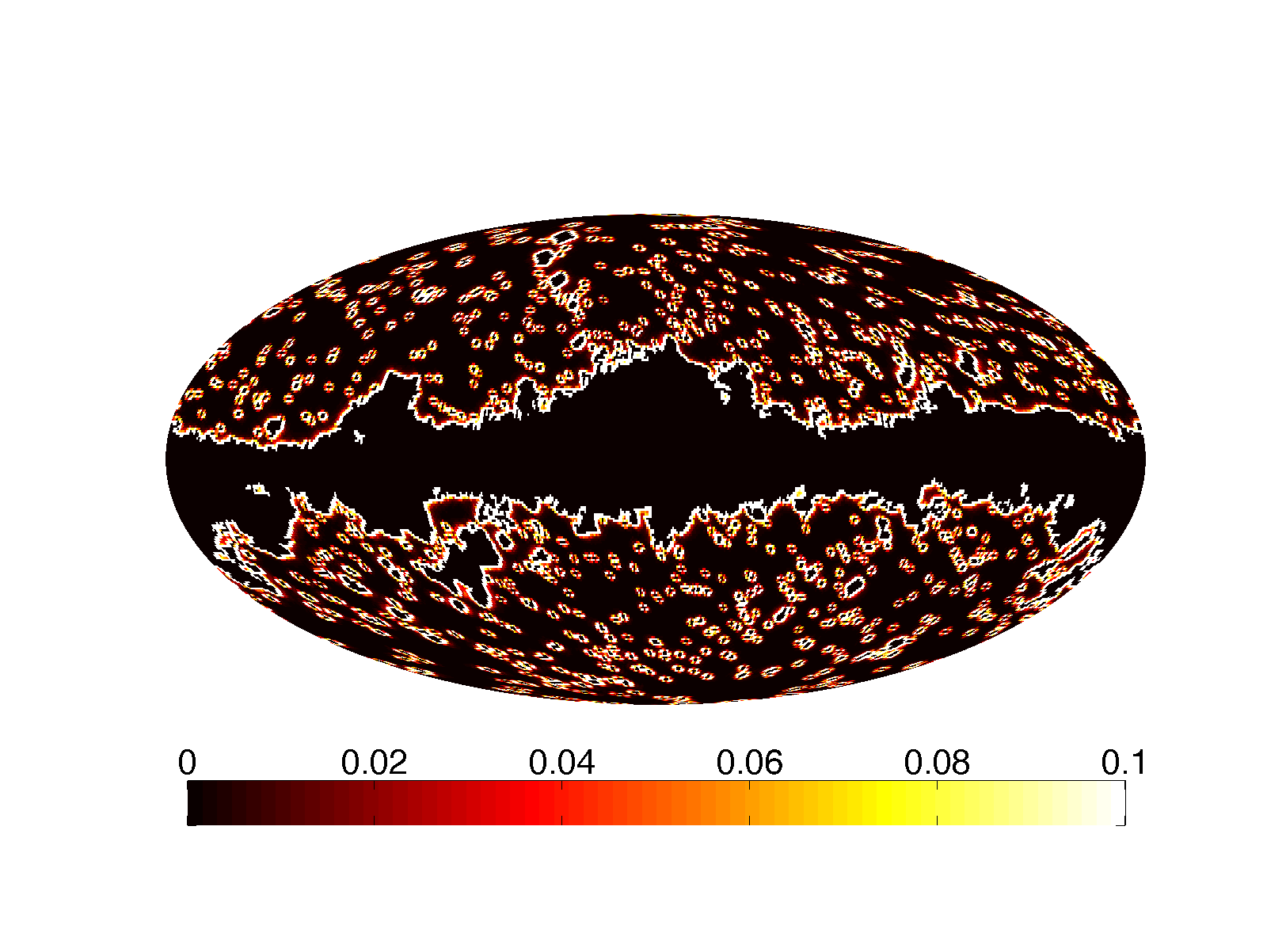}}\\
% \subfigure[a]{\includegraphics[viewport= 75 110 520 335,clip = true, width=.4\textwidth]{Figures/loc_L0256_J008_Jmin003_B2.00e+00_N003_nsim100_j007_n004}}
% \subfigure[a]{\includegraphics[viewport= 75 110 520 335,clip = true, width=.4\textwidth]{Figures/loc_L0256_J008_Jmin003_B2.00e+00_N003_nsim100_j007_n005}}
\includegraphics[viewport= 75 55 520 110,clip = true, width=.6\textwidth]{Figures/loc_L0256_J008_Jmin003_B2.00e+00_N003_nsim100_j007_n001}
\caption{Localisation statistic $\Delta^{{(\wscale)}}(\eul)$, plotted
  using a Mollweide projection for each orientation \eulc, computed from %100
 Monte Carlo simulations 
($\elmax=256,\ \dilparam=2,\ \mmax=3,\ \wscale=2$).}
%($\elmax=256, \dilparam=2, \mmax=3, \wscalemax=8, \wscalemin=3, \wscale=6$).)
\label{fig:localisation_2}
\end{figure}

To study localisation properties in the context of incomplete coverage
we compute the following localisation statistic:
\begin{equation}
  \Delta^{(\wscale)}(\eul) = 
  \frac{\expv \biggl[
    \Bigl\vert\widehat{\wcoeff}^{\wav^{(\wscale)}}(\eul) - \wcoeff^{\wav^{(\wscale)}}(\eul)\Bigr\vert^2
  \biggr]}
  {\expv \biggl[
    \Bigl\vert\wcoeff^{\wav^{(\wscale)}}(\eul)\Bigr\vert^2
    \biggr]
  }
  \spcend,
\end{equation}
where $\widehat{\cdot}$ denotes a quantity observed over incomplete
coverage (adopting the WMAP KQ75 mask illustrated in
\fig{\ref{fig:mask}}).  The localisation statistic
$\Delta^{{(\wscale)}}(\eul)$ computed from Monte Carlo
simulations is shown in \fig{\ref{fig:localisation_1}} and
\fig{\ref{fig:localisation_2}} for different wavelet parameters.
Notice that $\Delta^{{(\wscale)}}(\eul)$ is close to zero over the
majority of the sphere and only deviates significantly from zero along the
mask boundaries, highlighting the excellent spatial localisation
properties of scale-discretised wavelets.  
As expected, deviations from zero in $\Delta^{{(\wscale)}}(\eul)$ are induced when
the size of the gap in coverage is of a comparable or greater size than
the wavelet considered.  For example, small point source regions of the
mask have a minimal impact in \fig{\ref{fig:localisation_1}} but a
more significant impact in \fig{\ref{fig:localisation_2}}, where the
size of the wavelet is smaller.

%=============================================================================
\subsection{Correlation}
%=============================================================================

\begin{figure}
\centering
\subfigure[Analytic]{\includegraphics[viewport= 30 20 545 410,clip = true, width=.48\textwidth]{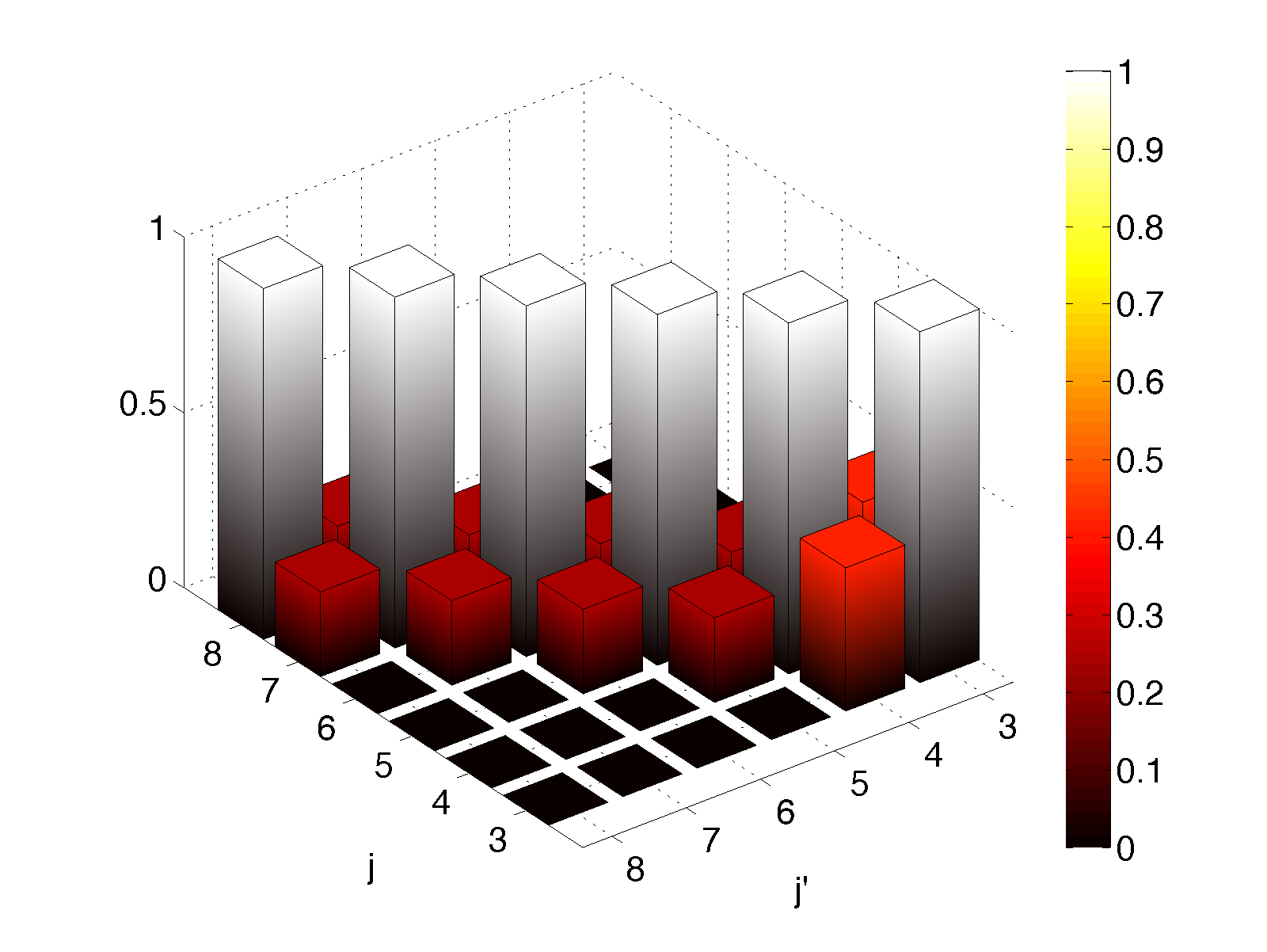}}\\
\subfigure[Empirical without mask]{\includegraphics[viewport= 30 20 545 410,clip = true, width=.48\textwidth]{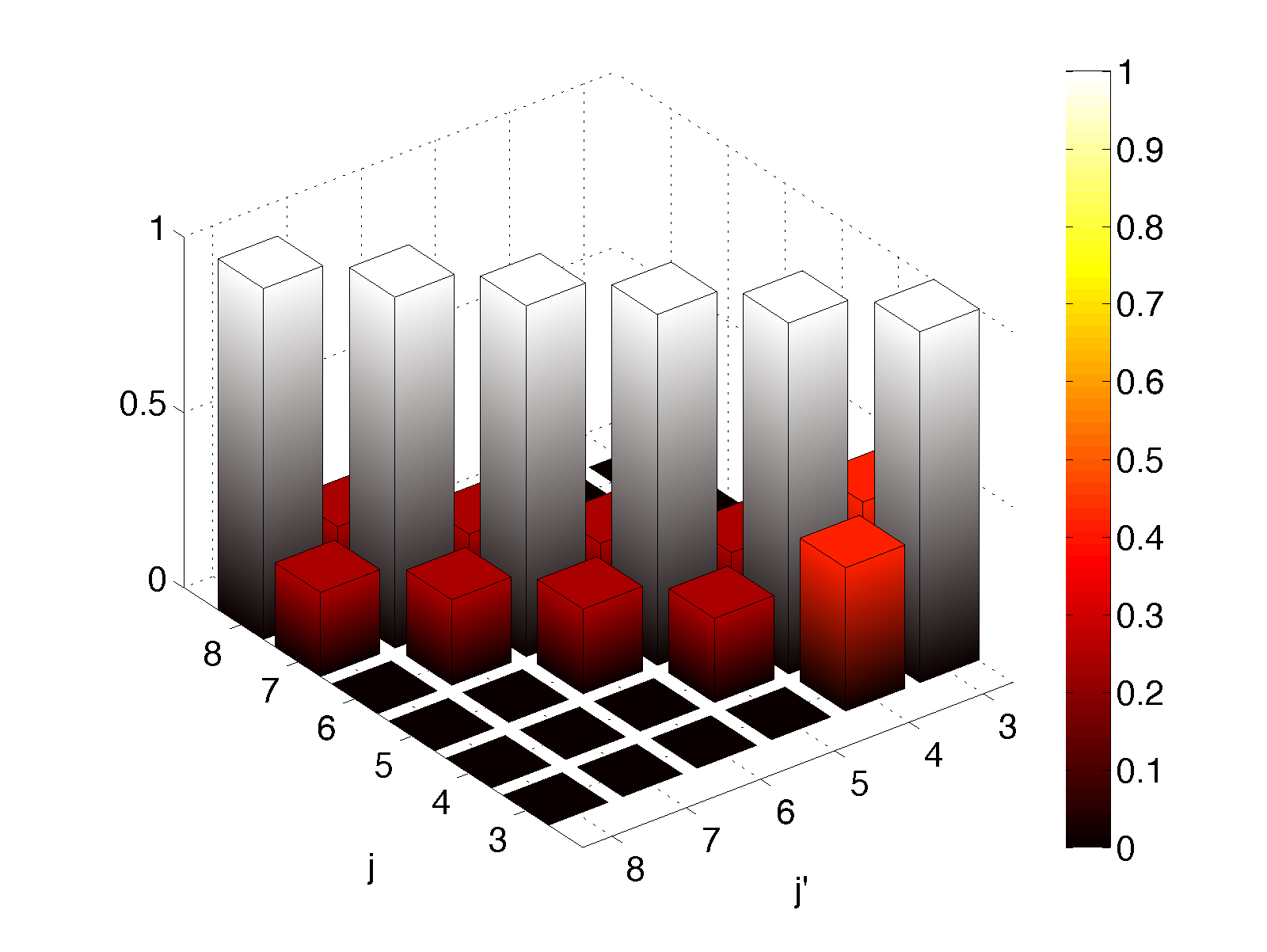}}
\subfigure[Empirical with mask]{\includegraphics[viewport= 30 20 545 410,clip = true, width=.48\textwidth]{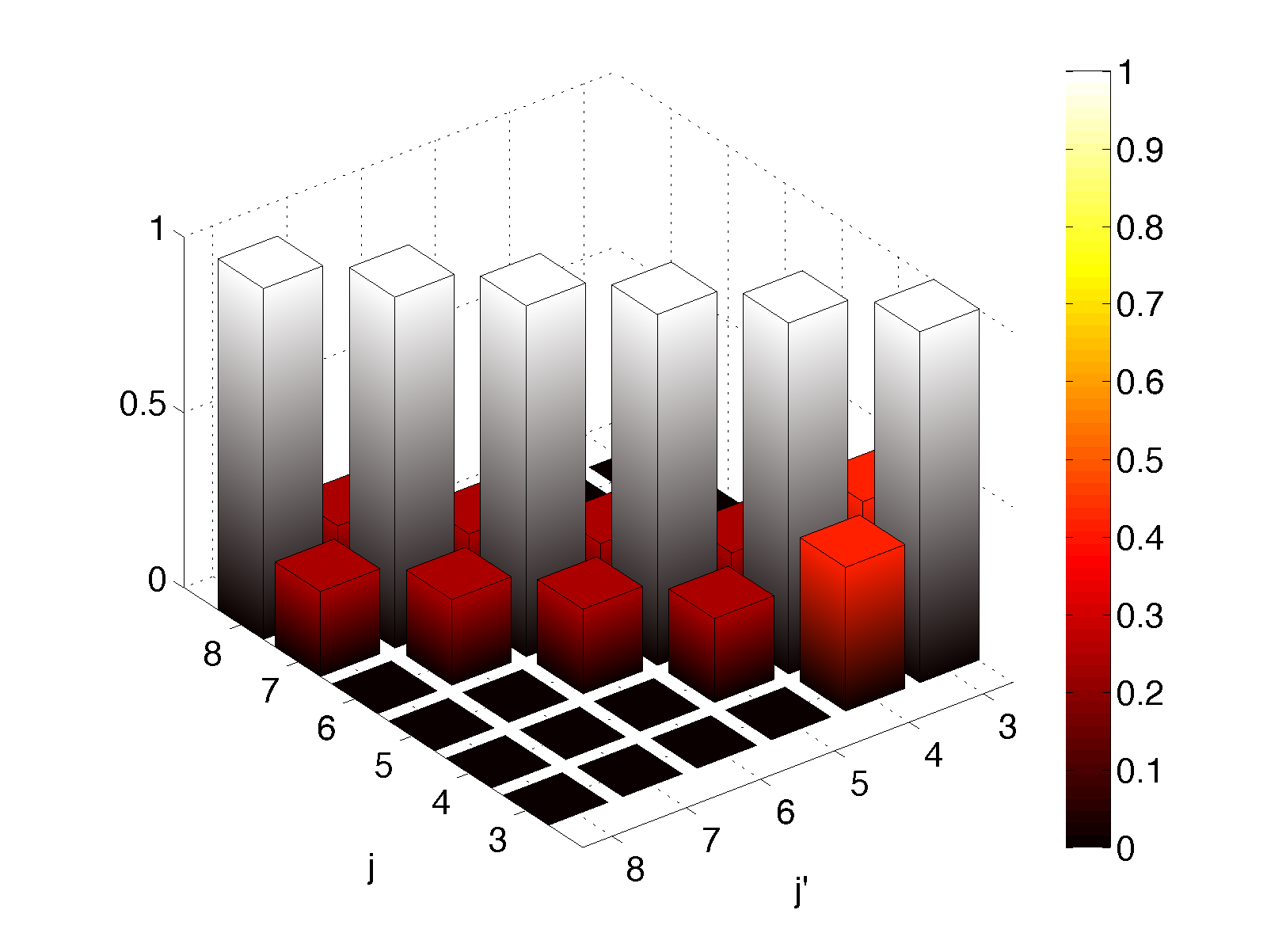}}
\subfigure[Difference between analytic and empirical]{\includegraphics[viewport= 30 20 545 410,clip = true, width=.48\textwidth]{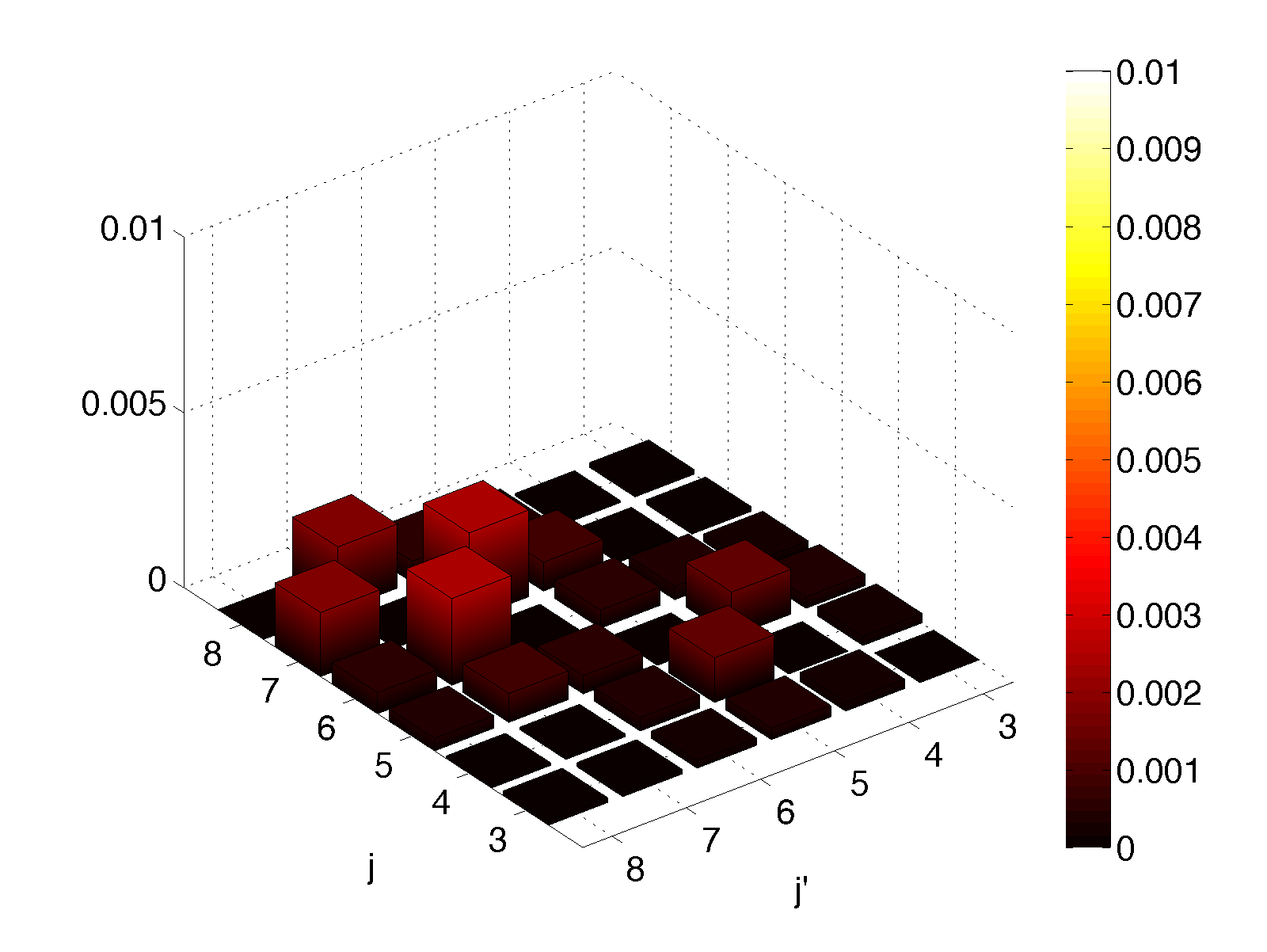}}
\subfigure[Difference between empirical masked and unmasked]{\includegraphics[viewport= 30 20 545 410,clip = true, width=.48\textwidth]{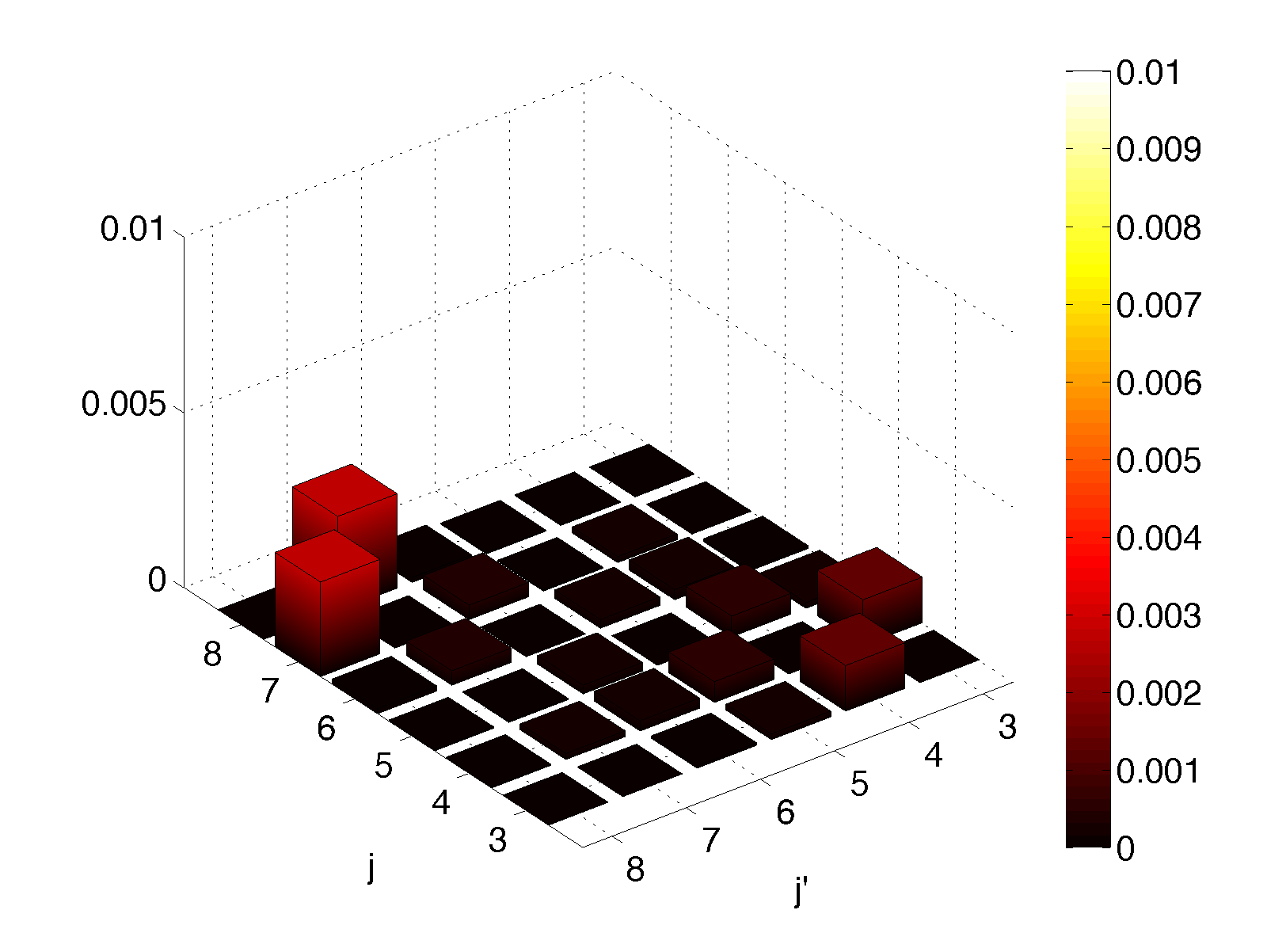}}
\caption{Correlation $\xi^{(\wscale\wscale\p)}(\eul,\eul)$ computed
  analytically and empirically (from Monte Carlo simulations) in the
  absence and presence of a mask 
  ($\elmax=256,\ \dilparam=2,\ \mmax=3)$.}
%  ($\elmax=256,\ \dilparam=2,\ \mmax=3, \wscalemax=8, \wscalemin=3, \wscale=6$).}
\label{fig:covariance_2}
\end{figure}

To study the correlation properties of scale-discretised wavelets we
compute the expected correlation
$\Xi^{(\wscale\wscale\p)}(\eul,\eul)$ defined by
\eqn{\ref{eqn:correlation}}.
  The correlation is computed empirically
from Monte Carlo simulations, for both complete and incomplete
coverage (adopting the WMAP KQ75 mask illustrated in
\fig{\ref{fig:mask}}), and also analytically by noting
\begin{equation}
  \xi^{(\wscale\wscale\p)}(\eul,\eul) 
  %= \expv \Bigl[ \wcoeff^{\wav^{(\wscale)}}(\euls) \:
  %\wcoeff^{\wav^{(\wscale\p)}}{}^{\cconj}(\euls) \Bigr]
  = 
  \sum_{\el=0}^{\elmax-1}\summ
  \noisecl_\el \:
  \shc{\wav}{\el}{\m}^{(\wscale) \cconj}
  \shc{\wav}{\el}{\m}^{(\wscale\p)}
  \spcend .
\end{equation}
Computed correlation values are illustrated in
\fig{\ref{fig:covariance_2}}.  Notice that the analytic calculation is
in close agreement with the empirical calculation for both complete
and incomplete coverage.
Since the implementation of the scale-discretised wavelet transform is
built on exact quadrature \cite{mcewen:fssht, mcewen:so3} and any
errors in computed wavelet transforms are of the order of machine
precision \cite{leistedt:s2let_axisym, mcewen:s2let_spin}, differences
between analytic and empirical computations are due to statistical
noise (100 Monte Carlo simulations were computed).
As expected the correlation is essentially zero for
$\vert \wscale - \wscale\p \vert > 1$.

% %=============================================================================
% \section{Conclusions}
% \label{sec:conclusions}
% %=============================================================================

% Directional scale-discretised wavelets on the sphere satisfy similar
% localisation and uncorrelation properties to axisymmetric needlets.
% We have derived local concentration and asymptotic uncorrelation
% properties for directional scale-discretised wavelets in both scalar
% and spin settings and performed numerical experiments to study the
% localisation and correlation of wavelet coefficients of homogenous and
% isotropic Gaussian random fields on the sphere.  Specifically, we
% studied simulations of the CMB observed over full and incomplete sky
% coverage on the sphere.  In addition, we presented for the first time
% the detailed derivation of the directional construction of
% scale-discretised wavelets and their directional steerability and
% auto-correlation properties.
% %
% Exact and efficient codes to compute directional scale-discretised
% wavelet transforms on the sphere are open-source and publicly available.
% %
% In summary, directional scale-discretised wavelets exhibit excellent
% concentration properties and can thus be considered as directional
% needlets.

%=============================================================================
\section*{Acknowledgements}
%=============================================================================

We thank Domenico Marinucci for his precious suggestions and
useful discussions.  We acknowledge use of the \healpix\ package
\cite{gorski:2005} and the \lambdaarchtext\ (\lambdaarch).  Support
for \lambdaarch\ is provided by the NASA Office of Space Science.

%=============================================================================
\appendix{}
\section{Auxiliary results}
\label{sec:appendix}
%=============================================================================

In this appendix we collect some useful technical results. In the
first subsection we recall well-known results related to spherical
harmonics, spin spherical harmonics, Wigner \dmatbig-functions and
some related properties. The latter subsections include results
pivotal for the exhaustive proofs of the localisation of the
directional scale-discretised wavelets $\Psi ^{( j) }$ and
the upper bound of the covariance between
$W^{\Psi ^{( j) }}( \rho _{1}) $ and
$W^{\Psi ^{( j^{\prime }) }}( \rho _{2}) $.
This appendix should be read in conjunction with the main text, where
symbols and expressions introduced already are defined explicitly.

%-----------------------------------------------------------------------------
\subsection{Spherical harmonics and Wigner \dmatbig-functions}
\label{sec:appendix:special_functions}

We succinctly recall the main definitions and properties of the spherical
harmonic functions and the Wigner \dmatbig-functions, which we make use of throughout the article.  For further
details and proofs we refer the reader to
\cite{marinucci:2011:book,varshalovich:1989}.

% Let the space of harmonic polynomials of degree $\ell$ over $\sphere$
% be denoted by $\mathcal{H}_{\ell}$. As well-known in the literature,
% $\mathcal{H}_{\ell}$ is spanned by the eigenfunctions of the spherical
% Laplacian operator $\Delta_\sphere$ associated with eigenvalue
% $-\ell(\ell+1)$, with multiplicity $2\ell+1$, (see \eg\
% \cite{marinucci:2011:book}). On the other hand, it is also a
% well-known fact that the space given by
% $\bigoplus_{\ell \geq 0} \mathcal{H}_{\ell}$ is dense in
% $\ltwo(\sphere)$ \revision{and, moreover, equal to it}: in this sense the set
% $\bigl\{Y_{\ell m}(\sa), \: \omega \in \sphere: \ell \geq
%   0, m=-\ell,\ldots,\ell\bigr\}$
% describes a basis for $\ltwo(\sphere)$ with respect to the uniform
% Lebesgue measure. 

The scalar spherical harmonic functions are
explicitly defined by
\begin{equation}
  \label{eqn:spherical_harmonic_function}
  \shfarg{\el}{\m}{\sas} = 
  (-1)^\m
  \sqrt{\frac{2\el+1}{4\pi} \elmfact} \:
  \aleg{\el}{\m}{\cos\saa} \:
  \exp{(\img \m \sab)} 
  \spcend ,
\end{equation}
for natural $\el\in\naturals$ and integer $\m\in\integers$,
$|\m|\leq\el$, where $\aleg{\el}{\m}{\cdot}$ are the associated Legendre
functions, which can be related to the Legendre polynomials
$\leg{\el}{\cdot}$ by
\begin{equation}
  \label{eqn:associated_legendre_function}
  \aleg{\el}{\m}{\cos\saa}
  \equiv 
  (\sin \saa )^{m}\frac{\dx^{m}}{\dx(\cos \saa )^{m}}
  \leg{\el}{\cos\saa}
  \spcend .
\end{equation}
Recall the definition of the $C^{\infty }$ differential operator $\Toperator_\m$
specified by \eqn{\ref{TT}}: it can be readily noted that 
\begin{equation} \label{Toper}
P_{\ell }^{m}(\cos \theta ) = \Toperator_\m P_{\ell}(\cos \theta ).
\end{equation}
Note that here we adopt the Condon-Shortley phase convention, with the
$(-1)^\m$ phase factor included in
\eqn{\ref{eqn:spherical_harmonic_function}} above.
%  ensuring that the
% conjugate symmetry relation
% $\shfargc{\el}{\m}{\sas} = (-1)^\m \: \shfargsp{\el}{-\m}{\sas}$
% holds.
%
The orthogonality and completeness relations for the spherical
harmonics read, respectively,
\begin{equation}
  \label{eqn:spherical_harmonic_ortho}
  \innerp{\shf{\el}{\m}}{\shf{\el\p}{\m\p}}
  = 
  \int_\sphere \dmu{\sas} \:
  \shfarg{\el}{\m}{\sas} \:
  \shfargc{\el\p}{\m\p}{\sas}
  =
  \kron{\el}{\el\p}
  \kron{\m}{\m\p}
\end{equation}
and
\begin{equation}
  \sumlm
  \shfarg{\el}{\m}{\saa,\sab} \:
  \shfargc{\el}{\m}{\saa\p,\sab\p} 
  =
  \delta(\cos\saa - \cos\saa\p) \:
  \delta(\sab - \sab\p)
  \spcend ,
\end{equation}
where $\kron{i}{j}$ is the Kronecker delta symbol and
$\delta(x)$ is the one-dimensional Dirac delta function.

% As far as the space $\ltwo(\sothree)$ is concerned, by the Peter-Weyl
% theorem, an irreducible representation is provided by the set of
% Wigner's $\dmatbig$-matrices
% $\left\{D^\ell(\rotmatarg{\eul}): \ell =0,1,2,\ldots\right\}$, of
% dimension $(2\ell+1)\times(2 \ell +1)$ (see
% \eg\ \cite{marinucci:2011:book}). Wigner $\dmatbig$-matrices
% describe a complete orthonormal basis for the space
% $\ltwo(\sothree)$. The $\dmatbig^\el(\rotmatarg{\eul})$ matrices operate irreducibly and equivalently
% on the $(2\ell +1)$ so-called isotypical spaces, each of
% them spanned by a column of the matrix representation
% itself. Furthermore, note that Wigner's $D$-matrices can be
% equivalently parametrised by the rotation $\rotmatarg{\eul}$ or the
% corresponding Euler angles $\eul=(\euls)$, where $\alpha \in [ 0, 2\pi
% )$, $\beta \in [0,\pi]$ and $\gamma \in [ 0, 2\pi )$.
% For $\el\in\naturals$, each element of $D^\ell$ is given by the
% corresponding 

The Wigner $\dmatbig$-function \Dlmn, for integer
$\m,\n\in\integers$, $|\m|,|\n|\leq\el$, may be decomposed, in
terms of Euler angles, by 
\begin{equation}
  \label{eqn:d_decomp}
  \dmatbig_{\m\n}^{\el}(\euls)
  = {\rm e}^{-\img \m\eula} \:
  \dmatsmall_{\m\n}^\el(\eulb) \:
  {\rm e}^{-\img \n\eulc}
  \spcend .
\end{equation}
% The real polar \dmatsmall-functions may be represented explicitly by 
% \begin{equation}
%   \label{eqn:wignerd_b}
%   \dlmnb = %&
%   \sqrt{\frac{(\el+\n)! (\el-\n)!}{(\el+\m)! (\el-\m)!}}
%   \left( \sin\frac{\eulb}{2} \right)^{\n-\m} 
%   \left( \cos\frac{\eulb}{2} \right)^{\n+\m} 
%   \jacobi{\el-\n}{\n-\m}{\n+\m}{\cos\eulb}
%   \spcend ,
% \end{equation}
% and can be computed by recursion (\eg\
% \cite{risbo:1996,trapani:2006}), where $\jacobi{\el}{a}{b}{\cdot}$ are
% the Jacobi polynomials.  
The orthogonality and completeness relations
for the Wigner $\dmatbig$-functions read, respectively,
\begin{equation}
  \label{eqn:wignerd_ortho}
  \innerp{\Dlmn}{ \dmatbig_{\m\p\n\p}^{\el\p}} 
  = \int_\sothree \deul{\eul} \:
  \Dlmnp \:
  \dmatbig_{\m\p\n\p}^{\el\p \cconj}(\eul)
  = \frac{8 \pi^2}{2\el+1} \kron{\el}{\el\p} \kron{\m}{\m\p} \kron{\n}{\n\p}   
  \spcend ,
\end{equation}
and 
\begin{equation}
  \sumlmn 
  \dmatbig_{\m\n}^{\el}(\eula,\eulb,\eulc) \:
  \dmatbig_{\m\n}^{\el\cconj}(\eula\p,\eulb\p,\eulc\p) 
   =
  \delta(\eula - \eula\p) \:
  \delta(\cos\eulb - \cos\eulb\p) \:
  \delta(\eulc - \eulc\p)
  \spcend .
\end{equation}
where $\innerp{\cdot}{\cdot}$ is used to denote inner products over
both the sphere and the rotation group (the case adopted can be
inferred from the context).

We note the additive property of the \dmatbig-functions is given by
\begin{equation}
  \dmatbig_{\m\n}^{\el}(\eula,\eulb,\eulc) 
  =
  \sum_{k=-\el}^\el
  \dmatbig_{\m k}^{\el}(\eula_1,\eulb_1,\eulc_1) \:
  \dmatbig_{k \n}^{\el}(\eula_2,\eulb_2,\eulc_2) 
  \spcend ,
\end{equation}
where $\eul=(\euls)$ describes the rotation formed by composing the
rotations described by $\eul_1=(\eula_1,\eulb_1,\eulc_1)$ and
$\eul_2=(\eula_2,\eulb_2,\eulc_2)$, \ie\ $\rotmatarg{\eul} =
\rotmatarg{\eul_1} \rotmatarg{\eul_2}$.  The Euler angles may be
related explicitly by 
\begin{equation}
  \cot ( \eula-\eula_2)  
  =
  \cos \eulb_2 \cot( \eula_1 +\eulc_2) 
  + \cot \eulb_1 \frac{\sin \eulb_2}{\sin ( \eula_1 +\eulc_2) }
  \spcend , 
\end{equation}
\begin{equation}
  \cos \eulb 
  = \cos \eulb_1 \cos \eulb_2 
  - \sin \eulb_1 \sin \eulb_2
  \cos ( \eula_1 +\eulc_2) , 
\end{equation}
and
\begin{equation}
  \cot ( \eulc-\eulc_1) 
  = \cos \eulb_1 \cot (\eula_1 +\eulc_2) 
  + \cot \eulb_2
  \frac{\sin \eulb_1}{\sin ( \eula_1 +\eulc_2) }
  \spcend .
\end{equation}
Thus
\begin{equation}  
  \label{eqn:wignerd_add_conj}
  \sum_{k=-\el}^\el
  \dmatbig_{k \m}^{\el \cconj}(\eula_1,\eulb_1,\eulc_1) \:
  \dmatbig_{k \n}^{\el}(\eula_2,\eulb_2,\eulc_2) 
  % =
  % \sum_{k=-\el}^\el
  % \dmatbig_{\m k}^{\el}(-\eulc_1,-\eulb_1,-\eula_1) \:
  % \dmatbig_{\n k}^{\el}(\eula_2,\eulb_2,\eulc_2)
  =
  \dmatbig_{\m\n}^{\el}(\eula,\eulb,\eulc) 
  \spcend ,
\end{equation}
where now $\eul=(\euls)$ describes the rotation formed by composing
the inverse of the rotation described by
$\eul_1=(\eula_1,\eulb_1,\eulc_1)$ and the rotation described by
$\eul_2=(\eula_2,\eulb_2,\eulc_2)$, \ie\
$\rotmatarg{\eul} = \rotmatarg{\eul_1}^{-1} \rotmatarg{\eul_2}$, since
\begin{equation}  
  \dmatbig_{\m \n}^{\el}(\eula,\eulb,\eulc) 
  =
  \dmatbig_{\n \m}^{\el \cconj}(-\eulc,-\eulb,-\eula) 
  \spcend .
\end{equation}
For the case $\eul_1=\eul_2$, 
\begin{equation}  
  \label{eqn:wignerd_add_conj_same_ang}
  \sum_{k=-\el}^\el
  \dmatbig_{k \m}^{\el \cconj}(\eula_1,\eulb_1,\eulc_1) \:
  \dmatbig_{k \n}^{\el}(\eula_1,\eulb_1,\eulc_1) 
  =
  \kron{\m}{\n}
  \spcend .
\end{equation}

Finally, we note that one can relate the Wigner $\dmatbig$-functions and
the spherical harmonics (for further details and discussion we
refer to \cite{marinucci:2011:book, newman:1966}):
% As well-known in the
% literature, the sphere $\sphere$ can be identified as the quotient
% space $\sothree/\sotwo$.  Consequently, the spectral representation of
% elements belonging to $\ltwo(\sphere)$ corresponds to a subset of
% coefficients of Wigner $\dmatbig$-functions, in particular to ones on
% the central column of the matrix representation, \ie\ $n=0$, so that

\begin{equation}  
Y_{\ell m} \left(\theta,\phi \right) = \sqrt{\frac{2\ell+1}{4 \pi}}\dmatbig_{\m 0}^{\el\:\cconj}(\sab,\saa  ,0).
\end{equation}
The Wigner \dmatbig-functions may also be related to the spin spherical
harmonics \cite{newman:1966}, $\sshf{\el}{\m}{\n} \in \ltwo(\sphere)$, which  can be constructed from the
scalar harmonics through repeated action of the differential spin
raising/lowering operators.  When applied to a spin-\n\ function, the
spin raising and lowering operators are defined by 
\begin{equation}  
  \spinup \equiv
  -\sin^\n \saa 
  \biggl ( 
  \frac{\partial}{\partial \saa} 
  + \frac{\img}{\sin\saa} \frac{\partial}{\partial \sab}
  \biggr)
  \sin^{-\n}\saa
\end{equation}
and
\begin{equation}  
  \spindown \equiv 
  -\sin^{-\n} \saa 
  \biggl ( 
  \frac{\partial}{\partial \saa} 
  - \frac{\img}{\sin\saa} \frac{\partial}{\partial \sab}
  \biggr)
  \sin^{\n}\saa
  \spcend ,
\end{equation}
respectively.  The spin-\n\ spherical harmonics can hence be expressed in terms of the
scalar (spin-zero) harmonics by 
\begin{equation}  \label{spintoscale1}
  \sshfarg{\el}{\m}{\sas}{\n} 
  =
  \biggl[ \frac{(\el-\n)!}{(\el+\n)!} \biggr]^{1/2} 
  \spinup^\n
  \shfarg{\el}{\m}{\sas}
  \spcend ,
\end{equation}
for $0 \leq \n \leq \el$,
and by 
\begin{equation} \label{spintoscale2} 
  \sshfarg{\el}{\m}{\sas}{\n} 
  =
  (-1)^\n
  \biggl[ \frac{(\el+\n)!}{(\el-\n)!} \biggr]^{1/2} 
  \spindown^{-\n}
  \shfarg{\el}{\m}{\sas}
  \spcend ,
\end{equation}
for $-\el \leq \n \leq 0$.
Spin spherical harmonics are related to
the Wigner \dmatbig-functions by \cite{goldberg:1967}
\begin{equation}
  \label{eqn:ssh_wigner}
  \sshfarg{\el}{\m}{\sas}{\n} = (-1)^\n 
  \sqrt{\frac{2\el+1}{4\pi} } \:
  \dmatbig_{\m,-\n}^{\el\:\cconj}(\sab,\saa  ,0)
  \spcend .
\end{equation}
% from which it follows
% \begin{equation}
%   \dmatbig_{\m\n}^{\el}(\eula,\eulb, \eulc) 
%   =
%   (-1)^\n 
%   \sqrt{\frac{4\pi}{2\el+1} } \:
%   \sshfargc{\el}{\m}{\eulb,\eula}{-\n} \:   
%   \exp{(\img \n \eulc)}
%   \spcend .
% \end{equation}
% In other words, spin spherical harmonics span the isotypical space corresponding to the column $-n$ of the Wigner $\dmatbig$-matrix. 
Expressing the spin harmonics as spin raised or lower scalar
harmonics, it follows that  
\begin{equation}\label{wigner2}
  \dmatbig_{\m\n}^{\el}(\eula,\eulb, \eulc) 
  =
  %(-1)^\n 
  \sqrt{\frac{4\pi}{2\el+1} } \:
  %(-1)^\n
  \biggl[ \frac{(\el-\n)!}{(\el+\n)!} \biggr]^{1/2} 
  \spindown^{\n}
  \shfargc{\el}{\m}{\eulb,\eula} \:
  \exp{(\img \n \eulc)}
  \spcend ,
\end{equation}
for $0 \leq \n \leq \el$,
and 
\begin{equation}\label{wigner3}
  \dmatbig_{\m\n}^{\el}(\eula,\eulb, \eulc) 
  =
  (-1)^\n 
  \sqrt{\frac{4\pi}{2\el+1} } \:
  \biggl[ \frac{(\el+\n)!}{(\el-\n)!} \biggr]^{1/2} 
  \spinup^{-\n}
  \shfargc{\el}{\m}{\eulb, \eula} \:
  \exp{(\img \n \eulc)}
  \spcend ,
\end{equation}
for $-\el \leq \n \leq 0$.

Before concluding this subsection, let us introduce the
Mehler-Dirichlet approximation for Legendre polynomials. Further
details can be found in \cite{narcowich:2006}. Let us start from
\eqn{32} in \cite[p.~177]{erdelyi:1981}, oncerning the integral representation of the
Gegenbauer polynomials: as well-known in the literature, Legendre polynomials correspond to Gegenbauer of parameter $\frac{1}{2}$, so that we obtain%
\begin{equation}
P_{\ell}( \cos \theta )  = C_{\ell}^{\frac{1}{2}}( \cos \theta
) =\frac{\sqrt{2}\Gamma ( 1) \Gamma ( \ell+1) }{\pi
^{\frac{1}{2}}\ell!\Gamma ( \frac{1}{2}) \Gamma ( 1) }%
\int_{0}^{\theta }\frac{\cos ( \ell+\frac{1}{2}) \phi }{\sqrt{\cos
\phi -\cos \theta }}\dx\phi 
= \frac{\sqrt{2}}{\pi }\int_{0}^{\theta }\frac{\cos ( \ell+\frac{1}{2}%
) \phi }{\sqrt{\cos \phi -\cos \theta }}\dx\phi 
\spcend .
\end{equation}
As suggested in \cite{erdelyi:1981}, we replace $\phi $ and $\theta $ with $%
\pi -\phi $ and $\pi -\theta $ respectively, to get
\begin{equation}
P_{\ell}( \cos ( \pi -\theta ) )  =\frac{\sqrt{2}}{\pi }\int_{\phi }^{\pi }\frac{\cos \bigl( ( \ell+\frac{1%
}{2}) \pi -\ell( \phi +\frac{1}{2}) \bigr) }{\sqrt{\cos
\theta -\cos \phi }}\dx\phi 
\spcend .
\end{equation}
On one hand, recall that%
\begin{equation}
\cos \Bigl( \Bigl( \ell+\frac{1}{2}\Bigr) \pi -\ell \Bigl( \phi +\frac{1}{2}%
\Bigr) \Bigr)  
= ( -1) ^{\ell}\sin \Bigl ( \ell \Bigl( \phi +\frac{1}{2}\Bigr) \Bigr) 
\spcend .
\end{equation}
On the other hand, using the symmetry property of Legendre polynomials, we
have
\begin{equation}
P_{\ell}\bigl( \cos ( \pi -\theta ) \bigr) =P_{\ell}( -\cos
( \theta ) ) =( -1) ^{\ell}P_{\ell}( \cos (
\theta ) )
\spcend .
\end{equation}
Combining together all these results, we obtain
\begin{equation}
P_{\ell}( \cos ( \theta ) ) =\frac{\sqrt{2}}{\pi }%
\int_{\phi }^{\pi }\frac{\sin (\ell( \phi +\frac{1}{2})
) }{\sqrt{\cos \theta -\cos \phi }}\dx\phi 
\spcend .
\label{mehdir}
\end{equation}%

%-----------------------------------------------------------------------------
\subsection{A general result on localisation over compact manifolds}
\label{sec:appendix:general_localisation}

Here we recall the general result established in \cite{geller:2009} as
Theorem 2.2, properly adapted to the sphere $\sphere$. Let
$g\in C^{\infty }(\mathbb{R})$ and let
$\left\{\lambda _{\ell }\right\}$ be the set of eigenvalues associated
to the Beltrami-Laplacian operator over $\sphere$. Let
$\Lambda (x,y,t;g)$ be given by
\begin{equation}
\Lambda (\omega _{1},\omega _{2},t;g)\equiv 
\sumlm
g(t\sqrt{\lambda_{\ell }}) \: Y_{\ell m}(\omega _{1}) \: {Y}_{\ell m}^\cconj(\omega _{2})
\spcend .
\end{equation}%
For $t\in \realsnz$, for any function $g\in C^{\infty }(\mathbb{R})$, for every
pair of differential operators on the unit sphere $\mathbb{S}%
^{2}$, $T_{1}$ and $T_{2}$, depending respectively on
$\omega_{1}\in \sphere$ and $\omega _{2}\in\sphere$, and defined such
that $\deg T_{1}=i_{1}$, $\deg T_{2}=i_{2}$, and for every
non-negative integer
$\tau \in \naturals$, there exists a constant $C_{\tau }\in \realsnz$ such that
\begin{equation}
\bigl\vert T_{1}\:T_{2}\:\Lambda ( \sa_{1},\sa_{2},t;g )
\bigr\vert 
\leq 
\frac{C_{\tau } \: t^{-2-i_{1}-i_{2}}}{\Bigl( 1+\frac{d(\sa_1,\sa_2) }{t}\Bigr)^{\tau }}
\text{ , for all }t \in \realsnz,\text{ }\sa_1,\sa_2\in \sphere
\spcend ,  
\label{gellermayeli1}
\end{equation}
where $d(\sa_1,\sa_2)=\arccos ( \vect{\hat{\sa}_1} \cdot \vect{\hat{\sa}_2} )$ denotes
the geodesic distance between $\sa_1,\sa_2\in \sphere$. In the
framework of directional wavelets, we have:
\begin{equation}
\Lambda ^{(j)}(x,y)\equiv \Lambda (x,y,\lambda ^{j}L^{-1};\kappa^{(j)}( \ell ) )
\spcend .
\end{equation}

%-----------------------------------------------------------------------------
\subsection{Upper bound of $\sqrt{(\el-m)! / (\el+m)!}$}
\label{sec:appendix:bound_el_m}

As far as the behaviour of $\sqrt{( \ell -m)! / ( \ell +m) !}$ is
concerned, we make use of the following approximation, which we show
here:
\begin{equation}
\sqrt{\frac{( \ell -m) !}{( \ell +m) !}}\leq \ell
^{-m}
\spcend .
\label{glm}
\end{equation}
Stirling's approximation leads to
\begin{align}
\frac{( \ell -m) !}{( \ell +m) !} 
&\approx \frac{
( \ell -m) ^{\frac{1}{2}}( \ell -m) ^{\ell
  -m}\exp{(-( \ell -m) }}{( \ell +m) ^{\frac{1}{2}}(
\ell +m) ^{\ell +m}\exp{(-( \ell +m) )}} \\
&=\ell ^{-2m}\exp{(2m)}\Biggl( \frac{( 1-\frac{m}{\ell }) }{( 1+%
\frac{m}{\ell }) }\Biggr) ^{\ell }\frac{( 1-\frac{m}{\ell }%
) ^{\frac{1}{2}-m}}{( 1+\frac{m}{\ell }) ^{^{\frac{1}{2}+m}}%
}
\spcend .
\end{align}%
Consider now the positive function $h \in C^{\infty }$, defined on the support $( -1,1)$ as 
\begin{equation}
h( x) =\biggl( \frac{1-x}{1+x}\biggr) ^{\frac{1}{2}}\frac{1}{( 1-x^{2}) ^{^{m}}}
\spcend .
\end{equation}
Its first derivative is given by $\frac{\dx}{\dx x}h( x) =(1+x) ^{-( m+\frac{3}{2}) }( 1-x) ^{^{-( m+%
\frac{1}{2}) }}( 2mx-1) $, so that we have $\frac{\dx}{\dx x}h( x) =0$
for $x=( 2m) ^{-1}$.  The function $h$ is
therefore monotonically decreasing in the interval $( -1,1/2m) $
and increasing for $x\in ( 1/2m,1) \,$. Consider $%
x\equiv m/\ell $: because $m=-M,...,M$, it follows that $\vert m/ \ell
\vert \leq \vert M/\ell \vert $.
Hence we obtain 
\begin{equation}
\biggl\vert \frac{M}{\ell }\biggr\vert <1 \Rightarrow \biggl\vert 1\pm \frac{m}{%
\ell }\biggr\vert >0
\spcend .
\end{equation}
Therefore, we have that, for\ any $\ell $, 
\begin{equation}
h\Bigl( \frac{m}{\ell }\Bigr) <+\infty 
\spcend .
\end{equation}
On the other hand, $h$ attains its minimum for $\ell =2m^{2}$, where%
\begin{equation}
h\biggl( \frac{1}{2m}\biggr) =\biggl( \frac{2m-1}{2m+1}\biggr) ^{\frac{1}{2}%
}\biggl( \frac{4m^{2}}{4m^{2}-1}\biggr) >0
\spcend .
\end{equation}
Because $\ell >(L\lambda ^{-( 1+j) }\vee m)$ and $\lim_{\ell
\rightarrow \infty }h( \frac{m}{\ell }) =1$, we have that 
\begin{equation}
\frac{( 1-\frac{m}{\ell }) ^{\frac{1}{2}-m}}{( 1+\frac{m}{
\ell }) ^{^{\frac{1}{2}+m}}}\leq \max \biggl( 1,h\biggl( \frac{m}{
(L\lambda ^{-( 1+j) }\vee m)}\biggr) \biggr) 
\spcend .
\end{equation}
Furthermore, for large $\ell$, it can be easily seen that 
\begin{equation}
\lim_{\ell \rightarrow +\infty }\frac{( 1-\frac{m}{\ell }) ^{\ell
}}{( 1+\frac{m}{\ell }) ^{\ell }}=\exp{(-2m)}
\spcend .
\end{equation}
Consequently, the approximation specified above holds for large
$\ell$. On the other hand, for small $\ell$, let us define
$F_{\ell, m}=\exp({2m})\frac{(1-\frac{m}{\ell})^\ell}{(1+\frac{m}{\ell})^\ell}$.
We must prove that $F_{\ell, m} \leq 1$. Let us compute
\begin{equation}
\log F_{\ell, m} = 2 m +\ell \Bigl(\log\Bigl(1-\frac{m}{\ell}\Bigr)-\log(1+\frac{m}{\ell}\Bigr)\Bigr)  
\spcend .
\end{equation}
Now, because $\log(1+x) = x-\frac{x^2}{2}+\frac{x^3}{3} + \order(x^4)$ and $\log(1-x) = -x-\frac{x^2}{2}-\frac{x^3}{3} + \order(x^4)$, we have
\begin{equation}
\log F_{\ell, m} = 2 m +\ell \biggl(-\frac{m}{\ell}-\frac{1}{2}\frac{m^2}{\ell^2}-\frac{1}{3}\frac{m^3}{\ell^3} -\frac{m}{\ell}+\frac{1}{2}\frac{m^2}{\ell^2}-\frac{1}{3}\frac{m^3}{\ell^3} + \order\Bigl(\Bigl(\frac{m}{\ell}\Bigr)^4\Bigr)\biggr),  
\end{equation}
so that 
\begin{equation}
\log F_{\ell, m}=-\frac{2}{3}\frac{m^3}{\ell^2}+\order\Bigl(\frac{m^4}{\ell^3}\Bigr) \leq 0.
\end{equation}
Thus, $F_{\ell, m} \leq 1$, and consequently the approximation specified above also holds for small $\ell$.

%-----------------------------------------------------------------------------
\subsection{Upper bound of $U^{(j)}_{m}( \cos \theta ) $}
\label{sec:appendix:bound_u}

As far as the kernel $U^{(j)}_{m}( \cos \theta)$, defined in
\eqn{\ref{Kkernel}}, is concerned, we obtain the following upper
bound: there exists $\xi \in \realsnn$ such that 
\begin{equation}
\bigl\vert U^{(j)}_{m}( \cos \theta ) \bigr\vert \leq \frac{C_{\xi
}/\varepsilon_\wscale ^{2}}{\Bigl( 1+\bigl\vert \frac{\theta }{\varepsilon_\wscale }%
\bigl \vert \Bigr) ^{\xi }}
\spcend .
\end{equation}

The construction of this bound strictly follows the procedure used to
establish the localisation property for the so-called spherical
standard needlets developed in \cite{narcowich:2006}. First of all,
observe that for large $\ell$,
$b_{m,\varepsilon_j }\left( \varepsilon_j \left(\ell+\frac{1}{2}\right)
\right) \equiv b_{m,\varepsilon_j }\left( \varepsilon_j \left(\ell\right)
\right)$.
Furthermore, using the Meher-Dirichlet formula for Legendre
polynomials \eqn{\ref{mehdir}}, we obtain
\begin{align}
U^{(j)}_{m}( \cos \theta ) 
&\leq \sum_{\ell =0}^{\infty
}b_{m,\varepsilon_j }\Bigl( \ell +\frac{1}{2}\Bigr) \Bigl( \ell
  +\frac{1}{2} \Bigr) P_{\ell }( \cos \theta ) \\
&=\frac{1}{2}\int_{\theta }^{\pi }\frac{G_{m,\varepsilon_j }( \alpha
) }{\sqrt{\cos \theta -\cos \alpha }}d\alpha 
\spcend ,
\end{align}
where 
\begin{equation}
G_{m,\varepsilon_j }( \alpha ) =\sum_{\ell =-\infty }^{\infty
}b_{m,\varepsilon_j }\Bigl( \ell +\frac{1}{2}\Bigr) \Bigl( \ell +\frac{1}{2}%
\Bigr) \sin \Bigl( \alpha \Bigl( \ell +\frac{1}{2}\Bigr) \Bigr) 
\spcend .
\end{equation}
Observe that $b_{m,\varepsilon_j }( x) x\sin ( \alpha x) 
$ is an even function. We now compute the Fourier transform, here
denoted $\mathcal{F}[\cdot]$, of $%
G_{m,\varepsilon_j }( \alpha ) $, considering three different cases
depending on the sign of $m$: 
\begin{enumerate}
\item $m=0$. In this case, the proof is trivially equivalent to \cite{narcowich:2006}.
\item $m<0$. In this case, we have
\begin{equation}
xb_{m,\varepsilon }( x) =\kappa^{(j)}(x) (
x) ^{\vert m\vert +1}=\kappa_\lambda(\varepsilon_j x) (
x) ^{\vert m\vert +1}
\spcend .
\end{equation}
For any integrable function $g$, let its Fourier transform be denoted by $\mathcal{F}\left[f\left(x\right)\right]\left(\nu\right)$,  $\nu \in \reals$. Furthermore we define $\widehat{\kappa}\left(\nu\right) = \mathcal{F}\left[\kappa_\lambda\left(x\right)\right]\left(\nu\right)$. Therefore, we obtain
\begin{align}
\mathcal{F}[ xb_{m}( \varepsilon_j x) ] ( \nu
) &=\mathcal{F}[ x^{\vert m\vert +1}\kappa_{\lambda}(
\varepsilon_j x) ] ( \nu ) \\
&=\frac{i^{\vert m\vert +1}}{\varepsilon_j }\widehat{\kappa}^{(
\vert m\vert +1) }\Bigl( \frac{\nu}{\varepsilon_j }%
\Bigr) 
\spcend ,
\end{align}%
where we adopt the notation here $\widehat{\kappa}^{( \vert m\vert +1)}(\cdot) = \frac{d^{\vert
m\vert +1}}{d\nu ^{\vert m\vert +1}}\mathcal{F}[\kappa_\lambda( x) ](\cdot) $.  Hence, we have 
\begin{equation}
  \int_{\reals}b_{m,\varepsilon }\Bigl( x+\frac{1}{2}\Bigr) \Bigl( x+\frac{1}{2}
  \Bigr) \sin \Bigl( \theta \Bigl(x +\frac{1}{2}\Bigr) \Bigr)
  \exp{(-\img\nu x)} \dx x=\frac{1}{2\pi i} \exp{ \Bigl(  \img\frac{\nu }{2}\Bigr)} 
  \frac{i^{\vert m\vert +1}}{\varepsilon_j }\widehat{\kappa}^{(
    \vert m\vert +1) }\Bigl( \frac{\nu }{\varepsilon_j }
  \Bigr) 
  \spcend ,
\end{equation}
which, by Poisson summation formula and some simplifications, leads to 
\begin{equation}
G_{m,\varepsilon_\wscale }(\cos \theta ) \leq \frac{1}{2\pi \varepsilon }
\biggl \vert \sum_{\nu =-\infty }^{+\infty }\widehat{\kappa}^{( \vert
m\vert +1) }\biggl( \frac{\theta +\nu }{\varepsilon_j }\biggr)
\biggr \vert 
\spcend .
\end{equation}
Now, following \cite{narcowich:2006}, standard Fourier properties yield 
\begin{equation}
\mathcal{F}\biggl[ \frac{\dx^{r}}{\dx x^{r}}\bigl( x^{1+\vert m\vert
}\kappa_\lambda( x) \bigr) \biggr] (\nu ) =i^{r+1}\nu ^{r}
\widehat{\kappa}^{( \vert m\vert +1) }( \nu )
\spcend .
\end{equation}
Hence, we obtain 
\begin{equation}
\biggl\vert \frac{\theta +\nu }{\varepsilon_j }\biggr\vert ^{r}\biggl\vert 
\widehat{\kappa}^{( \vert m\vert +1) }\Bigl( \frac{\phi
+\nu }{\varepsilon_j }\Bigr) \biggr\vert \leq \frac{1}{\varepsilon_j }
\biggl\Vert \frac{\dx^{r}}{\dx x^{r}}\bigl( x^{1+\vert m\vert }\kappa_\lambda(
x) \bigr) \biggr\Vert _{L^{1}}
\spcend .
\end{equation}
Therefore, for any given $\xi >M$, if 
\begin{equation}
V_{\xi ,\kappa_\lambda} \equiv \max_{0\leq r\leq \xi }\biggl\Vert \frac{\dx^{r}}{\dx x^{r}}\bigl(
x^{1+\vert m\vert }\kappa_\lambda( x) \bigr) \biggr\Vert _{L^{1}}
\spcend ,
\end{equation}
we have 
\begin{equation}
G_{m,\varepsilon_j}( \theta ) \leq \frac{1}{4\pi^2 \varepsilon_j }%
\biggl\vert \sum_{\nu=-\infty }^{+\infty }\frac{2 V_{\xi ,\kappa_\lambda}/\varepsilon_j }{%
1+\bigl\vert \frac{\theta +\nu}{\varepsilon_j }\bigr\vert ^{\xi }}%
\biggr\vert 
\spcend .
\end{equation}
The remainder of the proof is equivalent to \cite{narcowich:2006}.

\item $m>0$. By construction, \thinspace $b_{m,\varepsilon_j }( \cdot
) $ fulfils all the conditions necessary for boundedness as in \cite%
{narcowich:2006}.
\end{enumerate}
In all the three cases, we have therefore
\begin{equation}
\bigl\vert U^{(j)}_{m}( \cos \theta ) \bigr\vert \leq \frac{V_{\xi ,\kappa_\lambda}%
}{4\pi^2 \varepsilon_j ^{2}}\int_{\theta }^{\pi }\frac{1}{1+\bigl\vert \frac{%
\phi }{\varepsilon_j }\bigr\vert ^{\xi }}\frac{d\phi }{\sqrt{\cos \theta
-\cos \phi }} 
\spcend .
\end{equation}
According to \cite{narcowich:2006}, we obtain 
\begin{equation}
\bigl\vert U^{(j)}_{m}( \cos \theta ) \bigr\vert \leq \frac{C_{\xi
}/\varepsilon_j ^{2}}{( 1+\bigl\vert \frac{\theta }{\varepsilon_j }%
\bigr\vert ) ^{\xi }}
\spcend .
\end{equation}

%=============================================================================
\bibliographystyle{elsarticle-harv}
\bibliography{bib_journal_names_long,bib_myname,bib}
%=============================================================================

\end{document}

%% file: macros.tex
\newcommand{\eqn}[1]{(#1)}

\newcommand{\fig}[1]{Fig.~#1}

\newcommand{\sectn}[1]{Sec.~#1}

\newcommand{\appn}[1]{Appendix~#1}

\newcommand{\eg}{\mbox{\it e.g.}}
\newcommand{\ie}{\mbox{\it i.e.}}

\newcommand{\cf}{\mbox{\it cf.}}

\newcommand{\wmap}{{WMAP}}
\newcommand{\wmaptext}{{Wilkinson Microwave Anisotropy Probe}}

\newcommand{\healpix}{{\tt HEALPix}}

\newcommand{\sshtcode}{{\tt SSHT}}
\newcommand{\sothreecode}{{\tt SO3}}
\newcommand{\stwoletcode}{{\tt S2LET}}

\newcommand{\lambdaarch}{{LAMBDA}}
\newcommand{\lambdaarchtext}{{Legacy Archive for Microwave Background Data Analysis}}

\newcommand{\spcend}{\ensuremath{\:}}
\newcommand{\img}{\ensuremath{{\rm i}}}
\newcommand{\cconj}{\ensuremath{\ast}} 
 
\newcommand{\reals}{\ensuremath{\mathbb{R}}}
\newcommand{\realsnn}{\ensuremath{\mathbb{R^+}}}
\newcommand{\realsnz}{\ensuremath{\mathbb{R}^{+}_{\ast}}}
\newcommand{\integers}{\ensuremath{\mathbb{Z}}}
\newcommand{\naturals}{\ensuremath{\mathbb{N}}}

\newcommand{\ltwo}{\ensuremath{\mathrm{L}^2}}
\newcommand{\sphere}{\ensuremath{{\mathbb{S}^2}}}
\newcommand{\sothree}{\ensuremath{{\mathrm{SO}(3)}}}

\newcommand{\vect}[1]{\ensuremath{\mbox{\boldmath ${#1}$}}}

\newcommand{\dx}{\ensuremath{\mathrm{\,d}}}
\newcommand{\dmu}[1]{\ensuremath{\dx \Omega(#1)}}

\newcommand{\deul}[1]{\ensuremath{\dx \varrho(#1)}}

\newcommand{\innerp}[2]{\ensuremath{\langle {#1},\: {#2} \rangle}}

\newcommand{\sa}{\ensuremath{\omega}}
\newcommand{\saa}{\ensuremath{\theta}}
\newcommand{\sab}{\ensuremath{\varphi}}
\newcommand{\sas}{\ensuremath{\saa, \sab}}
\newcommand{\eul}{\ensuremath{\mathbf{\rho}}}
\newcommand{\euls}{\ensuremath{\eula, \eulb, \eulc}}
\newcommand{\eula}{\ensuremath{\alpha}}
\newcommand{\eulb}{\ensuremath{\beta}}
\newcommand{\eulc}{\ensuremath{\gamma}}

\newcommand{\eulci}{\ensuremath{g}}

\newcommand{\eulciang}{\ensuremath{\eulc_\eulci}}
\newcommand{\el}{\ensuremath{\ell}}
\newcommand{\m}{\ensuremath{m}}
\newcommand{\n}{\ensuremath{n}}

\newcommand{\spin}{\ensuremath{s}}

\newcommand{\elmax}{\ensuremath{{L}}}
\newcommand{\mmax}{\ensuremath{{M}}}
\newcommand{\nmax}{\ensuremath{{N}}}

\newcommand{\p}{\ensuremath{^\prime}}

\newcommand{\kron}[2]{\ensuremath{\delta_{{#1}{#2}}}}

\renewcommand{\exp}[1]{\ensuremath{{\rm e}^{#1}}}
\newcommand{\shfarg}[3]{\ensuremath{Y_{#1#2}({#3})}}
\newcommand{\shfargc}[3]{\ensuremath{Y_{#1#2}^\cconj({#3})}}

\newcommand{\sshfarg}[4]{\ensuremath{{{}_{#4} Y_{#1#2}({#3})}}}

\newcommand{\shf}[2]{\ensuremath{Y_{#1#2}}}

\newcommand{\shc}[3]{\ensuremath{{#1}_{{#2}{#3}}}}
\newcommand{\shcc}[3]{\ensuremath{{#1}_{{#2}{#3}}^\cconj}}

\newcommand{\sshf}[3]{\ensuremath{{{}_{#3} Y_{#1#2}}}}

\newcommand{\leg}[2]{\ensuremath{P_{{#1}}({#2})}}
\newcommand{\aleg}[3]{\ensuremath{P_{#1}^{#2}({#3})}}

\newcommand{\dmatbig}{\ensuremath{D}}
\newcommand{\Dlmn}{\ensuremath{ \dmatbig_{\m\n}^{\el} }}

\newcommand{\Dlmnp}{\ensuremath{ \dmatbig_{\m\n}^{\el}(\eul) }}
\newcommand{\Dlmnpc}{\ensuremath{ \dmatbig_{\m\n}^{\el\cconj}(\eul) }}

\newcommand{\dmatsmall}{\ensuremath{d}}
\newcommand{\dlmn}{\ensuremath{ \dmatsmall_{\m\n}^{\el} }}

\newcommand{\rot}{\ensuremath{\mathcal{R}}}
\newcommand{\rotarg}[1]{\ensuremath{\mathcal{R_{#1}}}}

\newcommand{\rotmatarg}[1]{\ensuremath{\mathbf{R}_{#1}}}

\newcommand{\spinup}{\ensuremath{\eth}}
\newcommand{\spindown}{\ensuremath{\bar{\eth}}}

\newcommand{\f}{\ensuremath{f}}

\newcommand{\wav}{\ensuremath{\psi}}
\newcommand{\swav}{\ensuremath{{}_\spin\psi}}

\newcommand{\wavs}{\ensuremath{\Phi}}

\newcommand{\wcoeff}{\ensuremath{W}}
\newcommand{\scoeff}{\ensuremath{W}}
\newcommand{\wscale}{\ensuremath{j}}
\newcommand{\wscalemax}{\ensuremath{J}}
\newcommand{\wscalemin}{\ensuremath{J_0}}

\newcommand{\dilparam}{\ensuremath{\alpha}}
\newcommand{\wavker}{\ensuremath{\kappa}}
\newcommand{\wavsteer}{\ensuremath{s}}
\newcommand{\steerinterp}{\ensuremath{z}}

\newcommand{\elmfact}{\ensuremath{\frac{(\el-\m)!}{(\el+\m)!}}}

\newcommand{\sumlm}{\ensuremath{\sum_{\el=0}^{\infty} \sum_{\m=-\el}^\el}}
\newcommand{\sumlmn}{\ensuremath{\sum_{\el=0}^{\infty} \sum_{\m=-\el}^\el} \sum_{\n=-\el}^\el}
\newcommand{\suml}{\ensuremath{\sum_{\el=0}^{\infty}}}
\newcommand{\sumlmbl}{\ensuremath{\sum_{\el=0}^{\elmax-1} \sum_{\m=-\el}^\el}}
\newcommand{\sumlmnbl}{\ensuremath{\sum_{\el=0}^{\elmax-1} \sum_{\m=-\el}^\el} \sum_{\n=-\el}^\el}

\newcommand{\summ}{\ensuremath{\sum_{\m=-\el}^\el}}

\newcommand{\sumn}{\ensuremath{\sum_{\n=-\el}^\el}}

\newcommand{\order}{\ensuremath{\mathcal{O}}}

\newcommand{\noisecl}{\ensuremath{C}}